\documentclass[alpha-refs]{wiley-article}

\usepackage[T1]{fontenc}
\usepackage{lmodern}
\usepackage{fix-cm}

\usepackage{graphicx}
\usepackage[space]{grffile}
\usepackage{latexsym}
\usepackage{textcomp}
\usepackage{longtable}
\usepackage{tabulary}
\usepackage{booktabs,array,multirow}
\usepackage{amsfonts,amsmath,amssymb}
\usepackage{placeins}
\usepackage{url}
\usepackage{hyperref}
\hypersetup{colorlinks=false,pdfborder={0 0 0}}
\usepackage{etoolbox}
\makeatletter

\newif\iflatexml\latexmlfalse

\AtBeginDocument{\DeclareGraphicsExtensions{.pdf,.PDF,.eps,.EPS,.png,.PNG,.tif,.TIF,.jpg,.JPG,.jpeg,.JPEG}}

\usepackage[utf8]{inputenc}
\usepackage[english]{babel}
\usepackage{siunitx}

\usepackage{hyperref}       
\usepackage{url}           
\usepackage{booktabs}      
\usepackage{amsfonts}       
\usepackage{nicefrac}      
\usepackage{microtype}      
\usepackage{lipsum}		
\usepackage{graphicx}
\usepackage{doi}
\usepackage{csquotes}
\usepackage{bm}
\usepackage{dsfont}
\usepackage{amsmath,amssymb,amsfonts}
\usepackage{easyReview}
\usepackage{multirow}
\usepackage{graphicx}
\usepackage{rotating}
\usepackage{tikz}
\usepackage{array}
\usepackage{colortbl}
\usepackage{lineno}
\usepackage{listings}
\usepackage{algorithm} 
\usepackage{algorithmicx} 
\usepackage{algpseudocode} 
\usepackage{adjustbox}
\usepackage[most]{tcolorbox}
\usepackage{longtable}
\usepackage{placeins}
\usepackage{enumitem}

\usepackage{hyperref}
\usepackage[nameinlink]{cleveref}
\usepackage{nameref}

\usepackage{xcolor}
\usepackage{hyperref}
\usepackage{cleveref}

\newcounter{myexample}

\makeatletter
\newenvironment{myexample}[1][]%
{%
  \refstepcounter{myexample}%
  \if\relax\detokenize{#1}\relax
    \def\@currentlabelname{Example \themyexample}%
  \else
    \def\@currentlabelname{#1}%
  \fi
  \par\vspace{3pt}%
  \noindent
  {\bfseries
    Example \themyexample%
    \if\relax\detokenize{#1}\relax
    \else : #1%
    \fi
    .%
  }%
  \hspace{0.5em}%
  \normalfont
}%
{%
  \par\vspace{3pt}%
}
\makeatother

\crefname{myexample}{Example}{Examples}
\Crefname{myexample}{Example}{Examples}

\crefformat{myexample}{\textbf{Example~#2#1#3:}}

\usepackage{lineno}

\paperfield{}
\abbrevs{REM, relational event model; US, United States; AIC, Akaike Information Criterion; FR, first record; MLE, Maximum Likelihood Estimation; PCH, Piecewise-Constant Hazard; NCC, Nested case-control; GAM, generalized additive model; GOF, Goodness-of-fit; WTC, World Trade Center.}
    
\corraddress{Martina Boschi, Faculty of Informatics, Università della Svizzera italiana, Lugano, 6900, Switzerland}
\corremail{martina.boschi@usi.ch}
\fundinginfo{Swiss National Science Foundation, 

Grant Number: 192549, 240065}

\papertype{Tutorial}

\title{Introduction to Relational Event Modelling}
\author[1]{Martina Boschi}
\author[1]{Ernst C. Wit}
\affil[1]{Università della Svizzera italiana}

\runningauthor{Boschi and Wit}

\iflatexml
\else
\makeatletter

\DeclareFontFamily{OMX}{cmex}{}
\DeclareFontShape{OMX}{cmex}{m}{n}{<-7.5>cmex7<7.5-8.5>cmex8<8.5->cmex10}{}
\DeclareSymbolFont{CMlargesymbols}{OMX}{cmex}{m}{n}
\DeclareSymbolFont{CMoperators}{OT1}{cmr}{m}{n}

\DeclareMathSymbol{\sumop}{\mathop}{CMlargesymbols}{"50}
\renewcommand{\sum}{\DOTSB\sumop\slimits@}

\DeclareMathDelimiter{(}{\mathopen} {CMoperators}{"28}{CMlargesymbols}{"00}
\DeclareMathDelimiter{)}{\mathclose}{CMoperators}{"29}{CMlargesymbols}{"01}
\DeclareMathDelimiter{[}{\mathopen} {CMoperators}{"5B}{CMlargesymbols}{"02}
\DeclareMathDelimiter{]}{\mathclose}{CMoperators}{"5D}{CMlargesymbols}{"03}

\DeclareMathSymbol{\intopCM}{\mathop}{CMlargesymbols}{"52}
\renewcommand{\int}{\DOTSI\intopCM\ilimits@}

\DeclareSymbolFont{CMoperators}{OT1}{cmr}{m}{n}
\DeclareMathSymbol{+}{\mathbin}{CMoperators}{"2B}

\DeclareMathSymbol{\prodop}{\mathop}{CMlargesymbols}{"51}
\renewcommand{\prod}{\DOTSB\prodop\slimits@}

\DeclareSymbolFont{CMoperators}{OT1}{cmr}{m}{n}

\DeclareMathAccent{\dot}{\mathalpha}{CMoperators}{"5F}
\DeclareMathAccent{\ddot}{\mathalpha}{CMoperators}{"7F}

\makeatother

\begin{document}

\maketitle

\begin{abstract}
Interactions and time shape many aspects of life. Everyday activities -- like conversations, emails, money transfers, citations, and even acts of violence -- are relational events: interactions between a sender and a receiver at a specific moment. At the intersection of event-history analysis and network modelling, relational event models (REMs) offer a powerful framework for studying when and why these events occur. Recent advances have made it possible to express REMs as generalized additive models, allowing researchers to capture complex, non-linear patterns over time.

A hands-on guide to REMs is still missing. This tutorial fills that gap. It provides a practical introduction to REMs, incorporating the latest developments in the field. It demonstrates how to simulate synthetic relational-event data and walks through several empirical applications, comparing different modeling and inference strategies.

By bringing together theory, simulation, and application, this tutorial lowers the barrier to entry and makes REMs a more accessible and practical tool.

\textbf{Keywords} --- relational events, relational event models, dynamic networks, generalized additive models, tutorial

\end{abstract}   

\newpage
\section{Introduction}\label{sec:introduction}
With recent advances in real-time data recording, the availability of temporal relational data has increased significantly. Examples include communication exchanges, citation processes, patient transfers between hospitals, and other interactions in social networks. The temporal component of relational data is reflected in the way relationships between entities appear, change, or disappear. This highlights the dynamic nature of the underlying relational processes.

Relational data take different forms. Relations may describe \textit{states}, where connections persist over a period of time, or instantaneous \textit{events} \citep{butts2023relational}. For example, consider a group of students in a class. As a relational state, we might observe friendship ties among individuals, while as a relational event, we might record the messages they exchange on a social platform. The former allows us to study the persistence of relationships between individuals, whereas the latter captures time-stamped instantaneous interactions. This paper focuses specifically on the latter.
Examples include a wide range of interactions involving different types of actors: from people sending emails or making phone calls, to connecting or following each other on social media, to users editing or commenting on articles in online platforms \citep{juozaitiene2024nodal, lerner2020reliability, uzaheta2023random}. Animal interactions \citep{tranmer2015using}, species dispersion processes \citep{boschi2025mixed}, financial transactions (such as money transfers between bank accounts \citep{bianchi2023ties}), patent citations \citep{filippi2023drivers}, and co-authorship in scientific research \citep{lerner2023micro} have all been studied using a relational event perspective.

Over the past two decades, research on relational event modelling has not only led to a wide range of applications but also several advancements beyond the original formulation by \citet{butts2008relational}. Originally conceived within the framework of event history models, \textit{relational event models} (REMs) serve to describe the rate of a relational event as a function of past events and other external information \citep{bianchi2024relational}. In this tutorial, we use the terms ``rate,'' ``intensity,'' and ``hazard'' interchangeably. Initially considering only fixed linear effects, REMs can now include sources of heterogeneity \citep{uzaheta2023random}, allow time-varying and non-linear effects of statistics on the rate  \citep{boschi2025mixed, filippi2023drivers}, and integrate global effects common to all actors in the system \citep{lembo2025relational, amati2019some, stadtfeld2017dynamic}.

From a computational perspective, significant efforts have been dedicated to making REMs scalable, facilitating their application to datasets with millions of events \citep{lerner2020reliability, artico2023fast, filippi2024stochastic, filippi2024modeling}. These advancements collectively contribute to the robustness, applicability, and efficiency of relational event modelling techniques, making them suitable for answering practical questions across diverse domains, such as communication \citep{perry2013point}, ecology \citep{juozaitiene2023analysing}, healthcare \citep{vu2017relational}, and political science \citep{brandenberger2019predicting}. 

A recent essay \citep{butts2023relational} and review \citep{bianchi2024relational} about REMs provide detailed technical insights, explore various applications, and highlight future directions for the development and use of REMs. \citet{butts2017relational} offers a practical guide to the relational event model, though it remains limited to the original linear formulation. \textbf{This tutorial aims to introduce readers to the basics of relational event modelling and equip them with skills to use flexible model formulations in practice}. We will focus on both monadic and bipartite
networks, but always considering only dyadic relational events. Relational event models have been extended to more complex scenarios, for example, to situations in which multiple actors are simultaneously involved in a single relational event. Such events are referred to as \textit{relational hyperevents}, involving  either multiple interacting entities without distinct sender and receiver roles \citep{lerner2021dynamic} or groups of senders interacting with groups of receivers \citet{lerner2025relational}. Another extension involves the introduction of multiple event types. In signed networks \citep{harary1953notion} -- where edges are interpreted as positive (e.g. like) or negative (e.g. dislike) -- the sign can be understood as an event type. \textit{Multiplex networks} \citep{lerner2017third} generalize this idea to more than two event types. \textit{Stratification} \citep{vu2015relational} can be used to infer parameters in such multiplex networks. The above extensions are mentioned but not discussed in detail in this tutorial paper. By the end of this tutorial, readers will understand what insights REMs can provide, when to use them, and how to evaluate their performance effectively.

The rest of this tutorial is organized as follows: Section \ref{sec:overview} offers the motivation for using relational event models, together with a brief overview of key tools for relational event modelling and model fitting. 
Section \ref{sec:modeling} introduces the basic formulation of relational event models and examines their main components. Section \ref{sec:inference} provides a concise overview of likelihood-based approaches used for parameter estimation. Key methodological and computational advances are also highlighted at the end of the respective sections. Section \ref{sec:applications} presents practical examples of REMs applied to both synthetic and real data, including three distinct empirical cases that demonstrate the model versatility. Finally, the tutorial concludes with a discussion of the main strengths of REMs, future research directions, and remaining challenges.

\section{When, Why and How to use Relational Event Models?}\label{sec:overview}

\textit{Network} representations of complex phenomena have become increasingly popular across scientific disciplines. Fields as diverse as biology \citep{jeong2000large}, neuroscience \citep{sporns2006human}, sociology \citep{granovetter1973strength}, communication \citep{onnela2007structure}, epidemiology \citep{pastor2001epidemic}, and computer science \citep{faloutsos1999power} have all exploited \textit{network} as a powerful tool for representing and analysing systems of interest.
Many problems, regardless of their domain-specific complexity, can be summarized into a common structure of entities (nodes) and the relationships or interactions between them (edges) -- that is, as a ``network'' -- and are thus subject to a set of laws and principles that govern networks. Understanding the network therefore provides a key to understanding the \textit{system} \citep{barabasi2016network-chapter1}. Furthermore, accounting for the ``network structure'' of a problem allows additional properties to emerge that would otherwise remain hidden if the \textit{relational nature} of the data is not properly accounted for.

\subsection{Motivation for using relational event models}

\textit{Network models} can be classified in several ways, but one distinction is particularly important: can the network of interest be considered \textit{static}, or is it more appropriate to represent it as \textit{dynamic}? There are many ways in which time might come into play: new entities might join the system, the attributes of entities or their connections might also change over time, or the nature of a connection might change (appear, disappear, or simply evolve). An accurate modelling framework should account for the \textit{timing} and \textit{duration} of each interaction \citep{barabasi2016network-chapter10}. Depending on what constitutes the observed data, how time enters into the definition of the phenomenon of interest, and how it is assumed to govern network dynamics, several temporal network models are available in the literature.

First, it is useful to distinguish between two common forms of \textit{temporal network data}. \textit{Longitudinal network data} \citep{snijders2001statistical} consist of repeated observations of the network at a finite set of time points. A classic example is a friendship network in a school class surveyed across two or more waves \citep{snijders2010introduction}. Such data allow us to determine whether a tie has persisted, emerged, or disappeared between successive observations, but they do not reveal the exact timing or sequence of the changes that occurred between observation times. In contrast, \textit{relational event data} \citep{butts2017relational} record each interaction between entities together with its timestamp, yielding an ordered sequence of events. An example is the private messages sent on an online social network \citep{zhu2022quantifying}. 

These two examples -- the friendship relations among students and the messages they exchange -- illustrate not only two different forms of temporal data, but also two fundamentally different types of connection. Friendships are an example of \textit{relational states}: they persist over an interval of time, i.e., they have a \textit{duration}. By contrast, sending a message is a \textit{relational event}: it is an \textit{instantaneous} interaction occurring at a specific point in time. It is thus important to understand which of these two types of connections best characterizes the phenomenon of interest. It is worth noting, however, that relational events are general enough to also represent relational states: by considering two event types, \textit{appearance} and \textit{dissolution}, one can model the beginning of a relational state as an appearance event and its end as a dissolution event \citep{juozaitiene2026inference}.

Another important distinction concerns how time is represented and how network dynamics are modelled. In \textit{discrete-time} models, which are typically applied to longitudinal network data, the network is observed at a finite sequence of time points, and its evolution is described through transitions between successive observations. An example is the \textit{Temporal Exponential Random Graph Model} \citep{hanneke2010discrete}, in which the probability of observing a network at a given time point is modelled conditional on one or more previous network states. 
Discrete-time models may also assume that the observed network is generated by an underlying \textit{latent process} evolving over time as a function of its previous state(s), where this latent state may take the form of either a \textit{latent position} \citep{sarkar2005dynamic} or a \textit{latent category} \citep{yang2011detecting}. These models may not be appropriate if the temporal resolution of the phenomenon is much finer than that of the observed time points. When this is the case, \textit{continuous-time} network models should be considered instead. In this kind of model, the network can change \textit{at any} instant, and so the timing of these changes -- unlike in discrete-time models -- can be modelled explicitly.

Several continuous-time models have been proposed and, as before, one can choose among them depending on the type of observed data and the underlying assumptions that are most compatible with the phenomenon of interest. \textit{Stochastic Actor-Oriented Models} \citep{snijders1996stochastic, snijders2001statistical} have been developed for longitudinal network data; however, they rely on an underlying continuous-time formulation: network snapshots are assumed to result from a sequential revision of choices made by the actors (the nodes of the network) according to their objective function. In contrast, in \textit{Hawkes processes} \citep{hawkes1971spectra, fox2016modeling} events occur in continuous time through a self-exciting intensity function. Like Hawkes models, relational event models \citep{butts2008relational, perry2013point, bianchi2024relational} model instantaneous interactions among entities in the network as realizations of a continuous-time process with a particular intensity function. However, rather than focusing on self-excitation, REMs express the intensity as a function of exogenous \citep{juozaitiene2023analysing}, global \citep{lembo2025relational}, and endogenous covariates \citep{juozaitiene2024s}, tracking the evolving history of the network in a flexible and computationally tractable way \citep{boschi2025mixed, filippi2024stochastic}.

\begin{tcolorbox}[colback=white!10,
colframe=black,
coltitle=black,
colbacktitle=yellow!40,
fonttitle=\bfseries,
coltext=black,
boxrule=0.5pt,
arc=2pt,
left=6pt,
right=6pt,
top=4pt,
bottom=4pt,
title=When to use relational event models?]
\textbf{Relational event models are designed to analyse sequences of interactions among entities in a network occurring in continuous time}. When interactions are dyadic, involving a sender and a receiver, dyadic relational event models are appropriate \citep{bianchi2024relational}. For interactions involving more than two actors, relational hyper event models provide the suitable extension \citep{lerner2021dynamic}.
\end{tcolorbox}

REM formulation to date has been developed with the primary goal of modelling instantaneous interactions between dyads. Consequently, relational event models are particularly well suited to addressing \textit{micro-dynamic} questions. For instance, they might examine whether and how quickly a citation to a paper tends to be followed by related citations in a scientific field. Although these mechanisms operate at the level of individual events, relational event studies can also speak to \textit{network-level} questions \citep{butts2023relational}, in line with traditional social network research. For example, one might ask whether collaborations among scientists tend to cluster within existing research groups. Furthermore, as anticipated, it is possible to generalize relational event models to more traditional social network settings by interpreting the formation and dissolution of ties as relational events, thereby representing the evolution of network states through an underlying sequence of events \citep{juozaitiene2026inference}. 

\begin{tcolorbox}[colback=white!10,
colframe=black,
coltitle=black,
colbacktitle=yellow!40,
fonttitle=\bfseries,
coltext=black,
boxrule=0.5pt,
arc=2pt,
left=6pt,
right=6pt,
top=4pt,
bottom=4pt,
title=Why use relational event models?]
\textbf{Relational event models enables to understand direction, strength, and significance of features and mechanisms driving the occurrence of events.} To this end, REMs incorporate explanatory variables representing exogenous information on the entities, their connections, or the network itself, as well as endogenous statistics summarizing the history of past events. Their flexibility makes it possible to examine how the effects of these mechanisms vary over time or across contexts.
\end{tcolorbox}

\subsection{How to use relational event models: an essential overview}

In this section, we give a brief overview of relational event modelling, focusing on the main steps needed to apply it effectively. We cover its formulation, inference methods, validation, and interpretation. These topics will be explained in more detail in Sections \ref{sec:modeling} and \ref{sec:inference}. Our approach follows six key phases, involving collecting and preprocessing the data, calculating the network statistics, and the iterative process of proposing a model, performing inference, considering goodness-of-fit and selecting a final model, which then, in the final phase, gets interpreted. To illustrate these steps, we will use a simple example with synthetic data, where we aim to uncover the drivers of academic visits by researchers between United States (US) states. More detailed examples using both empirical and synthetic data will be presented in Section \ref{sec:applications}. 

\textbf{What is the main goal of this overview?} For readers new to REMs, this overview introduces key ideas, which will be explained in more detail in the following sections. For readers already familiar with REMs, this section presents the framework used in the tutorial and situates it within existing literature.

\subsubsection*{Phase 1: Collect and/or identify relational event data}\label{phase1}
In the context of relational event modelling, data take the form of a \textit{relational event sequence} referred to as a \textit{dynamic network}, where time-stamped edges in \(E\) denote relational events among nodes in \(V\). Regardless of their form, each interaction requires three key pieces of information: the \textit{time} \( t \) when a directed interaction occurs from a \textit{sender} \( s \in V \) to a \textit{receiver} \( r \in V\). These statistical units, denoted by \( e = \left(s, r, t \right) \), constitute an \textit{event sequence} \( E \), comprising \( n \) interactions \( \{ e_i, \quad i=1,\ldots,n \} \). For certain inferential tasks, exact time information may not be strictly necessary, but the temporal ordering of events must be known. 

In relational event modelling, the raw data come in the form of an event sequence. However, we then need to augment these data with additional information. To construct our augmented dataset, we consider, for each time point $t_i, i = 1,\dots,n$, the \textit{risk set} at that time, i.e., $\mathcal{R}_{t_i}$. The risk set is a potentially time-varying set of pairs of senders and receivers $\mathcal{R}_t \subseteq V \times V$ that are at risk of interacting at time $t$. At an event time $t_i$, $\mathcal{R}_{t_i}$ consists of the sender-receiver pair $\left(s_i,r_i\right)$ that occurs and potentially many pairs $\left(s^\ast,r^\ast \right)\in V\times V$ that do not occur. For each observed event, we then store in the dataset information about the event itself, $(s_i, r_i, t_i)$, and one representative dyad, $(s_i^\ast, r_i^\ast)$, sampled randomly from the risk set. This dyad -- unlike the one involved in the observed event -- could have interacted but did not: this is why we refer to $(s_i^\ast, r_i^\ast, t_i)$ as a ``\textit{non-event}''. The reason for doing so is that it will allow us to consider more complex models, on the one hand, while at the same time simplifying and speeding up inference, as will be explained in detail later. An example of an augmented dataset, based on synthetic data, is shown in Table~\ref{tab:synt-RCCD}. Each relational event represents one or more academics working at a university in a US state (denoted as the origin \(s\)), visiting a university in another US state (denoted as the destination \(r\)), on a specific day \(t\).
\begin{table}[h]
\centering\begingroup\fontsize{9}{11}\selectfont
\resizebox{\linewidth}{!}{
\begin{tabular}[t]{|l|l|l|l|l|}
\hline
\textbf{Time} & \textbf{Origin (Event)} & \textbf{Destination (Event)} & \textbf{Origin (Non-Event)} & \textbf{Destination (Non-Event)}\\
\hline
... & ... & ... & ... & ...\\
21.12.2023 & New York & Texas & Texas & Virginia\\
... & ... & ... & ... & \vphantom{1} ...\\
17.01.2024 & Iowa & Indiana & Indiana & Florida\\
17.01.2024 & Vermont & Washington & South Carolina & Maryland\\
\hline
17.01.2024 & Colorado & Rhode Island & Nevada & Alabama\\
18.01.2024 & West Virginia & Texas & Washington & Washington\\
19.01.2024 & Maryland & Guam & Wisconsin & New Jersey\\
19.01.2024 & Arizona & Nebraska & Commonwealth of the Northern Mariana Islands & West Virginia\\
19.01.2024 & New York & Hawaii & Florida & Florida\\
\hline
19.01.2024 & Connecticut & Wisconsin & Guam & Florida\\
19.01.2024 & New York & Texas & Maryland & Nebraska\\
20.01.2024 & Florida & North Carolina & Oklahoma & Tennessee\\
20.01.2024 & Washington & Vermont & Minnesota & Louisiana\\
... & ... & ... & ... & ...\\
\hline
\end{tabular}}
\endgroup{}
\caption[Synthetic (augmented) dataset showing academic collaborations between universities in different US states.]{\label{tab:synt-RCCD} \textbf{Synthetic (augmented) dataset showing academic collaborations between universities in different US states}. The first three columns list observed events with their time, origin state, and destination state. The fourth and fifth columns show, for each observed event, a randomly sampled dyad that could have observed but did not.}
\end{table}

Given the temporal ordering of the events, it is possible to evaluate at each time point $t_i$ the \textit{prior history} $\mathcal{H}_{t_i^-}$, consisting of all information in the system up to $t_i$, including prior events and additional \textit{exogenous information}. 

\subsubsection*{Phase 2: Build network statistics}\label{phase2}

Continuing with the construction of the augmented dataset, we determine the values of the covariates at the event times $t_i$. For \textit{exogenous covariates}, this comes from external information, whereas for \textit{endogenous covariates}, it is computed as an explicit function of the event history $\mathcal{H}_{t_i^-}$. Traditional \textit{network statistics}, such as reciprocity and sender out-degree, are typical examples of endogeneity, but so are more complicated functions involving both past relational events and external covariates. Given the large variety as well as the importance of potential endogenous covariates as drivers for dynamic networks, Section~\ref{subsec:effects} is entirely dedicated to this topic. In this example, where the event is a visit by academic staff from state $s$ to a university in state $r$, we consider a dyadic statistic, \texttt{Reciprocity}, $x_{sr}^{\text{rec}}(t)$, which counts the number of events in the opposite direction that have already occurred. Additionally, we consider \texttt{Distance}, $x_{sr}^{\text{dist}}\left(t\right)$, representing a measure of physical distance between US state \(r\) and the last state visited by academic staff from state \(s\) before time $t$. Additional information about the distance computation can be found in the Supplementary Materials. \texttt{Reciprocity} is generated solely from relational event data, while \texttt{Distance} also uses exogenous information -- specifically, the distance between pairs of US states \citep{tigris, pebesma2023spatial, geosphereRpackage}. Table~\ref{tab:endogenous-simple} provides computational details, and Figure~\ref{tab:endogenous-data} shows covariate values for the events in Table~\ref{tab:synt-RCCD}.

\begin{table}[ht]		

	\centering

	\resizebox{0.9\textwidth}{!}{

		\begin{tabular}{|>{\columncolor{gray!10}}l|c|l|}

			\hline

			\rowcolor{white} 

			\textbf{Network Statistics} & \textbf{Representation} & \textbf{Computation}  \\

            \hline
			\texttt{Reciprocity} 
			& \begin{tikzpicture}[scale=0.45, baseline=(current bounding box.center)]
				\tikzset{every node/.style={draw=black, fill=white, shape=circle, inner sep=1.5pt, minimum size=0.35cm}}
				\node (a) at (0,0) [label={[label distance=-0.0cm]90:$s$}]{};
				\node (b) at (2.5,0)[label={[label distance=-0.0cm]90:$r$}]{};
				\draw[<-] (a.30)--(b.-210);
				\draw[dashed,<-] (b.210)--(a.-30);
			\end{tikzpicture}%
			& $x_{\textmd{sr}}^{\textmd{rec}}\left(t\right) = \sum_{t_i<t} \mathds{1}_{\{\left(s_i, r_i, t_i\right) = \left(r,s,t_i\right)\}} $\\
            
			\hline

			\texttt{Distance} 

			&

			\begin{tikzpicture}[scale=0.45, baseline=(current bounding box.center)]

				\tikzset{every node/.style={draw=black, fill=white, shape=circle, inner sep=1.5pt, minimum size=0.35cm}}

				\node (a) at (0.5,-1.5) [label={[label distance=-0.0cm]90:$s$}]{};

				\node (b) at (4,-1.5) [label={[label distance=-0.0cm]90:$r$}]{};

				\node (c) at (2,-0.5)[label={[label distance=-0.1cm]90:$r^{\textmd{last}}$}]{};

				\draw[dashed, ->] (a)--(b);

				\draw[->] (a)--(c);

				\draw[red, thick] (c)--(b);
			\end{tikzpicture}
			& 
			$\begin{aligned}
				& x_{sr}^{\textmd{dist}}\left(t\right) = \textmd{distance}\left( r, r^{\textmd{last}}(t) \right) \\
				& r^{\text{last}}\left(t\right) = r_i \iff i = \arg\max_{i'} \left\{ t_{i'} \mid t_{i'} < t \text{ and } \left(s_{i'}, r_{i'}, t_{i'}\right) = \left(s, r_{i'}, t_{i'}\right) \right\}
			\end{aligned}$
			\\
			
            \hline
		\end{tabular}
	}
    \vspace{1em}
	\caption[Definition and computation of network statistics.]{\label{tab:endogenous-simple} \textbf{Definition and computation of network statistics}. 
    \texttt{Reciprocity} counts past visits from academics from state \(r\) to state \(s\). \texttt{Distance} measures the distance between the receiver state and the last visited state by academics from the sender state.}
\end{table}

It is important to note that many endogenous covariates suffer from temporal boundary problems. For example, by definition, the reciprocity covariate is zero for each dyad at the beginning of the study. If the observation period does not correspond to the actual existence of the dynamic network itself, this value may potentially be incorrect. Similarly, if, among the observations, no researcher from a state $s$ has yet travelled anywhere, then the distance covariate $x_{sr}^{\textmd{dist}}$ is undefined and would require either some ad hoc imputation or, preferably, the removal of the entire relational event from the inference set. In this particular example, for each state we randomly sampled another state, assumed to be the last one visited.

\begin{table}[!h]
\centering
\begin{tabular}[t]{|l|l|l|l|l|l|}
\hline
\textbf{Time} & \textbf{Origin (Event)} & \textbf{Destination (Event)} & \textbf{Distance} & \textbf{Reciprocity} & \textbf{Weekday}\\
\hline
... & ... & ... & ... & ... & ...\\
21.12.2023 & \cellcolor[HTML]{ADD8E6}{New York} & \cellcolor[HTML]{FFFACD}{Texas} & 4.66 & 0 & 1\\
... & ... & ... & ... & ... & ...\\
17.01.2024 & Iowa & Indiana & 4.73 & 0 & 1\\
17.01.2024 & \cellcolor[HTML]{FFE4E1}{Vermont} & \cellcolor[HTML]{FFE4E1}{Washington} & 2.21 & 0 & 1\\
\hline
17.01.2024 & Colorado & Rhode Island & 0.68 & 1 & 1\\
18.01.2024 & West Virginia & Texas & 4.66 & 0 & 1\\
19.01.2024 & Maryland & Guam & 3.93 & 0 & 1\\
19.01.2024 & Arizona & Nebraska & 2.47 & 0 & 1\\
19.01.2024 & \cellcolor[HTML]{ADD8E6}{New York} & \cellcolor[HTML]{FFFACD}{Hawaii} & \cellcolor[HTML]{FFFACD}{3.93} & 0 & 1\\
\hline
19.01.2024 & Connecticut & Wisconsin & 4.16 & 0 & 1\\
19.01.2024 & \cellcolor[HTML]{ADD8E6}{New York} & \cellcolor[HTML]{FFFACD}{Texas} & \cellcolor[HTML]{FFFACD}{3.93} & 0 & 1\\
20.01.2024 & Florida & North Carolina & 2.14 & 0 & 1\\
20.01.2024 & \cellcolor[HTML]{FFE4E1}{Washington} & \cellcolor[HTML]{FFE4E1}{Vermont} & 2.54 & \cellcolor[HTML]{FFE4E1}{1} & 1\\
... & ... & ... & ... & ... & ...\\
\hline
\end{tabular}
\caption[Network statistics computation in practice.]{\label{tab:endogenous-data}\textbf{Network statistics computation in practice}. \texttt{Reciprocity} value is 1 on January 20th, indicating that a reciprocal event occurred earlier, on January 17th. \texttt{Distance} value for the event from New York to Texas on January 19th is 3.93, since the last state visited by New York academics at that point was Hawaii earlier on the same day, and distance(Texas, Hawaii) = 3.93. The same dyad may take a different value for \texttt{Distance}, for instance on December 21st. This is because the previously visited state at that time was farther away than Hawaii. The \texttt{Distance} value is equal to 0 when the last visited state is a neighbour of the current one. Information about whether the time corresponds to a weekday or not is also available. This is an example of a global covariate. Methods to estimate the impact of global covariates will be explored in Section~\ref{subsec:shifting}.}
\end{table}

\subsubsection*{Phase 3: Modelling relational events}\label{phase3} 

The core idea of REMs is to understand why certain events occur and others do not at a given time. This involves modelling the \textit{rate} of experiencing an event $(s,r,t)$ given the previous event history $\mathcal{H}_{t^-}$, which is assumed to contain all information necessary for describing the interaction to occur. In this simple synthetic example, the rate, denoted as $\lambda_{sr}(t)$, is assumed to have the following form:
\begin{equation}\label{eq:model-overview}
	\lambda_{sr}(t) = I_{sr}(t) \times \lambda_{0}(t) \times \exp{\Big\{ \beta^\textmd{rec} \cdot x_{\textmd{sr}}^{\textmd{rec}}(t) + f \left( x_{sr}^{\textmd{dist}}(t)\right)\Big\}},
\end{equation}
where $I_{sr}(t)$  is the \textit{risk indicator} indicating which events are at risk, $\lambda_{0}(t)$ is the \textit{baseline hazard} of occurrence (assumed to be equal across dyads in the system), while the third term captures the effect of the covariates on the event rate. We assume here a linear effect for reciprocity and a non-linear effect for distance. 

What does \(\beta^\textmd{rec}\) quantify? Consider the rate function in Equation~\eqref{eq:model-overview} and suppose that the value of reciprocity covariate increases by one unit, yielding
\(
\lambda^\ast_{sr}(t)
=
I_{sr}(t)\lambda_{0}(t)
\exp\!\left\{
\beta^{\textmd{rec}}\bigl(x_{sr}^{\textmd{rec}}(t)+1\bigr)
+
f\!\left(x_{sr}^{\textmd{dist}}(t)\right)
\right\}.
\)
The ratio between the two rates (commonly known as \textit{hazard ratio}) is, then,
\[
\frac{\lambda^\ast_{sr}(t)}{\lambda_{sr}(t)}
=
\exp\!\left(\beta^{\textmd{rec}}\right),
\quad \iff \quad 
\log\lambda^\ast_{sr}(t)-\log\lambda_{sr}(t)
=
\beta^{\textmd{rec}}.
\]
Hence, a one-unit increase in the reciprocity statistic multiplies the event rate by \(\exp(\beta^{\textmd{rec}})\). If \(\beta^{\textmd{rec}}\) is larger than zero, then \(\exp(\beta^{\textmd{rec}})\) is larger than one, indicating an increase in the event rate. Conversely, if \(\beta^{\textmd{rec}}\) is smaller than zero, \(\exp(\beta^{\textmd{rec}})\) is smaller than one, indicating a decrease in the event rate. 

What about \(f \left( x_{sr}^{\textmd{dist}}(t)\right)\)? To interpret the effect of the smooth term, consider two rate functions that differ only in the value of the distance covariate, namely
\(x_{sr}^{\mathrm{dist}}(t)\) and \(x_{sr}^{\mathrm{dist}}(t)+h\). The corresponding hazard ratio is
\[
\frac{\lambda_{sr}^\ast(t)}{\lambda_{sr}(t)}
=
\exp\!\left\{
f\!\left(x_{sr}^{\mathrm{dist}}(t)+h\right)
-
f\!\left(x_{sr}^{\mathrm{dist}}(t)\right)
\right\},
\]
where all remaining model components cancel in the ratio. Taking logarithms, dividing by \(h\), and letting \(h\to0\) gives
\[
\lim_{h\to0}
\frac{\log\lambda_{sr}^\ast(t)-\log\lambda_{sr}(t)}{h}
=
\lim_{h\to0}
\frac{
f\!\left(x_{sr}^{\mathrm{dist}}(t)+h\right)
-
f\!\left(x_{sr}^{\mathrm{dist}}(t)\right)
}{h}
=
f'\!\left(x_{sr}^{\mathrm{dist}}(t)\right).
\]

Therefore, \(f'\!\left(x_{sr}^{\mathrm{dist}}(t)\right)\) quantifies the instantaneous change in the log event rate with respect to the covariate. A positive derivative indicates that increasing the covariate locally increases the event rate, whereas a negative derivative indicates that increasing the covariate locally decreases the event rate. Moreover, the magnitude of the derivative reflects how rapidly the event rate changes: large absolute values correspond to regions where the rate changes quickly, while values close to zero indicate that the rate is approximately constant with respect to the covariate.

In real data settings, the relationship between covariates and the log-rate is not straightforward and should be carefully examined. REMs allow other flexible model specifications that are discussed in Section \ref{subsec:mod-extensions}.

\subsubsection*{Phase 4: Perform likelihood-based estimation}\label{phase4}

\begin{figure}[ht]
    \centering
    \begin{minipage}{0.55\textwidth}
\resizebox{\linewidth}{!}{
\begin{tabular}[t]{llllllllll}
\hline
& time & rec\_ev & rec\_non\_ev & delta.rec & dist\_ev & dist\_non\_ev & id\_ev & id\_non\_ev & y\\
\hline
& ... & ... & ... & ... & ... & ... & ... & ... & ...\\
720 & 21.12.2023 & 0 & 0 & 0 & 4.66 & 2.18 & 1 & -1 & 1\\
& ... & ... & ... & ... & ... & ... & ... & ... & ...\\
795 & 17.01.2024 & 0 & 0 & 0 & 4.73 & 2.77 & 1 & -1 & 1\\
796 & 17.01.2024 & 0 & 0 & 0 & 2.21 & 0 & 1 & -1 & 1\\
\hline
797 & 17.01.2024 & 1 & 2 & -1 & 0.68 & 2.14 & 1 & -1 & 1\\
798 & 18.01.2024 & 0 & 0 & 0 & 4.66 & 3.1 & 1 & -1 & 1\\
799 & 19.01.2024 & 0 & 0 & 0 & 3.93 & 3.55 & 1 & -1 & 1\\
800 & 19.01.2024 & 0 & 1 & -1 & 2.47 & 0 & 1 & -1 & 1\\
801 & 19.01.2024 & 0 & 0 & 0 & 3.93 & 2.81 & 1 & -1 & 1\\
\hline
802 & 19.01.2024 & 0 & 1 & -1 & 4.16 & 0 & 1 & -1 & 1\\
803 & 19.01.2024 & 0 & 0 & 0 & 3.93 & 4.67 & 1 & -1 & 1\\
804 & 20.01.2024 & 0 & 0 & 0 & 2.14 & 1.98 & 1 & -1 & 1\\
805 & 20.01.2024 & 1 & 0 & 1 & 2.54 & 1.64 & 1 & -1 & 1\\
& ... & ... & ... & ... & ... & ... & ... & ... & ...\\
\hline
\end{tabular}}
    \end{minipage}
    \hfill
    \begin{minipage}{0.4\textwidth}
        \centering
        \includegraphics[width=\linewidth]{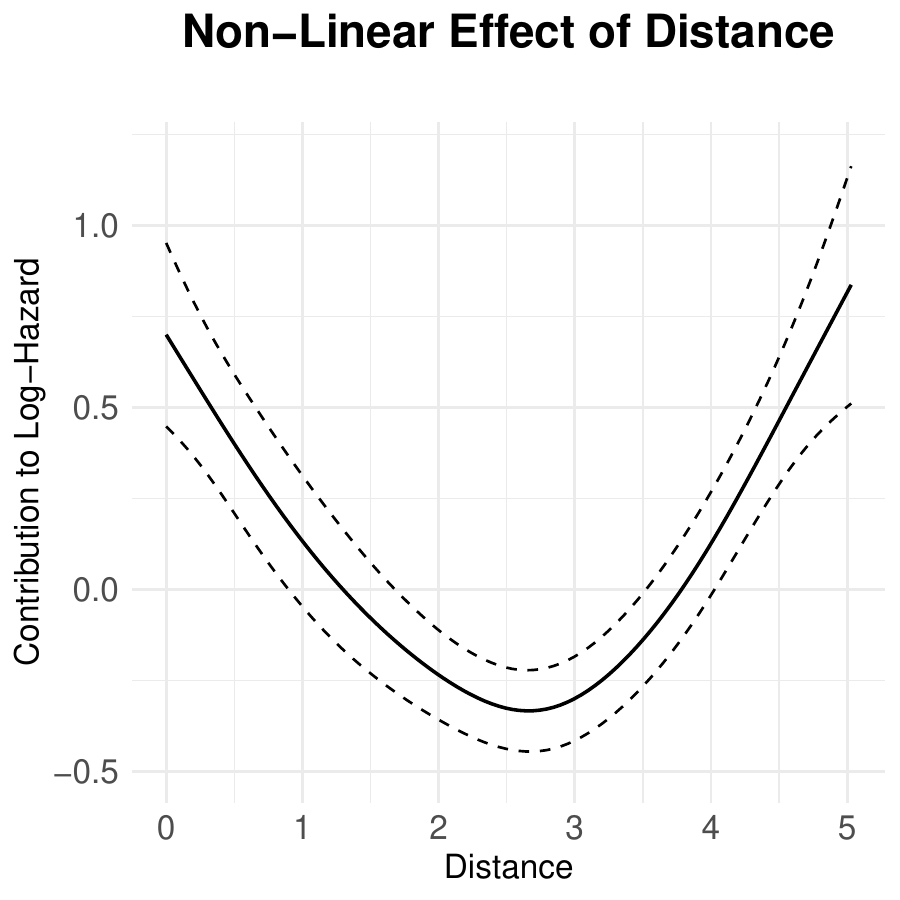}
    \end{minipage}
    % Row 2
    \vspace{0.5em}
    \begin{minipage}{\textwidth}
        \begin{center}
        \begin{tikzpicture}
            \node[draw, thick, rectangle, rounded corners=3pt, inner sep=3pt] (box) {
                \begin{minipage}{0.9\textwidth}
                    \centering
                    \begin{adjustbox}{max width=\textwidth}
                        \begin{lstlisting}[language=R]
library(mgcv)
dist_matrix <- cbind(ncc_data$dist_ev, ncc_data$dist_non_ev)
by_matrix <- cbind(ncc_data$id_ev, ncc_data$id_non_ev)
gam_fit <- gam(y ~ -1 + s(dist_matrix, by=by_matrix) + delta.rec,
               family="binomial"(link = "logit"), data = ncc_data)
plot(gam_fit)
                        \end{lstlisting}
                    \end{adjustbox}
                \end{minipage}
            };

            \node[draw, fill=blue!20, rounded corners=5pt, anchor=south west, font=\sffamily\bfseries\small, inner sep=4pt]
            at ([xshift=1em, yshift=-1em]box.north west) {R Code: REMs via logistic regression};
        \end{tikzpicture}
        \end{center}
    \end{minipage}
    \caption[Implementation of REM inference via logistic regression.]{\label{fig:distance_overview}\textbf{Implementation of REM inference via logistic regression}. \emph{Top Left}. Covariate columns of the augmented dataset, including values of the explanatory variable evaluated for both the event (\texttt{rec\_ev}, \texttt{dist\_ev}) and the corresponding non-event (\texttt{rec\_non\_ev}, \texttt{dist\_non\_ev}). For covariates assumed to have a linear effect, the dataset includes their difference (\texttt{delta.rec}). In cases where a covariate has a non-linear effect, two additional columns are included, with values set to 1 (\texttt{id\_ev}) and -1 (\texttt{id\_non\_ev}), respectively. The degenerate response variable (\texttt{y}) is fixed to 1. \emph{Top Right}. Estimated non-linear effect of distance on the log-rate of relational events. \emph{Bottom.} Code for fitting a logistic regression via \texttt{gam} in \texttt{R} package \texttt{mgcv} \citep{wood2011fast, wood2017generalized} with a smooth effect {\tt s()} of distance, using the prepared relational event data structure. The syntax with the matrices {\tt dist\_matrix} and {\tt by\_matrix} is such that it evaluates the difference of the smooth term for the event and non-event, $f(x_{sr}^{\textmd{dist}})-f(x_{s^*r^*}^{\textmd{dist}})$. More details about this logistic regression are provided in Section~\ref{subsec:general-inference}.}
\end{figure}

Inference is performed using \textit{maximum likelihood}. Likelihood-based techniques commonly used in the REMs literature are introduced in Section~\ref{sec:inference}. However, standard logistic regression is sufficient to infer the linear and non-linear effects of the covariates on the event log-rate (Section~\ref{subsubsec:logistic}). Figure~\ref{fig:distance_overview} illustrates how to construct the related dataset, how to perform the estimation procedure in \texttt{R}, and the fitted non-linear effect. 

\subsubsection*{Phase 5: Model comparison and goodness of fit}\label{phase5}

Based on the available information, multiple formulations of the model are conceivable, affording flexibility in both coefficients and covariates. When confronted with multiple model options, we can turn to information criteria such as the \textit{Akaike Information Criterion} (AIC), balancing model fit and complexity \citep{wit2012all}. 
Consider a list of candidate models $\mathbb{M} = \{\mathcal{M}_1,\ldots, \mathcal{M}_M\}$ with the estimates $(\bm{\hat\beta}^{(j)}, \bm{\hat f}^{(j)})$ for model $\mathcal{M}_j$. Because of the presence of smooth term $\bm{f}$, model selection is performed using the \textit{corrected conditional AIC} as follows:
\begin{equation}\label{eq:PLsampled-overview}
	\mathcal{M}_\textmd{best} = \arg \min_{\mathcal{M} \in \mathbb{M}} AIC(\mathcal{M}),  \quad  \mbox{ with } AIC(\mathcal{M}_j) = - 2 \log{\mathcal{L}\left(\hat{\bm{\beta}}^{(j)},{\bm \hat f}^{(j)}\right)} + 2 \hat{\kappa}_{\mathcal{M}_j}
\end{equation}
where \(\mathcal{L}(\hat{\bm{\beta}},{\bm \hat f})\) is the likelihood function computed in the previous phase, evaluated at the maximum likelihood estimate, and \(\hat{\kappa}_{\mathcal{M}}\) is an estimate of \textit{degrees of freedom} of the model \(\mathcal{M}\). More details can be found in Section \ref{subsec:selgof}. 

In the running example, we evaluate three candidate models: one incorporating only \texttt{Distance} non-linearly (AIC=$1304$), another focusing solely on \texttt{Reciprocity} (AIC=$1347$), and the third integrating both variables (AIC=$1268$) as in Equation \eqref{eq:model-overview}. With hindsight, this was expected because the synthetic data were generated from this model. In general, model selection procedures allow us to select the best model among a list of candidates. Nonetheless, while it may be the most favourable among the selected models, it is important to note that this does not guarantee its adequacy. In Section \ref{subsec:selgof}, techniques for \textit{model selection and diagnosis} are described. This includes goodness-of-fit tests, based on cumulative sums of martingale residuals, proposed by \citet{boschi2026goodness}. Goodness of fit for \texttt{Reciprocity} and \texttt{Distance} are both evaluated as adequate. 

\subsubsection*{Phase 6: Interpret the results}\label{phase6}

Once the model is validated, we can interpret the results. \textbf{Concerning linear effects}, we can examine the \textit{sign}, \textit{magnitude}, and \textit{statistical significance} of each estimated coefficient. In this context, the estimated coefficient for \texttt{Reciprocity} is positive (+0.62), indicating that the event rate increases as the number of previously reciprocated events grows. Specifically, a one-unit increase in the reciprocity statistic leads to an 86\% increase in the rate (the new rate is multiplied by \(\exp(0.62)=1.86\)).

\begin{tcolorbox}[colback=white,           
    colframe=black,          
    coltitle=black,            
    colbacktitle=red!30,    
    fonttitle=\bfseries,
    coltext=black,
    boxrule=0.5pt,
    arc=2pt,
    left=6pt,
    right=6pt,
    top=4pt,
    bottom=4pt,
    title=How to interpret a linear effect in a REM?]
    \textbf{The estimated coefficient \(\hat{\beta}\) represents the estimated change in the instantaneous log event rate associated with a one-unit increase in the corresponding covariate, all other model components being held constant}. Equivalently, \(\exp(\hat{\beta})\) is the multiplicative change in the event rate: values greater than one indicate an increase in the rate, whereas values smaller than one indicate a decrease.
\end{tcolorbox}

\textbf{Concerning non-linear effects}, Figure~\ref{fig:distance_overview} \emph{Top Right} shows the estimated effect of distance. It is important to note that the sign of this effect is not directly interpretable. This is because the absolute level of the smooth effect is not identifiable. Indeed, if \(f\!\left(x_{sr}^{\textmd{dist}}(t)\right)\) is a valid smooth effect, then \(f^\ast\!\left(x_{sr}^{\textmd{dist}}(t)\right) = f\!\left(x_{sr}^{\textmd{dist}}(t)\right)-c,\) for any constant \(c\), yields exactly the same model fit, as the constant can be absorbed into the baseline. The estimation procedure cannot distinguish between \(f\) and \(f^\ast\): by means of a centring constraint, only the \emph{shape} of the estimated function -- not its absolute level -- is identifiable and therefore interpretable. For example, two states $r^{\text{last}}$ and $r$ that are 3 distance units apart are separated by approximately $0.7 - (0.3) \approx 1$ additional log-hazard unit, as can be seen in Figure~\ref{fig:distance_overview}. This means that an academic researcher from state $s$ is $e^1=2.7$ times more likely to visit $r^{\text{last}}$ again rather than the state $r$ that is $3$ units further away.

\begin{tcolorbox}[colback=white,           
    colframe=black,          
    coltitle=black,            
    colbacktitle=red!30,    
    fonttitle=\bfseries,
    boxrule=0.5pt,
    arc=2pt,
    left=6pt,
    right=6pt,
    top=4pt,
    bottom=4pt,
    title=Interpreting estimated non-linear effects]
    The sign of an estimated non-linear effect -- such as the one shown in the Figure~\ref{fig:distance_overview} \emph{Top Right} -- cannot be interpreted because it is determined only up to an additive constant. Instead, relative changes in the effect are meaningful and indicate whether the rate increases or decreases as the curve rises or falls.
\end{tcolorbox}
	
\section{Statistical modelling of Relational Events}\label{sec:modeling}
Whenever an instantaneous directed interaction is initiated by a sender $s$ to a receiver $r$ at time $t$, it may be conceptualised as a relational event \citep{perry2013point} and mathematically represented as a tuple $(s, r, t)$. Relational events are the statistical units of the model explained in this tutorial -- or, using the words of \citet{butts2017relational}, \enquote{atomic units of social interaction}. Although we will be using a directed semantics throughout this tutorial, given the \textit{dyad-oriented focus} of our approach -- as opposed to an actor-oriented focus \citep{stadtfeld2017dynamic} -- all the methods are directly applicable to undirected dyadic interactions, such as an encounter between two people, or a contract signed by two economic actors. The only thing that changes are the semantic phrasing of sender and receiver, and the definition of relevant undirected network statistics,  as, e.g., reciprocity is not a relevant concept in an undirected interaction.

A relational event represents an occurrence -- something that has happened. To understand why certain events occur and to predict their recurrence, we first need to characterise them. A helpful way to do this is by distinguishing actual events from the many possibilities that could have occurred but did not. This highlights the need to differentiate (observed) events from non-events. A non-event, defined at the same time $t$ as the event of interest, is a tuple $(s^*, r^*, t)$. Sender $s$, non-sender $s^*$, receiver $r$, and non-receiver $r^*$ are entities capable of initiating or receiving an interaction at time $t$\footnote{Since a non-event differs by definition from an event, we require that $(s,r)$ differs from $(s^*,r^*)$. However, it is possible for the sender (or the receiver) of the event and non-event to coincide: an actor may have contacted (or been contacted by) one person, and not (by) another.}. Senders and receivers belong to the set of potential participants at time \(t\), $V_t$. If senders and receivers come from distinct groups, we can distinguish these as $V^S_t$ and $V^R_t$, respectively. We can thus distinguish between \textit{one-mode networks} where $V^S_t = V^R_t = V_t$ (where each actor in the system may initiate or receive an interaction) from \textit{bipartite networks} where $V^S_t$ and $V^R_t$ are disjoint and relational events intrinsically start from one type of node and go towards the other.
The set of entities in the system can change over the time window $[0, T]$. New entities (or dyads) may enter or leave the system, affecting the set of possible interactions. The risk set at time $t$, $\mathcal{R}_t$, captures these possibilities and is a subset of the Cartesian product $V^S_t \times V^R_t$.

Earlier dynamic network models, based on repeated static network measurements, relied on aggregating these occurrences into subintervals within the time frame \citep{snijders1996stochastic, robins2001random}. While such aggregation may be adequate when focusing on relational states and their evolution over time, it results in a significant loss of information when the instantaneous nature of events is available and crucial. First, the choices and assumptions required for time aggregation can significantly impact the results of the analysis \citep{butts2017relational}. Second, different sequences of relational events can potentially end up in the same network structure when time is aggregated, even though their differences may be relevant for determining differences in the generating process \citep{mulder2022using}.

\subsection{Counting process modelling for relational events}\label{subsec:cp-theory}

It is a statistician's natural inclination to count occurrences as they take place \citep{aalen2008survival}. 

In the context of relational events, assume the existence of a \textit{marked point process} $ \mathbb{P} = \{(t_i, (s_i, r_i)); \, i \ge 1\} $  generating time points \( t_i \) in the interval $[T_0,T]$  at which an interaction $(s_i \rightarrow r_i)$ occurs. We associate a \textit{counting process} \(\bm{N}\) with \(\mathbb{P}\). This process counts, for each mark \((s, r)\), the number of interactions \( s \rightarrow r \) that have occurred prior to each time point $t\in [T_0,T]$,
\begin{equation}\label{eq:counting}
	N_{sr}(t) = c_{sr} +\sum_{i \ge 1} 1{\{T_0 < t_i \le t, (s_i, r_i) = (s, r)\}},
\end{equation}
where $c_{sr} := N_{sr}(T_0)$ is the initial count before the process started. This is often assumed to be zero, but as shown in the Example~\ref{ex:alienspecies} below, this does not have to be the case. We make several assumptions.
First, events cannot occur simultaneously. Furthermore, \(N_{sr}\) is adapted with respect to the increasing history or \textit{filtration} \(\mathbb{H} = \{\mathcal{H}_t\}_{t \ge T_0}\), where \(\mathcal{H}_t\) is a sub-\(\sigma\) field generated by the events occurring until time \(t\). The fact that the process is adapted expresses that the history is generated by the counting process itself and possibly exogenous information as well \citep{aalen2008survival}. Finally, the counting process is well-defined in the sense that \(N_{sr}(t) < \infty \), almost surely. 

Under the assumptions above, due to the non-decreasing nature of the counting process, $N_{sr}$ is a continuous sub-martingale and can be therefore decomposed as a result of the \textit{Doob-Meyer theorem}:
\begin{equation}\label{eq:doob}
	M_{sr}(t) = N_{sr}(t)  - \Lambda_{sr}(t) = N_{sr}(t) -  \int_{T_0}^t \lambda_{sr}(u) \mathrm{d}u
\end{equation}
where $\Lambda_{sr}(t)$ is the \textit{predictable} part of the process $ N_{sr}(t) $ and $ M_{sr}(t) $ is a zero-mean \textit{martingale}. Given $ \mathcal{H}_{t^-} $, $\Lambda_{sr}(t)$ is a predictable and left-continuous process. The \textit{rate} $\lambda_{sr}(t)$ is defined as the derivative of  $\Lambda_{sr}(t)$, if it exists \citep{perry2013point, aalen2008survival}. In contrast, the martingale $M_{sr}(t)$ can be seen as the \textit{noisy} variation around the predictable part of the process.  

Intuitively, the \textit{intensity process} \(\lambda_{sr}(t)\) can be interpreted as the conditional ``risk'' of an event \(s \rightarrow r\) occurring at time \(t\), given the history of the process up to just before \(t\) \citep{daley2003introduction}. Equivalently, it can be viewed as the expected number of events of type \(s \rightarrow r\) per unit time (the rate), evaluated at time \(t\).

REMs are aimed at modelling the instantaneous rate of occurrence of a relational event involving $(s,r)$ at time $t$ as a function of (i) \textit{monadic covariates} $\bm{x}_{s}(t)$ or $\bm{x}_{r}(t)$, such as sender degree or receiver popularity, (ii) \textit{dyadic covariates} $\bm{x}_{sr}(t)$, such as  reciprocity, homophily but also triadic effects, and (iii) \textit{global covariates} $\bm{x}(t)$, such as the amount of precipitation. In a very general formulation, they can be expressed as follows:
\begin{eqnarray}
	\lambda_{sr}(t) &=& I_{sr}(t) \times \lambda_{0}(t) \times \exp\{f\left[\bm{x}_{sr}(t)\right]\} \label{eq:model} \\
	\lambda_0(t) &=& \lambda_0 \times \exp\{g_0(t) + g\left[\bm{x}(t)\right]\}\label{eq:baseline}
\end{eqnarray}
Equation \eqref{eq:model} involves three components: first, the edge-specific risk indicator $I_{sr}(t)$. It is equal to $1$ if the dyad $(s \rightarrow r)$ can occur at time $t$ and $0$ otherwise, i.e.  $I_{sr}(t) = \mathds{1} \{(s,r) \in \mathcal{R}_t\}$. Intuitively, the hazard of a relational event at time $t$ is equal to $0$ when it is currently impossible \citep{butts2017relational}.

Second, the risk of an interaction occurring depends on a baseline hazard. In the original formulation of REMs by \citet{butts2008relational}, the baseline hazard function was assumed to be constant, \(\lambda_{0}(t) =  \lambda_{0}\). However, it is preferable to allow \(\lambda_{0}(t)\) to vary over time, reflecting how the rate of interactions may change. Sources of heterogeneity in the baseline hazard may be governed by global, time-dependent covariate processes, denoted as \(\bm{x}(\cdot)\), which are adapted to the filtration $\mathbb{H}$. The effects of these \textit{global covariates} could be incorporated through a function $g\left[\bm{x}(t)\right]$. Under an \textit{additive model}, $g$ is expressed as the sum of smooth functions \(g[\bm{x}(t)] = \sum_{k=1}^{K_g} g_k[x_k(t)]\). Any remaining global time effects are captured by an additional smooth function, $g_0(t)$. When using this global-effect formulation of the baseline hazard, it is still necessary to include the constant term $\lambda_{0}$. Generally speaking, this term reflects the time unit used in the model formulation. For example, the data shown in Table \ref{tab:synt-RCCD} are expressed in days; if we were to transform time into hours, we would need to divide $\lambda_{0}$ by 24, but all the other effects would stay the same. 

We assume that Equation~\ref{eq:model} follows an additive structure as well, where \(f[\bm{x}_{sr}(t)] = \sum_{k=1}^{K_e} f_k[x_{sr,k}(t)]\) is expressed as the sum of several (potentially different) functions of the covariate processes in \(\bm{x}_{sr}(\cdot)\). These covariates are adapted to the filtration $\mathbb{H}$ and are therefore predictable. Section~\ref{subsec:effects} will describe the types of covariates typically used in REMs. In general, \( f[\bm{x}_{sr}(t)] \) captures the \textit{edge-specific risk determinants} of the model. While global determinants take the same value at time $t$ for any pair of sender and receiver, edge-specific determinants are specific for the considered dyad.

\begin{myexample}[Alien species invasions]\label{ex:alienspecies}
A practical example illustrating the building blocks of the model in Equation~\eqref{eq:model} is the spread of alien species \citep{seebens2021around}, where the relational event is a \textit{first record}. A first record (FR) is defined as the first year (\(t\)) in which a species (\(s\)) is detected in a region (\(r\)) where it is not native \citep{seebens2017no, seebens2018global}. As such, first records can be naturally represented as relational events. Using this perspective, \citet{juozaitiene2023analysing, boschi2025mixed} analyse first records of alien species recorded between 1880 and 2005. The sets of species and regions define the node sets \( V^S_t \) and \( V^R_t \), forming a bipartite relational event network. The \textit{native range} of a species is the set of regions where it is indigenous. For simplicity, the native range includes all countries $r$ where the species $s$ was present at the start of the study, in 1880, i.e., $N_{sr}(1880)=1$. The risk set \(\mathcal{R}_{1880}\) is therefore the complement of the native ranges of all the species. Since FRs are non-recurrent, once the dyad \((s, r)\) is observed at time \( t \), the dyad is removed from the risk set for all later times. In other words, \( I_{sr}(t^+) = 0 \) for \( t^+ > t \), while \( I_{sr} = 1 \) up to time \( t \), indicating that the dyad is indeed possible until the moment it is observed. The instantaneous rate of species \( s \) reaching region \( r \) at time \( t \) can be modelled as a function of risk factors. One such factor is the \texttt{Distance} \( x^{\textmd{dist}}_{sr}(t) \) from region \( r \) to the nearest region already invaded by species \( s \) before time \( t \). This covariate is computed as a function of the relational event history and evaluated at time \( t \). Its contribution to the log-rate is given by \(f^{\textmd{dist}}_{sr}\left[x^{\textmd{dist}}_{sr}(t)\right] = \beta \cdot x^{\textmd{dist}}_{sr}(t)\). If long-distance invasions are rare, we expect a negative coefficient \( \beta \). \texttt{Distance} and other endogenous covariates are defined in Table~\ref{tab:explanatory_variables}. The rate of invasion can also depend on global factors that affect all dyads in the risk set. For example, an international agreement to regulate invasive species, adopted at time \( t_{\text{agr}} \), may reduce the baseline hazard after that time: \( \lambda_0(t) < \lambda_0(t_{\text{agr}}) \) for \( t > t_{\text{agr}} \). Recent efforts to strengthen border surveillance further illustrate such time-dependent effects \citep{hulme2021unwelcome}. The empirical illustration in Section \ref{subsec:application} includes a complete analysis of these data.
\end{myexample}

\begin{table}[h]

\centering
\small
\begin{tabular}{|p{2.5cm}|p{1cm}|p{5.5cm}|p{2.8cm}|}
\hline
\textbf{Explanatory Variable} & \textbf{Formula} & \textbf{Computation} & \textbf{Source} \\
\hline
\texttt{Diff\_Temperature} &  \( x_{sr}^{\textmd{d\_temp}}(t) \) & Min absolute difference in near-surface air temperature between $r$ and regions invaded by $s$ before $t$. & \citep{watanabe2011miroc} \\
\hline
\texttt{Log\_Trade} &   \( x_{sr}^{\textmd{trade}}(t) \) & Log-sum of annual trade flows (in current US dollars) between $r$ and regions invaded by $s$ before $t$. & \citep{barbieri2009trading} \\
\hline
\texttt{Log\_Distance} &  \( x_{sr}^{\textmd{dist}}(t) \) & Log-distance between $r$ and the nearest region invaded by $s$ before $t$. & \citep{geosphereRpackage} \\
\hline
\end{tabular}
\caption[Explanatory variables for alien species invasions.]{\textbf{Explanatory variables for alien species invasions}. \textit{Distance}, \textit{Trade}, and \textit{Temperature} are three well-established drivers of alien species invasions \citep{seebens2018global}. This table illustrates how these variables can be operationalised as endogenous covariates by combining exogenous information with the temporal network of previously observed first records.}
\label{tab:explanatory_variables}

\end{table}

\subsection{Exogenous and endogenous covariates}\label{subsec:effects}

This section describes various approaches to defining and measuring \textit{covariate processes} \( \bm{x}_{sr}(\cdot) \), which serve as key drivers in the formulation of REMs. Consider a relational event network where time-stamped edges dynamically link entities within a system. The evolution of this network is shaped by a range of factors. Some of these factors are external to the system (\enquote{\textit{exogenous}}), while others arise from the system itself (\enquote{\textit{endogenous}}), specifically from the occurrence of past events. The definition, nature, and computation of covariate processes critically influence the model, shaping both its explanatory power and the types of dynamics that can be captured through REMs. In particular, these processes should help explain why certain events are more likely than others by identifying and modelling their sources of heterogeneity. One key source is \textit{emergence}, where the probability of future events depends on past occurrences -- a phenomenon effectively captured by endogenous statistics. However, endogenous covariates alone may be insufficient to explain other forms of heterogeneity, such as \textit{extrinsic} heterogeneity arising from latent, time-invariant features of the entities involved \citep{bianchi2024relational}. Moreover, understanding how time shapes both the definition of covariates and their effects on the rate is essential. Exploring this temporal dimension is often crucial for our understanding of the evolving dynamics driving relational event networks.

\begin{myexample}[Email communication]\label{ex:communication}
	We will consider email communication as a running example throughout Section~\ref{sec:modeling}, in order to clarify the definition of covariates as they are introduced. Sending an email is an archetypical instance of a relational event. The relational event network is one-mode, with a single node set \( V^S_t = V^R_t = V_t \) consisting of the employees of a manufacturing company based in Poland \citep{michalski2014seed, michalski2011matching, juozaitiene2022non, juozaitiene2024nodal}. The empirical illustration in Section \ref{subsec:application} includes a complete analysis of these data.
\end{myexample}
 
\subsubsection{Exogenous covariates}

\textit{Exogenous processes} external to the relational event network may give rise to exogenous covariates. These can include features at the node level, such as the duration of employment for an individual in the email example, or at the dyad level, where characteristics pertain to both the sender and receiver, such as working in the same department. A specific type of dyadic exogenous information is represented by \textit{relational covariates}, which aim to capture the nature of interactions between pairs, providing context for relational events. These covariates can reflect aspects such as \textit{affinity} (e.g., friendship) or \textit{flow} (measures of connectedness, such as the amount of communication between sender and receiver in other channels, such as social media) \citep{pilny2016illustration}. Exogenous covariates are not the only type of external information that can influence the relational event network. A particular type of relational covariate is used in the study of \emph{homophily}, i.e., the increased rate of interaction between individuals sharing similar attributes \citep{perry2013point}. In such cases, the exogenous relational covariate can be defined as the difference of a monadic covariate of the sender and the receiver, such as the difference in time of employment.  Often \emph{global exogenous covariates} are used to describe the baseline hazard \citep{lembo2025relational}.

Exogenous processes can also play a role in explaining the relational dynamics by affecting its structure. For example, a crisis involving the company that leads to the firing of several employees would be considered an exogenous event. Such an event would imply that the fired individuals would no longer appear in the company's email network. Such external information can therefore affect the risk set, where certain possibilities have been restricted or expanded due to circumstances outside the relational event network \citep{butts2017relational}. 

Exogenous process information at time $t$ needs always to be tracked in the history $\mathcal{H}_t$ for modelling the relational event hazard. Exogenous processes are thus recorded at the time of interest, but their evolution is not explained by means of the relational events directly. Using the words of \citet{stadtfeld2017interactions}, \enquote{exogenous processes are those that relate to information updates without being modelled explicitly}.

\subsubsection{Endogenous covariates}\label{sec:endougenous}

Endogenous covariates are a defining feature of relational event modelling \citep{brandenberger2019predicting}. These covariates summarise time-ordered sequences of past relational events into network statistics that help explain the dynamics of the relational event network. In particular, endogenous covariates enable the explicit incorporation of temporal interdependence among events directly into the model formulation \citep{meijerink2022dynamic}. They also allow researchers to investigate whether, and how, past interaction behaviours can affect future events \citep{perry2013point}. For example, in the context of email exchanges -- where the term \enquote{exchange} itself suggests an expectation of reciprocation, reflecting the conversational principle of \textit{turn-taking} \citep{stadtfeld2017interactions} -- we can assess whether an individual is responding to a prior message. Specifically, by examining the history of previously sent emails, incorporated in the filtration $\mathcal{H}_{t^-}$, we can check whether there exists a past email from $r$ to $s$, that is, an event involving $(r \rightarrow s)$ occurring before time $t$. In such cases, the event $(s \rightarrow r)$ at time $t$ is considered reciprocal, and the corresponding reciprocity endogenous indicator takes the value 1 at time $t$. 
 
There are numerous ways of defining network statistics due to an almost inexhaustible variety of ways of selecting sub-sequences from the event history \citep{butts2017relational}. Key factors include temporal and sequential dimensions, ordering, and closure of relational structures. Similar to exogenous factors, endogenous covariates can be considered at various levels: node, dyad, triad, and up to the network level. According to the \textit{hierarchy principle}, when including a higher-order network statistic, it is necessary to also include the related lower-order structures, such as node and dyad statistics \citep{pattison2002neighborhood, bianchi2024relational}. Furthermore, the type of relational event network may impose constraints on the network statistics that can be computed. For instance, in a bipartite network where the sender and receiver sets are intrinsically different, reciprocity cannot be evaluated; in the alien species invasion context (Example~\ref{ex:alienspecies}), there is not logical meaning for the reciprocation of an invasion by a species $s$ of a region $r$.

Given these considerations, we propose four generic endogenous covariates structures, which we describe considering the concept of reciprocity. We might simply want to check if an event is reciprocal, but we could also incorporate time-related information, such as evaluating only emails within a working week or giving more weight to the last-received email. This tutorial does not aim to provide an exhaustive account of all possible methods for computing endogenous covariates. Instead, it offers a flexible, generic approach to help readers understand how to adapt these methods to their specific context of interest. Specifically, we can employ different \textit{building blocks} for the computation of $x^\square_{sr}(t)$ for some \textit{endogenous mechanism} $\square$; some of them are listed below: 
\begin{enumerate}
	\item \textit{Indicator}: checking whether there is any prior interaction $(s \rightarrow r)$;
	\begin{displaymath}
		w_{sr}^{1}(t) = \mathds{1}_{\{ \exists t_i<t: (s_i, r_i, t_i) = (s,r,t_i)\}}
	\end{displaymath}
	\item \textit{Volume}: counting the total number of interactions $(s \rightarrow r)$ before time $t$;
	\begin{displaymath}
		w_{sr}^{V}(t) = \sum_{t_i<t} \mathds{1}_{\{(s_i,r_i,t_i)=(s,r,t_i)\}}
	\end{displaymath}
	\item \textit{Exponential Decay}: volume statistics that weight more recent interactions more highly, according to a given half-life parameter \citep{brandenberger2019predicting}. 
	\begin{displaymath}
		% w_{sr}^{E}(t; T_{\frac{1}{2}}) =  \sum_{t_i<t: s_i = s, r_i=r} \frac{\ln{2}}{T_{\frac{1}{2}}} \cdot \exp{\left[-(t-t_i)\cdot\frac{\ln{2}}{T_{\frac{1}{2}}}\right]}  
        w^{E}(t, t_i; T_{\frac{1}{2}}) =  \frac{\ln{2}}{T_{\frac{1}{2}}} \cdot \exp{\left[-(t-t_i)\cdot\frac{\ln{2}}{T_{\frac{1}{2}}}\right]} 
	\end{displaymath}

	\item \textit{Temporal}: network statistics measuring the time elapsed since a particular interaction of interest $(s \rightarrow r)$ occurred,
	\begin{equation*}
		w_{sr}^{\Delta T}(t; \min) 
        =  1-\exp{[-(t-\min_{t_i<t: s_i = s, r_i=r}t_i)]},\qquad 
        w_{sr}^{\Delta T}(t; \max) 
        =  1-\exp{[-(t-\max_{t_i<t: s_i = s, r_i=r}t_i)]},
	\end{equation*}
    which map the elapsed time to the interval $[0,1]$, where $1$ corresponds to the fact that the particular interaction of interest has never happened before. 
    
\end{enumerate}
\begin{table}
	\centering
    \resizebox{\textwidth}{!}{
	\begin{tabular}{l|c|c|c|c|c}
		\hline
		\textbf{} & \textbf{Representation} 
		& \textbf{Indicator} & \textbf{Volume} & \textbf{Exponential decay} & \textbf{Interarrival time} \\
		\hline
		\textbf{Sender activity} 
		& 
		\begin{tikzpicture}[scale=0.45, baseline=(current bounding box.center)]
			\tikzset{every node/.style={draw=black, fill=white, shape=circle, inner sep=1.5pt, minimum size=0.35cm}}
			\node (c) at (0.5,-1.5) [label={[label distance=-0.0cm]90:$s$}]{};
			\node (d) at (2,-0.5){};
			\node (e) at (2,-1.5) {};
			\node (f) at (2,-2.5) {};
			\draw[->] (c)--(d);
			\draw[->] (c)--(e);
			\draw[->] (c)--(f);
		\end{tikzpicture}
		
		& 
		$\displaystyle \max_r{\ w_{sr}^{1}(t)}$
		&
		$\displaystyle \sum_r{\ w_{sr}^{V}(t)}$
		&
		$\displaystyle \sum_r \sum_{t_{sr} < t} \ w^{E}(t, t_{sr}; T_{\frac{1}{2}})$
		&  
		$\displaystyle \min_r{\ w_{sr}^{\Delta T}(t; \text{max})}$
		\\
		\hline
		\textbf{Receiver popularity} 
		& 
		\begin{tikzpicture}[scale=0.45, baseline=(current bounding box.center)]
			\tikzset{every node/.style={draw=black, fill=white, shape=circle, inner sep=1.5pt, minimum size=0.35cm}}
			\node (c) at (0.5,-1.5) [label={[label distance=-0.0cm]90:$r$}]{};
			\node (d) at (2,-0.5){};
			\node (e) at (2,-1.5) {};
			\node (f) at (2,-2.5) {};
			\draw[->] (d)--(c);
			\draw[->] (e)--(c);
			\draw[->] (f)--(c);
		\end{tikzpicture}%
		
		& 
		$\displaystyle \max_s{\ w_{sr}^{1}(t)}$
		&
		$\displaystyle \sum_s{\ w_{sr}^{V}(t)}$
		&
		$\displaystyle \sum_s \sum_{t_{sr} < t} \ w^{E}(t, t_{sr}; T_{\frac{1}{2}})$
		&  
		$\displaystyle \min_s{\ w_{sr}^{\Delta T}(t; \text{max})}$
		\\
		\hline
		\textbf{Reciprocity} 
		& 
		\begin{tikzpicture}[scale=0.45, baseline=(current bounding box.center)]
			\tikzset{every node/.style={draw=black, fill=white, shape=circle, inner sep=1.5pt, minimum size=0.35cm}}
			\node (a) at (0,0) [label={[label distance=-0.0cm]90:$s$}]{};
			\node (b) at (2.5,0)[label={[label distance=-0.0cm]90:$r$}]{};
			\draw[->] (b.-210)--(a.30);
			\draw[dashed,->] (a.-30)--(b.210);
		\end{tikzpicture}%
		
		& 
		$\displaystyle w_{rs}^{1}(t)$
		&
		$\displaystyle w_{rs}^{V}(t)$
		&  
		$\displaystyle \sum_{t_{rs} < t} \ w^{E}(t, t_{rs}; T_{\frac{1}{2}})$
		&  
		$\displaystyle w_{rs}^{\Delta T}(t; \text{max})$
		\\
		\hline
		\textbf{Repetition} 
		& 
		\begin{tikzpicture}[scale=0.45, baseline=(current bounding box.center)]
			\tikzset{every node/.style={draw=black, fill=white, shape=circle, inner sep=1.5pt, minimum size=0.35cm}}
			\node (a) at (0,0) [label={[label distance=-0.0cm]90:$s$}]{};
			\node (b) at (2.5,0)[label={[label distance=-0.0cm]90:$r$}]{};
			\draw[<-] (b.-210)--(a.30);
			\draw[dashed,<-] (b.210)--(a.-30);
		\end{tikzpicture}%
		
		& 
		$\displaystyle w_{sr}^{1}(t)$
		&
		$\displaystyle w_{sr}^{V}(t)$
		&  
		$\displaystyle \sum_{t_{sr} < t} \ w^{E}(t, t_{sr}; T_{\frac{1}{2}})$
		&  
		$\displaystyle w_{sr}^{\Delta T}(t; \text{max})$
		\\
		\hline
		\textbf{Transitive closure} 
		& 
		\begin{tikzpicture}[scale=0.45, baseline=(current bounding box.center)]
			\tikzset{every node/.style={draw=black, fill=white, shape=circle, inner sep=1.5pt, minimum size=0.35cm}}
			\node (a) at (0,0) [label={[label distance=-0.0cm]180:$s$}]{};
			\node (b) at (2,0)[label={[label distance=-0.0cm]0:$r$}]{};
			\node (c) at (1,1.45) [label={[label distance=-0.0cm]90:$a$}]{};
			\draw[dashed,->] (a)--(b);
			\draw[->] (a)--(c);
			\draw[->] (c)--(b);
		\end{tikzpicture}%
		
		& 
		$ \begin{array}{l}
			\textmd{If order is irrelevant:} \\
			\vspace{0.2em}
			\displaystyle \max_a{\ w_{sa}^{1}(t) \times w_{ar}^{1}(t)} \\
			\vspace{0.2em}
			\textmd{If order is relevant:} \\
			\vspace{0.2em}
			\displaystyle \max_a{\ w_{sa}^{1}(t) \times w_{ar}^{1}(t)} \\
			\vspace{0.2em}
			\textmd{\footnotesize over $a$ with $s\to a$ before $a\to r$}
		\end{array} $
		&
		$ \begin{array}{l}
			\textmd{If order is irrelevant:} \\
			\vspace{0.2em}
			\displaystyle \sum_a{\ w_{sa}^{1}(t) \times w_{ar}^{1}(t)} \\
			\vspace{0.2em}
			\textmd{If order is relevant:} \\
			\vspace{0.2em}
			\displaystyle \sum_a{\ w_{sa}^{1}(t) \times w_{ar}^{1}(t)} \\
			\vspace{0.2em}
			\textmd{\footnotesize over $a$ with $s\to a$ before $a\to r$}
		\end{array} $
		&  
		$ \begin{array}{l}
			\textmd{If order is irrelevant:} \\
			\vspace{0.2em}
			\displaystyle \sqrt{\sum_{t_{sa}, t_{ar}<t} \ w^{E}(t, t_{sa}; T_{\frac{1}{2}}) \times w^{E}(t, t_{ar}; T_{\frac{1}{2}})} \\
			\vspace{0.2em}
			\textmd{If order is relevant:} \\
			\vspace{0.2em}
			\displaystyle \sqrt{\sum_{t_{sa}<t_{ar}<t}\ w^{E}(t, t_{sa}; T_{\frac{1}{2}}) \times w^{E}(t, t_{ar}; T_{\frac{1}{2}})}
		\end{array} $
		&  
		$ \begin{array}{l}
			\textmd{If order is irrelevant:} \\
			\vspace{0.2em}
			\displaystyle \min_a{\{w_{sa}^{\Delta T}(t; \text{max}), w_{ar}^{\Delta T}(t; \text{max})\}} \\
			\vspace{0.2em}
			\textmd{If order is relevant:} \\
			\vspace{0.2em}
            \displaystyle \min_a{\ w_{ar}^{\Delta T}(t; \text{max})} \\
			\vspace{0.2em}
			\textmd{\footnotesize over $a$ with $s\to a$ before $a\to r$}
		\end{array} $
		\\
		\hline
		\textbf{Cyclic closure} 
		& 
		\begin{tikzpicture}[scale=0.45, baseline=(current bounding box.center)]
			\tikzset{every node/.style={draw=black, fill=white, shape=circle, inner sep=1.5pt, minimum size=0.35cm}}
			\node (a) at (0,0) [label={[label distance=-0.0cm]180:$s$}]{};
			\node (b) at (2,0)[label={[label distance=-0.0cm]0:$r$}]{};
			\node (c) at (1,1.45) [label={[label distance=-0.0cm]90:$a$}]{};
			\draw[dashed,->] (a)--(b);
			\draw[->] (c)--(a);
			\draw[->] (b)--(c);
		\end{tikzpicture}%
		
		& 
		$ \begin{array}{l}
			\textmd{If order is irrelevant:} \\
			\vspace{0.2em}
			\displaystyle \max_a{\ w_{ra}^{1}(t) \times w_{as}^{1}(t)} \\
			\vspace{0.2em}
			\textmd{If order is relevant:} \\
			\vspace{0.2em}
			\displaystyle \max_a{\ w_{ra}^{1}(t) \times w_{as}^{1}(t)} \\
			\vspace{0.2em}
			\textmd{\footnotesize over $a$ with $r\to a$ before $a\to s$}
		\end{array} $
		&
		$ \begin{array}{l}
			\textmd{If order is irrelevant:} \\
			\vspace{0.2em}
			\displaystyle \sum_a{\ w_{ra}^{1}(t) \times w_{as}^{1}(t)} \\
			\vspace{0.2em}
			\textmd{If order is relevant:} \\
			\vspace{0.2em}
			\displaystyle \sum_a{\ w_{ra}^{1}(t) \times w_{as}^{1}(t)} \\
			\vspace{0.2em}
			\textmd{\footnotesize over $a$ with $r\to a$ before $a\to s$}
		\end{array} $
		&  
		$ \begin{array}{l}
			\textmd{If order is irrelevant:} \\
			\vspace{0.2em}
			\displaystyle \sqrt{\sum_{t_{ra}, t_{as}<t} \ w^{E}(t, t_{ra}; T_{\frac{1}{2}}) \times w^{E}(t, t_{as}; T_{\frac{1}{2}})} \\
			\vspace{0.2em}
			\textmd{If order is relevant:} \\
			\vspace{0.2em}
			\displaystyle \sqrt{\sum_{t_{ra}<t_{as}<t}\ w^{E}(t, t_{ra}; T_{\frac{1}{2}}) \times w^{E}(t, t_{as}; T_{\frac{1}{2}})}
		\end{array} $
		&
        $ \begin{array}{l}
			\textmd{If order is irrelevant:} \\
			\vspace{0.2em}
			\displaystyle \min_a{\{w_{ra}^{\Delta T}(t; \text{max}), w_{as}^{\Delta T}(t; \text{max})\}} \\
			\vspace{0.2em}
			\textmd{If order is relevant:} \\
			\vspace{0.2em}
            \displaystyle \min_a{\ w_{as}^{\Delta T}(t; \text{max})} \\
			\vspace{0.2em}
			\textmd{\footnotesize over $a$ with $r\to a$ before $a\to s$}
		\end{array} $
		\\
		\hline
	\end{tabular}}
	\caption[Network statistics used to incorporate nodal, dyadic, and triadic relational mechanisms.]{\textbf{Network statistics used to incorporate nodal, dyadic, and triadic relational mechanisms}. These network statistics are computed as a function of the event sequence $E = (e_1,...,e_i,...,e_n)$, where each event $e_i$ is represented as $(s_i, r_i, t_i)$. In particular, if a statistic is computed at time $t$, it depends only on events that occurred previously, i.e. at times $t^-<t$. Each row in the table corresponds to a specific endogenous statistic $x_{sr,k}(t)$. These statistics are computed from the available event history and may employ various building blocks $w^T_{sr}(t)$ of a particular type $T$. Node \(a\) represents the \textit{alter}, namely the third actor distinct from nodes $s$ and $r$. The dotted arrow represents the current interaction at the time in which the network statistic is computed, while the continuous arrows represent previously occurred relational events. In the last column, 
    the operators \(\min\) and \(\max\) can be replaced by various pairs of the set \(\{ \min, \max, \mathrm{sum}, \mathrm{mean}\}\) depending on the context.}
	\label{tab:endogenous}
\end{table}

Each of the four covariate components aim to capture, with increasing complexity, how previous occurrences in the event sequence of the event $s\rightarrow r$ may affect the dynamics now. The first simply checks whether the event of interest has happened in the past. The second counts how many. The third is a weighted sum of $s\rightarrow r$ events, where the weights decay relative to a half-time decay factor $T_{\frac{1}{2}}$. The fourth records the first or last time since event involving $s\rightarrow r$ happened. Whereas the first three building blocks are typically part of covariates with linear effects, the covariates incorporating the fourth component require a non-linear formulation, as shown, for example, in Figure~\ref{fig:reciprocity-non-linear} \emph{Left}. 

Table~\ref{tab:endogenous} outlines how to compute several network statistics commonly used to describe relational dynamics in the literature on relational event modelling. In the context of the email communication (Example~\ref{ex:communication}), the endogenous covariates listed in Table~\ref{tab:endogenous} are intuitively interpretable in light of the underlying social behaviours. For instance, reciprocity captures the expectation that individuals will respond to emails they receive, reflecting polite conversational norms. Repetition reflects the tendency to maintain established communication patterns by following up with previously contacted individuals. Additionally, transitive closure plays a crucial role in \textit{path shortening}: once it is recognised that a third party is not relevant to communication, indirect communication is streamlined into a direct one. In the context of the alien species invasions (Example~\ref{ex:alienspecies}), \( x^{{\textmd{dist}}}_{sr}(t) \), defined as the minimum distance from region \( r \) to the nearest region invaded by species \( s \) before time \( t \), is constructed endogenously based on exogenous information. Specifically, information about distances between regions is externally sourced and independent of the relational event model. However, the set of regions invaded by the species \(s\) before time \( t \) can be represented as \(\{r: w^1_{sr}(t) = 1\}\). As these two examples suggest, endogeneity can manifest in various forms depending on the context, and network statistics can capture these nuances. These endogenous factors can potentially be combined with exogenous information whenever available. The selection of covariates to include in the model formulation should be guided by the research question and assessed through model selection criteria. We will examine these aspects in detail in Section~\ref{subsec:selgof}.

\subsection{Recent progress in relational event modelling}\label{subsec:mod-extensions}

In the previous section -- and in many REM formulations in the literature -- effects are assumed to be fixed: a single parameter $\beta$ is used to describe the impact of a endogenous mechanism or an exogenous feature. This, of course, presents some limitations, and in this section we aim at presenting how to incorporate (and, in the following sections, how to implement) more flexible forms of effects. There are two main reasons why we may want to go \textit{beyond linearity} \citep{boschi2026beyond}. First, we may want to allow our effect to change over time, according to external changes that we may not be able to measure directly but that modify the impact of the explanatory variable under consideration. Second, assuming a fixed effect implies that the effect is monotonic -- increasing or decreasing without reaching a maximum or a minimum. For this reason, we may want to allow our effect to be not only non-linear, but, in particular, non-monotonic, able to represent patterns that first increase and then decrease, or vice versa, depending on the underlying mechanism driving the relational dynamics. 

We will discuss these aspects in Section~\ref{subsubsec:TVE-NLE}, while in Section~\ref{subsubsec:random-effects} we will see how it may be useful to capture latent features, either as an alternative to endogenous covariates or in addition to them. Dedicated boxes highlight the use cases of these model extensions, indicating when a particular feature may -- or may not -- be beneficial.

\subsubsection{Flexible REMs with time-varying and non-linear effects}\label{subsubsec:TVE-NLE}

To motivate the use of \textbf{time-varying effects}, we consider the context of alien species invasions (Example~\ref{ex:alienspecies}). In existing literature, international trade has been recognised as a key factor in the spread of alien species \citep{seebens2018global}. The \texttt{Trade} network statistic \( x^{\textmd{trade}}_{sr}(t) \) can be defined as the sum of annual trade flows (\(\$\)) between region \( r \) and other countries that have been invaded by species \( s \) before time \( t \). Similar to \texttt{Distance}, \texttt{Trade} is computed based on the set \(\{r: w^1_{sr}(t) = 1\}\), found endogenously, together with exogenous trade information (\(\$\)) provided by \cite{barbieri2009trading}. Definition and further information about \texttt{Distance} and \texttt{Trade} can be found in Table~\ref{tab:explanatory_variables}. Given the long time span of 125 years between 1880 and 2005, one may ask if it is reasonable to assume that the effect of these covariates should be constant over time. Indeed, the nature of trade has been changing dramatically over the past century, and for this reason its impact may also have changed. This idea is supported statistically, when fitting a REM to alien species invasions that allows a time-varying effect for trade. \citet{boschi2025mixed} reports that the effect of trade on insects and plants over time is positive, but decreasing, meaning that the presence of some trade among countries is still favouring alien species invasions, but its impact has decreased over time. The reason may be found in the reduced percentage of traded goods with a relevant impact on the phenomenon \citep{juozaitiene2023analysing}. 

\begin{tcolorbox}[colback=white,
    colframe=black,
    coltitle=black,
    colbacktitle=yellow!30,
    fonttitle=\bfseries,
    coltext=black,
    boxrule=0.5pt,
    arc=2pt,
    left=6pt,
    right=6pt,
    top=4pt,
    bottom=4pt,
    title=When to include a time-varying effect?]
    Time-varying effects should be used \textbf{whenever a covariate's or a mechanism's influence is expected to change over time due to external factors that are not explicitly captured by the model}; if the effect is in fact constant, the time-varying specification naturally reduces to the time-homogeneous case.
\end{tcolorbox}

We have given an intuitive idea of why a \textit{time-varying effect} can be relevant in some contexts. We now need to define it formally. We focus on the effect $f_k$ of the $k$-th covariate. By defining it as a linear function with a time-varying slope,  $f_k[x_{sr,k}(t), t] =  \beta_k(t) \cdot  x_{sr,k}(t)$, the effect is modulated over calendar time, where $ \beta_k(t) $ corresponds to the evaluation of the effect at time \(t\).

\begin{figure}[t]
    \centering
    \begin{minipage}{0.45\textwidth}
        \centering
        \includegraphics[width=\linewidth]{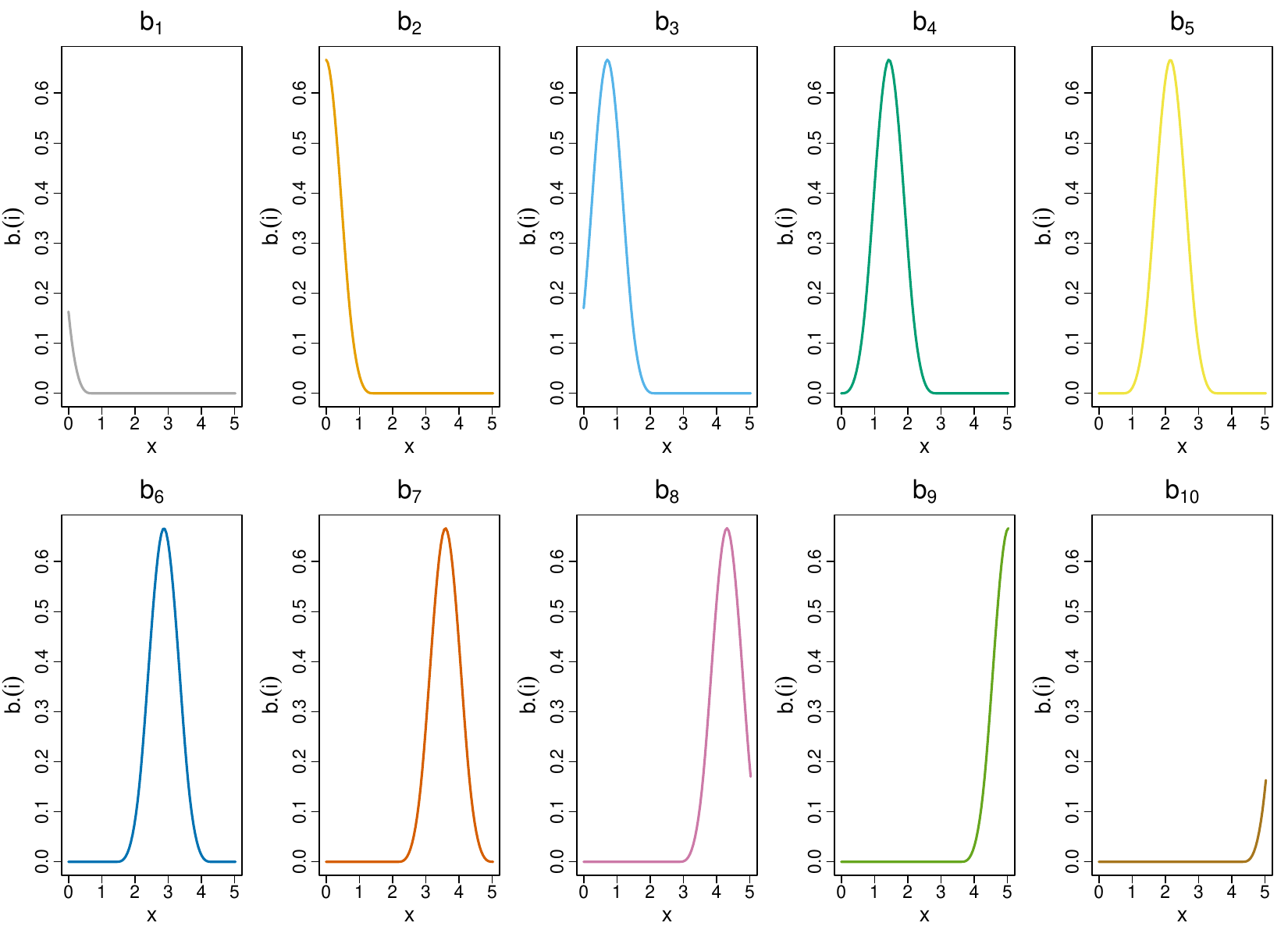}
    \end{minipage}
    \begin{minipage}{0.45\textwidth}
        \centering
        \includegraphics[width=\linewidth]{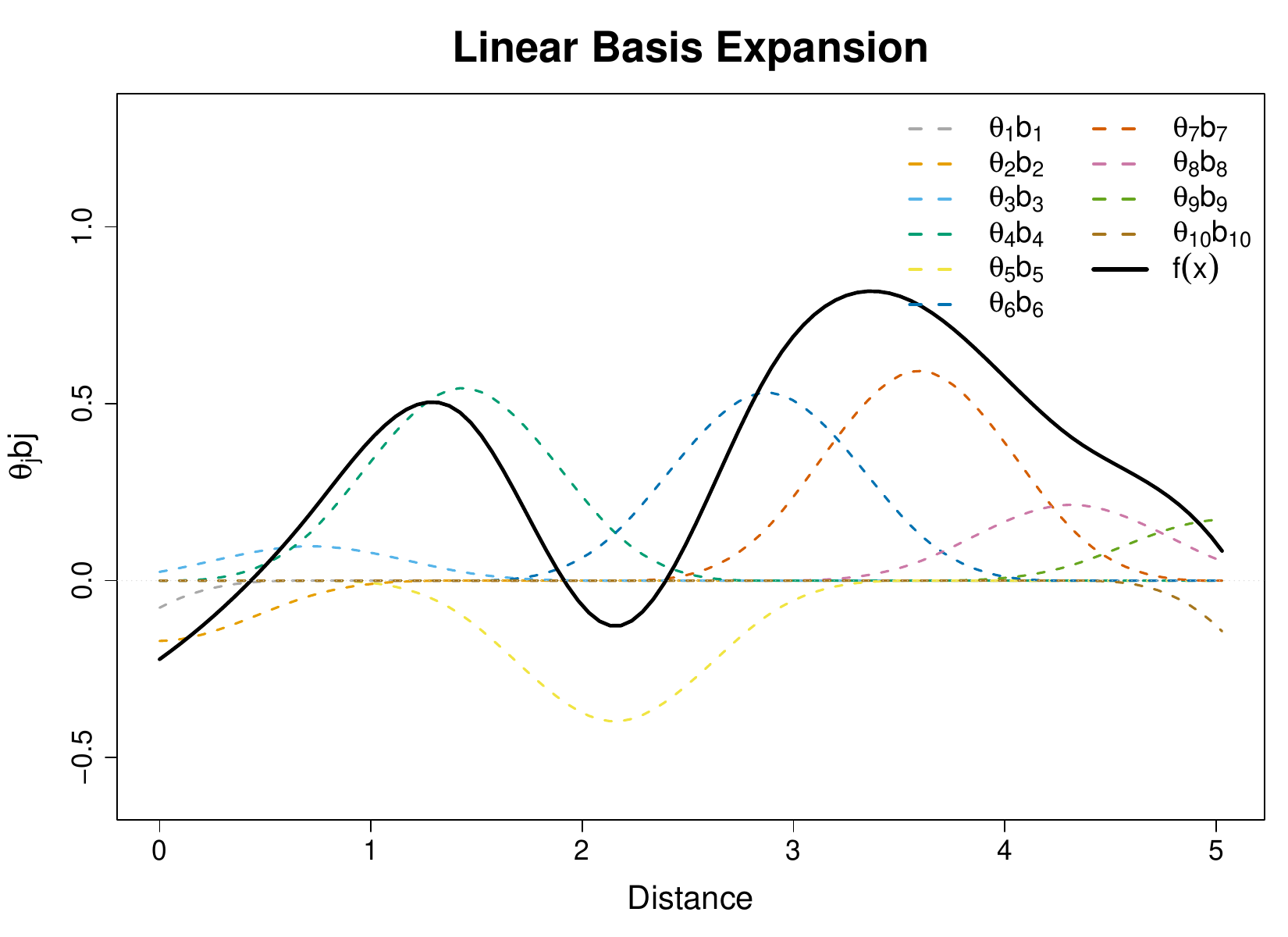}
    \end{minipage}
    \caption[Construction of non-linear effects via basis expansion.]{\label{fig:splines}\textbf{Construction of non-linear effects via basis expansion}. A smooth function $f(x)$ can be represented as a linear combination of non-linear \textit{basis functions}, $f(x) = \sum_{j=1}^{q} \theta_j b_j(x)$, where $q$ is the basis dimension (degrees of freedom). Here we use B-splines, defined via the recursive de Boor formula \citep{bauer2022smooth, filippi2024stochastic}. \textit{Left.} The $q=10$ individual B-spline basis functions $b_j(x)$, $j=1,\dots,q$, are evaluated over the range of the \texttt{Distance} covariate considered in the Overview and Simulation Study of this tutorial. Each basis function $b_j(x)$ is constructed recursively: a B-spline of order $D$ is a weighted combination of two B-splines of order $D-1$, with weights given by linear functions of $x$ determined by the positions of the associated knots. Applying the recursion up to order $D=2$, with $10$ degrees of freedom and $q + D + 2$ knots, yields the cubic B-splines (degree $3$) shown here. \textit{Right:} the weighted basis functions $\theta_j b_j(x)$ (dashed) and their sum $f(x) = \sum_{j=1}^q \theta_j b_j(x)$ (solid black), the resulting non-linear log-event rate contribution of \texttt{Distance}. Since the data here are synthetic, the values of the parameters are known; otherwise, they would need to be estimated (see Section \ref{subsec:further-inference}).}
\end{figure}

Various methods may be used to express such a non-linear function of time. In general, in its simplest form, a function can be expressed as a linear combination of non-linear functions (a \textit{linear basis expansion}),
\begin{equation}\label{eq-tve}
    \beta(t) = \sum_{j=1}^q \theta_{j} \cdot b_j(t),
\end{equation}
and this expansion is not unique. Several types of spline functions admit a linear basis expansion, such as \textit{B-splines} and \textit{thin plate splines} \citep{wood2017generalized}. We give an intuitive illustration of B-splines in Figure~\ref{fig:splines}.

The increased flexibility comes at a computational cost. Specifically, for each covariate, the corresponding column in the model matrix is replaced by \(q\) columns containing the evaluations of the chosen basis functions over time. Consequently, a single time-homogeneous coefficient, \(\beta_k\), is replaced by \(q\) basis coefficients, \(\theta_1,\ldots,\theta_q\). Although this increase in dimensionality is typically not problematic when the number of observed events is sufficiently large, it should nevertheless be taken into account when specifying the model.

We refer to e-mail communication (Example~\ref{ex:communication}) for motivating \textbf{non-linear effects}. Consider an employee sending an email to a colleague. Intuitively, receiving a reply within one hour is much more likely than receiving one after a year. This is an example of a \emph{forgetfulness} effect: as the time since the previous interaction increases, its influence on the likelihood of a future interaction gradually diminishes. Now suppose you receive an email from a colleague for the first time. This initial interaction may substantially increase the likelihood that you will send an email in return. As additional emails are exchanged, the probability of replying may continue to increase, but at a progressively slower rate. For instance, the increase in the rate of replying after receiving the 100th email is likely to be much smaller than after receiving the first. This is an example of a \emph{saturation} effect: repeated interactions continue to strengthen the relationship, but with diminishing marginal impact. Consider now a different scenario: the colleague may be particularly busy today, but plans to check email again tomorrow morning and reply to the flagged messages left unanswered. If we were to model this tendency, we would expect the rate of replying to first decrease throughout the day, reach a minimum, and then increase again the following morning, as the colleague catches up on pending emails. This illustrates a \emph{non-monotonic} effect: unlike linear effects and the two non-linear effects mentioned above, the rate does not move in a single direction, but first declines and then rises again.

\begin{tcolorbox}[colback=white,
    colframe=black,
    coltitle=black,
    colbacktitle=yellow!30,
    fonttitle=\bfseries,
    coltext=black,
    boxrule=0.5pt,
    arc=2pt,
    left=6pt,
    right=6pt,
    top=4pt,
    bottom=4pt,
    title=When to include a non-linear effect?]
    Non-linear effects should be considered \textbf{whenever the influence of a covariate or mechanism is expected to change non-linearly as a function of the covariate value}; for example, when the covariate represents an interarrival time and its impact decreases over time due to \textit{forgetfulness}, or when it represents the volume of previous events and its impact increases at a decreasing rate due to \textit{saturation}. If the relationship is in fact linear, the non-linear specification naturally reduces to a linear effect.
\end{tcolorbox}

The idea that not all values of a covariate have the same impact on the log-rate can be captured by using covariates relying on an exponential decay building block, $w^{E}(t; T_{\frac{1}{2}})$. However, methods of this kind require several \emph{ad hoc} choices, such as the shape of the decay function and the half-life parameter $T_{\frac{1}{2}}$. \textit{Non-linear effects} \citep{bauer2022smooth, filippi2024stochastic} avoid such a priori choices, for instance by using \textit{B-splines}. As discussed for time-varying effects, we can consider non-linear functions of a particular variable and express them in their linear basis expansion formulation:
\begin{equation}
	f\left(x_{sr}(t)\right) =  \sum_{j = 1}^q  \theta_{j} \cdot b_{j}(x_{sr}(t))
    \label{eq-nle}
\end{equation}
See Figure~\ref{fig:splines} to get how this spline function can be constructed.

Whereas we started Section~\ref{sec:endougenous} by considering the simplest way in which the effect of reciprocity can be considered, namely as a fixed constant $ \beta^\textmd{rec} \cdot w_{rs}^{1}(t) $, with this non-linear framework we are now able to provide a flexible way to incorporate a temporally non-linear reciprocity effect, via $f^\textmd{rec}[w^{\Delta T}(t; \text{max})] $, where $f^\textmd{rec}(\cdot)$ is a smooth function of the time since the last reciprocal event \citep{juozaitiene2024s}.

As for the time-varying effect, this increased flexibility comes at a cost, represented by a larger number of parameters to be estimated. Furthermore, this also poses the risk of \textit{overfitting}. How to deal with this has been explained in Section~\ref{subsubsec:penalisation}. The code in order to implement time varying and non-linear effects in practice is provided in Section \ref{subsubsec:beyond} in a dedicated paragraph for going beyond linearity in relational event models. Practical examples in Section \ref{subsec:application} are also helpful ways in order to get the potentials of this kind of effects.

\subsubsection{Mixed relational event models}\label{subsubsec:random-effects}

In Section~\ref{subsec:effects}, we discussed how \textit{observable} information -- whether exogenous or endogenous -- can be incorporated into relational event models to explain part of the heterogeneity observed in the data. However, some sources of heterogeneity remain \textit{unobserved}, and accounting for them can further improve the model's ability to capture the underlying relational dynamics. Among the possible approaches for accounting for unobserved heterogeneity, \textit{random effects} provide a flexible framework for capturing latent sources of variation that are not directly measured by the available covariates.

Among the dynamics of interest, \textit{popularity} and/or \textit{virality} provide a representative example. From one perspective, popularity can be interpreted as the result of endogenous mechanisms. For instance, on a social media platform, an individual may become increasingly popular because they have already been followed by many other users. In this case, the number of previous followers captures an observable source of endogenous heterogeneity. However, popularity may also be driven by unobserved factors \citep{bianchi2024relational} -- and, in this case, we would speak about \emph{extrinsic heterogeneity} . When these factors are observable, they can be incorporated into the model through exogenous covariates. Otherwise, they can be accommodated using random effects, for example by including sender-specific random intercepts to account for persistent differences in actors' propensity to initiate interactions that cannot be explained by the observed covariates. In the alien species invasions context (Example~\ref{ex:alienspecies}), we can define a latent species-specific invasiveness effect, which reflects heterogeneity not solely attributable to the observed number of past invasions. By including a species-specific random effect alongside the previously introduced covariates, the log-rate function becomes \(\log{\left[\lambda_{sr}(t)\right]} =  \beta \cdot x^{{\textmd{dist}}}_{sr}(t) + \beta(t) \cdot x^{{\textmd{trade}}}_{sr}(t) + \gamma^{{\textmd{inv}}}_s\). Similarly, a region-specific random effect could be incorporated to account for latent invasibility of a region, allowing us to assess whether unobserved or unmeasured characteristics make certain regions systematically more prone to alien species invasions.

Random effects are not limited to capturing \textit{first-order} heterogeneity. Similar to latent space models, they can also account for \textit{second-order} effects, representing latent dependencies between pairs of entities that cannot be explained by the observed covariates. As a practical application related to Example~\ref{ex:alienspecies}, we can refer to \citet{boschi2025mixed}, where, together with first-order species invasiveness and region popularity effects, overall \textit{co-invasion patterns among species} are investigated to assess whether the presence of one species in a region influences the invasion rate of subsequent species. To capture unobserved heterogeneity in these co-invasion dynamics, the authors introduce a dyadic random intercept. Specifically, the random effect is associated with pairs consisting of the currently invading species and the species that most recently invaded the region, thereby accounting for latent pair-specific propensities for co-invasion that are not explained by the observed covariates.

\begin{tcolorbox}[colback=white,
    colframe=black,
    coltitle=black,
    colbacktitle=yellow!30,
    fonttitle=\bfseries,
    coltext=black,
    boxrule=0.5pt,
    arc=2pt,
    left=6pt,
    right=6pt,
    top=4pt,
    bottom=4pt,
    title=When to include random effects?]
    Random effects should be considered \textbf{whenever there is reason to think that unobserved heterogeneity influences event occurrence beyond what can be explained by the available explanatory variables}; for example, to account for latent differences among entities (first-order effects) or among pairs of them (second-order effects). If no meaningful unobserved heterogeneity is in fact present, the estimated random-effect variance naturally shrinks towards zero, reducing to the fixed-effects model.
\end{tcolorbox}

For all the mentioned reasons, the incorporation of random effects into REMs has recently gained attention \citep{uzaheta2023random, juozaitiene2023analysing, juozaitiene2024nodal}. 
% due to their potential to capture and interpret \textit{unobserved heterogeneity} among actors and dyads that influences the dynamics of relational event networks, particularly in cases where such variability cannot be fully explained by endogenous effects alone
In this context, the complexity of the model formulation in Equation~\eqref{eq:model} increases as follows:
\begin{equation}\label{eq:model-re}
	\lambda_{sr}(t) = I_{sr}(t) \times \lambda_{0}(t) \times \exp\left\{ f\left[\bm{x}_{sr}(t)\right] + \bm{\gamma}^\top\bm{z}_{sr}(t) \right\}, \quad \bm{\gamma} \sim \mathcal{N}\left(\bm{0}, \Sigma(\bm{\phi})\right),
\end{equation}
where $\bm{\gamma}$ represents random effects drawn from a multivariate Gaussian distribution with zero mean and variance-covariance matrix $\Sigma(\bm{\phi})$. For example, for a sender and receiver random intercept model as described in Figure~\ref{fig:random_effects}, the covariance matrix \(\Sigma(\bm{\phi})\) of $(\gamma^{\text{s\_act}},\gamma^{\text{r\_pop}})$ is given as a $2\times 2$ matrix with diagonal elements $\boldsymbol{\phi}=(\phi^{\text{s\_act}},\phi^{\text{r\_pop}})$. The covariate processes $\bm{z}_{sr}(\cdot)$, adapted to the filtration $\mathbb{H}$, may fully, partially, or not at all overlap with the covariates included in $\bm{x}_{sr}(\cdot)$. The code to implement random effects in practice is provided in Section~\ref{subsec:further-inference}, in a dedicated paragraph on estimating random effects as smooth terms. A practical example is reported in Section~\ref{subsubsec:re-inference}.

\section{Inference Techniques}\label{sec:inference}
Several estimation techniques can be used for relational events: \textit{Maximum Likelihood Estimation} (MLE) \citep{butts2008relational, perry2013point, bianchi2024relational}, Bayesian estimation \citep{arena2024bayesian}, Method of Moments \citep{arastuie2020chip}. The purpose of this section is to provide a comprehensive overview of MLE approaches, as they currently provide the simplest and most extensive implementation.

\subsection{Maximum likelihood inference techniques for relational event models}\label{subsec:general-inference}
We collect all model parameters in \(\bm{\theta}\) and the corresponding model matrix entries in \(\bm{h}_{sr}(t)\). Given the additivity of the model and that non-linear effects can be expressed as linear combinations of basis functions, the model formulation can be written as:
\begin{equation}\label{eq:model-simple} 
	\lambda_{sr}(t; \bm{\theta}) = I_{sr}(t) \times \lambda_0(t) \times \exp\left\{ \bm{\theta}^\top \bm{h}_{sr}(t) \right\}.
\end{equation}

\subsubsection{Full likelihood.}

Relational event data are observed as sequences of events, denoted by \(E\). The \textit{full likelihood} is constructed as the joint probability of observing the entire event sequence \(E\) under the model specified in Equation~\eqref{eq:model-simple}. Under the non-homogeneous Poisson process framework \citep{bianchi2024relational}, the full likelihood can be written as:
\begin{eqnarray}
    \mathcal{L}^F(\bm{\theta}) &=& \prod_{i=1}^n p\left(t_i|\mathcal{H}_{t_i^-}, \bm{\theta}\right) \, p\left((s_i, r_i)\mid t_i, \mathcal{H}_{t_i^-}, \bm{\theta}\right) \label{eq:full_likelihood-1} \\
    &=& \prod_{i=1}^n \left[ \sum_{(s,r) \in \mathcal{R}_{t_i}} \lambda_{sr}(t_i; \bm{\theta}) \exp\left( - \sum_{(s,r) \in \mathcal{R}_{t_i}} \int_{t_{i-1}}^{t_i} \lambda_{sr}(u; \bm{\theta}) \, \text{d}u \right) \times 
    \frac{\lambda_{s_ir_i}(t_i; \bm{\theta})}{\sum_{(s,r) \in \mathcal{R}_{t_i}} \lambda_{sr}(t_i; \bm{\theta})} \right] \label{eq:full_likelihood-2}
\end{eqnarray}
where \( \mathcal{R}_{t_i} \) denotes the risk set of possible events at time \( t_i \). 

From Equation~\ref{eq:full_likelihood-2}, it is possible to note that the interarrival time is modelled according to a \textit{generalized exponential distribution} with rate equal to the sum of the competing rates, and that the observed dyad is modelled as the result of a \textit{multinomial experiment} with as many outcomes as the number of competing dyads. The expression can be further simplified, leading to the product of the hazard of the observed event and -- according to a survival analysis expression -- the \textit{survival} of all the dyads within the interarrival time.

Estimating the parameter vector \( \bm{\theta} \) by maximizing Equation~\eqref{eq:full_likelihood-2} is challenging due to two main computational issues: (i) the need to compute integrals over unknown time-varying rate functions and (ii) the summation across large risk sets at each time point. A common strategy to address the first challenge is to assume a \textit{Piecewise-Constant Hazard} (PCH) function, which allows the full likelihood to be rewritten as:
\begin{equation}\label{eq:poisson_regression}
    \mathcal{L}^F(\bm{\theta}) = \prod_{i=1}^n \prod_{(s,r) \in \mathcal{R}_{t_i}} \left( \lambda_{sr}(t_i; \bm{\theta}) \right)^{\Delta N_{sr}(t_i)} \exp\left[ - \lambda_{sr}(t_i; \bm{\theta}) \left( t_i - t_{i-1} \right) \right],
\end{equation}
where \( \Delta N_{sr}(t_i) = N_{sr}(t_i) - N_{sr}(t_{i-1}) \) is an indicator variable equal to 1 if the dyad \((s,r)\) occurs at time \(t_i\), and 0 otherwise. 
Equation~\eqref{eq:poisson_regression} is mathematically equivalent to the likelihood of a \textit{Poisson regression}, where the rate $\lambda_{sr}(t_i)$ is multiplied by the interarrival time $(t_i-t_{i-1})=e^{\log(t_i-t_{i-1})}$, which corresponds to an offset of $\log(t_i-t_{i-1})$. The Poisson regression formulation allows the estimation of parameters \( \bm{\theta} \) using generalized linear modelling techniques \citep{vieira2024fast}.

\begin{figure}[t]
    \centering
\begin{minipage}{\linewidth}
\begin{center}
    \begin{tabular}[t]{llllll>{\raggedright\arraybackslash}p{1.4cm}}
\toprule
sender & receiver & start & stop & event & x & log\_exposure\\
\hline
... & ... & ... & ... & ... & ... & ...\\
26 & 2 & 44 & 44.42 & 1 & 1 & -0.87\\
1 & 1 & 44 & 44.42 & 0 & 0 & -0.87\\
2 & 1 & 44 & 44.42 & 0 & 0 & -0.87\\
3 & 1 & 44 & 44.42 & 0 & 0 & -0.87\\
\addlinespace
4 & 1 & 44 & 44.42 & 0 & 0 & -0.87\\
5 & 1 & 44 & 44.42 & 0 & 0 & -0.87\\
6 & 1 & 44 & 44.42 & 0 & 0 & -0.87\\
7 & 1 & 44 & 44.42 & 0 & 0 & -0.87\\
... & ... & ... & ... & ... & ... & ...\\
\hline
\end{tabular}
\end{center}
    \end{minipage}
    
    \vspace{0.5em}
    \begin{minipage}{\textwidth}
        \begin{center}
        \begin{tikzpicture}
            \node[draw, thick, rectangle, rounded corners=3pt, inner sep=5pt] (box) {
                \begin{minipage}{0.9\textwidth}
                    \centering
                    \begin{adjustbox}{max width=\textwidth}
                        \begin{lstlisting}[language=R]
glm(event ~ x + offset(log_exposure), 
    family = poisson(link = "log"), data = poisson_data)
                        \end{lstlisting}
                    \end{adjustbox}
                \end{minipage}
            };

            \node[draw, fill=blue!20, rounded corners=5pt, anchor=south west, font=\sffamily\bfseries\small, inner sep=4pt]
            at ([xshift=1em, yshift=-1em]box.north west) {R Code: Full likelihood};
        \end{tikzpicture}
        \end{center}
    \end{minipage}
    \caption[Full likelihood inference via Poisson regression.]{\textbf{Full likelihood inference via Poisson regression}. \emph{Top}. Snapshot of relational event data structured for fitting a Poisson regresresion with one covariate \(x\). For each event, \(t_i\) is recorded in column \texttt{stop} and the preceding event \(t_{i-1}\) in \texttt{start}. The \texttt{event} column encodes whether an event occurred, \(\Delta N_{sr}(t_i) = 1\), or not \(\Delta N_{sr}(t_i) = 0\), distinguishing events from non-events. The column \texttt{log\_exposure} contains the logarithm of the interarrival time, \(\log(t_i - t_{i-1})\), and serves as the offset. \emph{Bottom.} Code for fitting a Poisson regression via \texttt{glm} in \texttt{R}, using the prepared relational event data structure.}
    \label{fig:poisson_regression}
\end{figure}
\begin{tcolorbox}[colback=white,           
    colframe=black,          
    coltitle=white,            
    colbacktitle=black!40,   
    fonttitle=\bfseries,
    boxrule=0.5pt,
    arc=2pt,
    left=6pt,
    right=6pt,
    top=4pt,
    bottom=4pt,
    title=Full likelihood inference via Poisson regression in practice]
{
Full likelihood inference on \(\bm{\theta}\) in Model~\eqref{eq:model-simple} is equivalent to a Poisson regression, assuming
\[\Delta N_{sr}(t_i)| \bm{h}_{sr}(t_i) \stackrel{\mbox{iid}}{\sim} \text{Poisson}\left( \mu_{sr}(t_i)\right) \qquad \log(\mu_{sr}(t_i)) = \bm{\theta}^\top\bm{h}_{sr}(t_i) + \log(t_i - t_{i-1}) \]
Figure~\ref{fig:poisson_regression} illustrates how relational event data can be structured and analysed using Poisson regression in \texttt{R}.}
\end{tcolorbox}
However, this simplification comes at a cost: the assumption that the rate function \( \lambda_{sr}(\cdot; \bm{\theta}) \) remains constant between events may be overly restrictive in some applications. Moreover, despite this assumption, the computational demands associated with the product over the entire risk set of order $O(|V^S|\times|V^R|)$ may remain substantial, particularly when the risk set is large.
\begin{figure}[t]
    \centering
    \begin{minipage}{\linewidth}
        \begin{center}
            \begin{tabular}[t]{llllll>{}p{1.4cm}}
\hline
sender & receiver & start & stop & event & x\\
\hline
... & ... & ... & ... & ... & ...\\
26 & 2 & 44 & 44.42 & 1 & 1\\
1 & 1 & 44 & 44.42 & 0 & 0\\
2 & 1 & 44 & 44.42 & 0 & 0\\
3 & 1 & 44 & 44.42 & 0 & 0\\
\addlinespace
4 & 1 & 44 & 44.42 & 0 & 0\\
5 & 1 & 44 & 44.42 & 0 & 0\\
6 & 1 & 44 & 44.42 & 0 & 0\\
7 & 1 & 44 & 44.42 & 0 & 0\\
... & ... & ... & ... & ... & ...\\
\hline
\end{tabular}
        \end{center}
    \end{minipage}
    
    \vspace{0.5em}
    \begin{minipage}{\textwidth}
        \begin{center}
        \begin{tikzpicture}
            
            \node[draw, thick, rectangle, rounded corners=3pt, inner sep=3pt] (box) {
                \begin{minipage}{0.9\textwidth}
                    \centering
                    \begin{adjustbox}{max width=\textwidth}
                        \begin{lstlisting}[language=R]
library(survival)                      
coxph(Surv(start,stop,event) ~ x, data = full_data)
                        \end{lstlisting}
                    \end{adjustbox}
                \end{minipage}
            };

            \node[draw, fill=blue!20, rounded corners=5pt, anchor=south west, font=\sffamily\bfseries\small, inner sep=4pt]
            at ([xshift=1em, yshift=-1em]box.north west) {R Code: Partial likelihood};
            
        \end{tikzpicture}
        \end{center}
    \end{minipage}
    \caption[Partial likelihood inference via Cox regression.]{\textbf{Partial likelihood inference via Cox regression}. \emph{Top}. Snapshot of relational event data structured for fitting a Cox regression with one covariate \(x\). For each event, \(t_i\) is recorded in column \texttt{stop} and the preceding event time \(t_{i-1}\) in \texttt{start}. If the covariate \(x\) remains constant within the interval, no additional adjustment is needed. However, if \(x\) changes within the interval, these changes must be tracked by inserting additional records that reflect the updated values of \(x\) over time. The \texttt{event} column encodes whether an event occurred, distinguishing events from non-events. \emph{Bottom.} Code for fitting a Cox regression in \texttt{R} via the \texttt{coxph} function in the \texttt{R} package \texttt{survival} \citep{survivalRpackage, therneau2000modeling}.}
    \label{fig:coxph_regression}
\end{figure}
\subsubsection{Partial likelihood.}  
A more flexible alternative to the PCH assumption consists of focusing only on the second term of Equation~\eqref{eq:full_likelihood-1}, namely on the product of multinomial probabilities of observing specific dyads \((s_i, r_i)\) among those at risk, conditional on the occurrence of an event at time \(t_i\) \citep{perry2013point}. This leads to the expression for \textit{partial likelihood}:
\begin{equation}\label{eq:partial_likelihood}
    \mathcal{L}(\bm{\theta}) = \prod_{i=1}^n \dfrac{\lambda_{s_ir_i}(t_i; \mathcal{H}_{t_i^-}, \bm{\theta})}{\sum_{(s,r) \in \mathcal{R}_{t_i}} \lambda_{sr}(t_i; \mathcal{H}_{t_i^-}, \bm{\theta})} = \prod_{i=1}^n \dfrac{\exp{\{ \bm{\theta}^\top \bm{h}_{s_ir_i}(t_i)\}}}{\sum_{(s,r) \in \mathcal{R}_{t_i}} \exp{\{ \bm{\theta}^\top \bm{h}_{sr}(t_i)\}}}.
\end{equation}
\begin{tcolorbox}[colback=white,           
    colframe=black,          
    coltitle=white,            
    colbacktitle=black!40,   
    fonttitle=\bfseries,
    boxrule=0.5pt,
    arc=2pt,
    left=6pt,
    right=6pt,
    top=4pt,
    bottom=4pt,
    title=Partial likelihood inference via Cox regression in practice]
    We perform inference on \(\bm{\theta}\) in Model \eqref{eq:model-simple} within a survival analysis framework, where
    \[\Big{\{}\Delta N_{sr}(t_i)| \bm{h}_{sr}(t_i)\Big{\}} \stackrel{\mbox{iid}}{\sim} \text{Multinomial}\left( \{\pi_{sr}(t_i)\}\right) \qquad \pi_{sr}(t_i) = \dfrac{\exp\{\bm{\theta}^\top\bm{h}_{sr}(t_i)\}}{\sum_{s'r' \in \mathcal{R}_{t_i}} \exp\{\bm{\theta}^\top\bm{h}_{s'r'}(t_i)\}} \]
    Figure~\ref{fig:coxph_regression} demonstrates how relational event data can be structured and analysed using Cox regression in \texttt{R}.
\end{tcolorbox}
Relying on this approach has two main consequences. First, it results in a slight loss of efficiency of the parameter estimates $\bm{\hat\theta}$, but, as \citet{cox1975partial} showed, this is mainly related to the baseline hazard \(\lambda_0\). However, this also leads to an important simplification: the baseline hazard $\lambda_0(t)$ is eliminated \citep{cox1975partial} and no assumptions have to be made about the value of the hazard in between observations. Furthermore, this framework permits REMs estimation even when only the order of events is known.
\begin{tcolorbox}[colback=white,
    colframe=black,
    coltitle=black,
    colbacktitle=yellow!30,
    fonttitle=\bfseries,
    coltext=black,
    boxrule=0.5pt,
    arc=2pt,
    left=6pt,
    right=6pt,
    top=4pt,
    bottom=4pt,
    title=Full or partial likelihood?]
    \textbf{The full likelihood approach relies on the (sometimes unrealistic) piecewise-constant hazard assumption, whereas the partial likelihood does not}. However, \textbf{when the number of events per parameter is small, it is worth using the full likelihood}, accepting some bias in order to reduce the variance of the estimates. The standard 10-events-per-variable rule \citep{peduzzi1996simulation}, or newer empirical guidelines \citep{van2019sample}, can be used in practice. Whereas the \textbf{partial likelihood typically loses all information on the baseline hazard or any global covariates, the full likelihood does not}. The full likelihood is a convenient way to estimate the effects of these global covariates \citep{stadtfeld2017dynamic}. \textbf{For both, number of entities must be sufficiently small} to allow evaluation of the risk set at each event time. 
\end{tcolorbox}

\subsubsection{Sampled partial likelihood.}  
The computational cost of evaluating $\mathcal{L}(\bm{\theta})$ remains substantial, as the denominator scales as $|V^S| \times |V^R|$. Moreover, it also requires full information on the covariate process for all the non-events at risk. \emph{Nested case-control} (NCC) sampling \citep{borgan1995methods} has been adapted to REMs \citep{vu2015relational, lerner2020reliability}. At each event time, a \emph{sampled risk set} is constructed that includes the observed dyad $(s, r)$, the \textit{case}, along with a reduced set of $m$ unobserved but possible dyads, the \textit{controls}, also at risk at the time of the event. Each time point is now associated with both the observed event and the corresponding sampled risk set $\dot{\mathcal{R}}_{t}$, where $\dot{\mathcal{R}}_{t} \subset \mathcal{R}_{t}$. The resulting point process is denoted by $\dot{\mathbb{P}} = \Big\{ \left[t_i, \left( (s_i, r_i), \dot{\mathcal{R}}_{t_i} \right) \right]; i \geq 1 \Big\}$. The associated counting process \(\dot{\bm{N}}\) and the corresponding intensity are given by:
\small
\begin{equation*}\label{counting-sampling}
	\dot{N}_{sr,\dot{\mathcal{R}}}(t) = \sum_{i \geq 1} \mathds{1}\left\{ t_i \leq t, \ (s_i, r_i) = (s, r), \ \dot{\mathcal{R}}_{t_i} = \dot{\mathcal{R}} \right\}, \quad \lambda_{sr,\dot{\mathcal{R}}}(t) = \lambda_{sr}(t) \cdot \frac{m}{|\mathcal{R}_t| - 1} \cdot 1_{\{(s,r) \in \dot{\mathcal{R}}, \ \dot{\mathcal{R}} \subseteq \mathcal{R}_t\}}.
\end{equation*}
\normalsize

The corresponding filtration $\dot{\mathbb{H}} = \{ \dot{\mathcal{H}}_t \}_{t \geq 0}$ incorporates both the event history $\mathcal{H}_t$ and the sequence of sampled risk sets $\sigma(\dot{\mathcal{R}}_{t_i}; t_i \leq t)$. Under the assumption of \textit{independent sampling}, sampling of non-events is independent of the timing of events \citep{borgan1995methods}. This modification leads to a revised version of the partial likelihood where the full risk set is replaced by the sampled risk set, and conditioning is performed with respect to the sampling history:
\begin{equation}\label{eq:sampled_partial_likelihood}
    \dot{\mathcal{L}}_m(\bm{\theta}) = \prod_{i=1}^n \dfrac{\lambda_{s_ir_i}(t_i; \dot{\mathcal{H}}_{t_i^-}, \bm{\theta})}{\sum_{(s,r) \in \dot{\mathcal{R}}_{t_i}} \lambda_{sr}(t_i; \dot{\mathcal{H}}_{t_i^-}, \bm{\theta})} = \prod_{i=1}^n \dfrac{\exp\left\{ \bm{\theta}^\top \bm{h}_{s_ir_i}(t_i) \right\}}{\sum_{(s,r) \in \dot{\mathcal{R}}_{t_i}} \exp\left\{ \bm{\theta}^\top \bm{h}_{sr}(t_i) \right\}}.
\end{equation}
\begin{figure}[t]
    \centering
    \begin{minipage}{\linewidth}
        \begin{center}
            \begin{tabular}[t]{llllll>{}p{1.4cm}}
\hline
sender & receiver & start & stop & event & x\\
\hline
... & ... & ... & ... & ... & ...\\
26 & 2 & 44 & 44.42 & 1 & 1\\
33 & 56 & 44 & 44.42 & 0 & 0\\
8 & 21 & 44 & 44.42 & 0 & 0\\
33 & 19 & 44 & 44.42 & 0 & 0\\
\addlinespace
35 & 10 & 44 & 44.42 & 0 & 0\\
20 & 13 & 44 & 44.42 & 0 & 0\\
9 & 21 & 44.42 & 45.47 & 1 & 0\\
51 & 52 & 44.42 & 45.47 & 0 & 0\\
... & ... & ... & ... & ... & ...\\
\hline
\end{tabular}
        \end{center}
    \end{minipage}
    
    \vspace{0.5em}
    \begin{minipage}{\textwidth}
        \begin{center}
        \begin{tikzpicture}
            \node[draw, thick, rectangle, rounded corners=3pt, inner sep=3pt] (box) {
                \begin{minipage}{0.9\textwidth}
                    \centering
                    \begin{adjustbox}{max width=\textwidth}
                        \begin{lstlisting}[language=R]
library(survival)   
clogit(event ~ x + strata(stop), data = sampled_data)
                        \end{lstlisting}
                    \end{adjustbox}
                \end{minipage}
            };

            \node[draw, fill=blue!20, rounded corners=5pt, anchor=south west, font=\sffamily\bfseries\small, inner sep=4pt]
            at ([xshift=1em, yshift=-1em]box.north west) {R Code: Sampled partial likelihood};
        \end{tikzpicture}
        \end{center}
    \end{minipage}
    \caption[Sampled partial likelihood via conditional logistic regression.]{\textbf{Sampled partial likelihood via conditional logistic regression}. \emph{Top}. Snapshot of relational event data formatted for fitting a conditional logistic regression with one covariate \(x\). For each event, \(t_i\) is recorded in column \texttt{stop} and the preceding event \(t_{i-1}\) in \texttt{start}. As in Figure \ref{fig:coxph_regression}, if \(x\) changes during the interval, these changes must be tracked. \(m=5\) \enquote{controls} are sampled per \enquote{case}. \texttt{event} column indicates if an event occurred. \emph{Bottom.} Code for fitting conditional logistic regression in \texttt{R} with \texttt{clogit}, using the prepared data.}
    \label{fig:clogit_regression}
\end{figure}
We refer to \eqref{eq:sampled_partial_likelihood} as \textit{sampled partial likelihood}. This seemingly small change substantially reduces the computational complexity: the denominator only requires the computation of \(m\) terms rather than $|V^S| \times |V^R|$. In practice, maximizing the sampled partial likelihood in Equation~\eqref{eq:sampled_partial_likelihood} is equivalent to estimating a \textit{conditional logistic regression}.

\begin{tcolorbox}[colback=white,           
    colframe=black,          
    coltitle=white,            
    colbacktitle=black!40,   
    fonttitle=\bfseries,
    boxrule=0.5pt,
    arc=2pt,
    left=6pt,
    right=6pt,
    top=4pt,
    bottom=4pt,
    title=Sampled partial likelihood via conditional logistic regression in practice]
    We perform inference on \(\bm{\theta}\) in Model \eqref{eq:model-simple} within a conditional logistic regression setting:
    \footnotesize
    \begin{equation*}
        \Bigg{\{}\Delta N_{sr}(t_i) |  \bm{h}_{sr}(t_i), \sum_{(s',r') \in \dot{\mathcal{R}}_{t_i}} \Delta N_{s'r'}(t_i) = 1 \Bigg{\}} \sim \text{Multinomial} \left( \{\pi_{sr}(t_i)\}\right),\qquad \pi_{sr}(t_i) =  \dfrac{\exp{\{ \bm{\theta}^\top \bm{h}_{sr}(t_i)\}}}{\sum_{(s',r') \in \dot{\mathcal{R}}_{t_i}} \exp{\{ \bm{\theta}^\top \bm{h}_{s'r'}(t_i)\}}}
    \end{equation*}
    \normalsize
    Figure~\ref{fig:clogit_regression} illustrates how relational event data can be structured and analysed using conditional logistic regression in \texttt{R}.    
\end{tcolorbox}
\begin{tcolorbox}[colback=white,
    colframe=black,
    coltitle=black,
    colbacktitle=yellow!30,
    fonttitle=\bfseries,
    coltext=black,
    boxrule=0.5pt,
    arc=2pt,
    left=6pt,
    right=6pt,
    top=4pt,
    bottom=4pt,
    title=Partial or sampled partial likelihood?]
    The sampled partial likelihood should be considered \textbf{when the number of entities is too large to evaluate the complete risk set at each event time}.
\end{tcolorbox}
In practice, choosing to use a sampled partial likelihood inference approach also implies choosing $m$. The choice of $m$ does not affect the consistency of the estimator, but it does affect its variance. A smaller $m$ comes, indeed, at the price of a somewhat larger variance. If computational resources allow fitting a model with a larger value of $m$, the researcher should consider doing so. On the other hand, the particular case of $m=1$ allows for convenient model extensions that enrich the formulation of relational event models, as presented in the following paragraph.

\subsubsection{Case-control partial likelihood.}\label{subsubsec:logistic}
By sampling only \(m=1\) non-event and dividing each term in the likelihood by the rate evaluated at the sampled non-event, we have that \eqref{eq:sampled_partial_likelihood} coincides with the likelihood of a logistic regression \citep{boschi2025mixed, filippi2024stochastic} with a fixed response equal to $1$ whose explanatory variables correspond to the difference between the explanatory variables of the event and those of the sampled non-event. The logistic formulation does not include the intercept: if the intercept term is present in model formulation~\eqref{eq:model-simple}, it cancels out when the difference is computed. We thus obtain the \textit{case-control partial likelihood}:
\begin{equation}\label{partial-GAM}
	\dot{\mathcal{L}}_1(\bm{\theta}) = \prod_{i=1}^{n} \left[ 1+\exp{\left( -  \bm{\theta}^\top \Delta \bm{h}_i \right)} \right]^{-1}
\end{equation}
Consider $\Delta \bm{H} = (\Delta \bm{h}_1,\dots,\Delta \bm{h}_{n}) \in \mathbb{R}^{n \times P}$,
where the $i$-th row is \(\Delta \bm{h}_{i} = \bm{h}_{s_i r_i}(t_i) - \bm{h}_{s_i^\ast r_i^\ast}(t_i),\)
i.e., the difference between the model matrix entry for the observed event 
$(s_i,r_i)$ and that of the corresponding non-event $(s_i^\ast,r_i^\ast)$, both evaluated at time $t_i$.
For covariates entering linearly, this is simply the difference of the two covariate values \(\bm{x}_{s_i r_i}(t_i) - \bm{x}_{s_i^\ast r_i^\ast}(t_i)\).
Whenever a covariate or time enters the linear predictor non-linearly, the
corresponding entries of $\Delta \bm h_i$ correspond to the difference of the basis-function evaluations.

For example, if a covariate \(x\) enters as a smooth term $\sum_{j=1}^{q} \theta_j b_j(x)$, where \(\{b_j(x)\}\) are the $q=10$ cubic B-spline basis functions in Figure~\ref{fig:splines} \textit{Left}, the block of $\Delta \bm h_i$ corresponding to this term is
\[
\Big(b_1(x_{s_ir_i}) - b_1(x_{s_i^\ast r_i^\ast}),\ \dots,\ b_q(x_{s_ir_i}) - b_q(x_{s_i^\ast r_i^\ast})\Big),
\]
and the corresponding $q$ elements of $\bm\theta$ are the coefficients $\theta_1,\dots,\theta_q$
weighting these basis functions; once estimated, they reconstruct the non-linear
log-event-rate contribution of \(x\), shown as the solid curve in Figure~\ref{fig:splines} \textit{Right}. To make the model identifiable, these basis-function blocks of the model matrix are
centred \citep{wood2017generalized}. More details on the implementation of non-linear REMs via case-control partial likelihood are reported in Section~\ref{subsubsec:beyond}.

\begin{tcolorbox}[colback=white,           
    colframe=black,          
    coltitle=white,            
    colbacktitle=black!40,   
    fonttitle=\bfseries,
    boxrule=0.5pt,
    arc=2pt,
    left=6pt,
    right=6pt,
    top=4pt,
    bottom=4pt,
    title=Case-control partial likelihood via logistic regression in practice]
    We perform inference on \(\bm{\theta}\) in Model \eqref{eq:model-simple} within a generalized additive model (for binary outcomes) setting:
    \[ Y_i|\Delta\bm{h}_i\stackrel{\mbox{iid}}{\sim} \text{Bernoulli}\left( \pi_i \right), \quad
    \text{logit}(\pi_i) = \bm{\theta}^\top\Delta\bm{h}_i, \quad \text{no intercept}, \quad y_i = 1~\forall i=1,\ldots,n \]
    Figure~\ref{fig:logistic_regression} illustrates how to construct the case-control dataset and how to fit a logistic regression to it in \texttt{R}.  
\end{tcolorbox}
\begin{figure}[t]
    \centering
    \begin{minipage}{\linewidth}
        \begin{center}
\resizebox{\linewidth}{!}{\begin{tabular}[t]{llllll>{\raggedright\arraybackslash}p{1.4cm}llll}
\hline
time & sender\_ev & receiver\_ev & sender\_non\_ev & receiver\_non\_ev & x\_ev & x\_non\_ev & delta\_x & id\_ev & id\_non\_ev & y\\
\hline
... & ... & ... & ... & ... & ... & ... & ... & ... & ... & ...\\
44.42 & 26 & 2 & 1 & 51 & 2.50 & 0.00 & 2.50 & 1 & -1 & 1\\
45.47 & 9 & 21 & 5 & 13 & 2.39 & 2.24 & 0.15 & 1 & -1 & 1\\
46.07 & 40 & 34 & 13 & 11 & 2.57 & 0.50 & 2.07 & 1 & -1 & 1\\
46.47 & 50 & 4 & 21 & 35 & 4.72 & 1.97 & 2.75 & 1 & -1 & 1\\
\addlinespace
46.52 & 16 & 40 & 14 & 32 & 3.10 & 3.02 & 0.09 & 1 & -1 & 1\\
... & ... & ... & ... & ... & ... & ... & ... & ... & ... & ... \\
\hline
\end{tabular}}
\end{center}
    \end{minipage}
    
    \vspace{0.5em}
    \begin{minipage}{\textwidth}
        \begin{center}
        \begin{tikzpicture}
            \node[draw, thick, rectangle, rounded corners=3pt, inner sep=3pt] (box) {
                \begin{minipage}{0.9\textwidth}
                    \centering
                    \begin{adjustbox}{max width=\textwidth}
                        \begin{lstlisting}[language=R]
library(mgcv)   
glm(y ~ -1 + delta_x, family="binomial"(link = "logit"), data = ncc_data)
                        \end{lstlisting}
                    \end{adjustbox}
                \end{minipage}
            };

            \node[draw, fill=blue!20, rounded corners=5pt, anchor=south west, font=\sffamily\bfseries\small, inner sep=4pt]
            at ([xshift=1em, yshift=-1em]box.north west) {R Code: Case-control partial likelihood};
        \end{tikzpicture}
        \end{center}
    \end{minipage}
    \caption[Case-control partial likelihood via logistic regression.]{\textbf{Case-control partial likelihood via logistic regression}. \emph{Top}. Snapshot of case-control relational event data formatted for fitting a logistic regression with one covariate \(x\). For each observed event, the corresponding \texttt{time} \(t_i\) as well as the associated \texttt{sender\_ev} and \texttt{receiver\_ev} are recorded. \texttt{sender\_non\_ev} and \texttt{receiver\_non\_ev} are also stored for related non-events. For both events and non-events, the value of the covariate \(x\) is evaluated in \texttt{x\_ev} and \texttt{x\_non\_ev} and their difference is stored in \texttt{delta\_x}. \emph{Bottom.} Code for fitting a logistic regression via \texttt{gam} in \texttt{R}, using the prepared relational event data structure.}
    \label{fig:logistic_regression}
\end{figure}
Expression~\eqref{partial-GAM} is relevant for two key reasons: (i) performing inference on \(\bm{\theta}\) within a logistic framework allows for the incorporation of the more complex types of effects discussed in Section~\ref{subsubsec:TVE-NLE} and~\ref{subsubsec:random-effects}; and (ii) it ensures that the computational cost of the entire likelihood only scales linearly with the number of observed events and does not depend on the size of the sender and receiver sets.
\begin{tcolorbox}[colback=white,
    colframe=black,
    coltitle=black,
    colbacktitle=yellow!30,
    fonttitle=\bfseries,
    coltext=black,
    boxrule=0.5pt,
    arc=2pt,
    left=6pt,
    right=6pt,
    top=4pt,
    bottom=4pt,
    title = Sampled or case-1-control partial likelihood?]
    The case-control partial likelihood is particularly well-suited \textbf{when fitting complex model, including non-linear or mixed effects}.
\end{tcolorbox}

\subsubsection{Beyond linearity: time-varying and non-linear effects.}\label{subsubsec:beyond}

In Section~\ref{subsubsec:TVE-NLE}, we motivate the use of time-varying and non-linear effects in relational event modelling and show how to include them as smooth terms. By means of the case-control partial likelihood inference method reported in Equation~\eqref{partial-GAM}, a relational event model can be fitted using a logistic regression; consequently, the smooth terms representing the time-varying or non-linear effects can be fitted by estimating a \textit{generalized additive model} (GAM) for binary outcomes \citep{boschi2025mixed, filippi2024stochastic}. Specifically, we rely on the \texttt{R} package \texttt{mgcv} \citep{wood2011fast, wood2017generalized} to fit these smooth terms, as for the linear effect shown in Figure~\ref{fig:logistic_regression}.

For a time-varying effect, we model the coefficient as a smooth function of time multiplied by the covariate difference. Hence, the effect remains linear in the covariate values while allowing the corresponding regression coefficient to vary smoothly over time. In \texttt{mgcv} \citep{wood2011fast, wood2017generalized}, this specification is implemented using a \textit{varying-coefficient smooth} through the \texttt{by} argument,

\begin{tikzpicture}
            \node[draw, thick, rectangle, rounded corners=3pt, inner sep=3pt] (box) {
                \begin{minipage}{0.9\textwidth}
                    \centering
                    \begin{adjustbox}{max width=\textwidth}
                        \begin{lstlisting}[language=R]
library(mgcv)   
gam(y ~ -1 + s(time, by=delta_x),
        family="binomial"(link = "logit"), data = ncc_data)
                        \end{lstlisting}
                    \end{adjustbox}
                \end{minipage}
            };

            \node[draw, fill=blue!20, rounded corners=5pt, anchor=south west, font=\sffamily\bfseries\small, inner sep=4pt]
            at ([xshift=1em, yshift=-1em]box.north west) {R Code: Time-varying effects};
        \end{tikzpicture}

For a non-linear effect, we model the contribution to the log-event rate as a smooth function of the covariate itself. Hence, the effect remains fixed in time during the observed time window. In \texttt{mgcv},

        \begin{tikzpicture}
            \node[draw, thick, rectangle, rounded corners=3pt, inner sep=3pt] (box) {
                \begin{minipage}{0.9\textwidth}
                    \centering
                    \begin{adjustbox}{max width=\textwidth}
                        \begin{lstlisting}[language=R]
library(mgcv)   
x_matrix <- cbind(ncc_data$x_ev, ncc_data$x_non_ev)
by_matrix <- cbind(ncc_data$id_ev, ncc_data$id_non_ev)
gam(y ~ -1 + s(x_matrix, by=by_matrix),
        family="binomial"(link = "logit"), data = ncc_data)
                        \end{lstlisting}
                    \end{adjustbox}
                \end{minipage}
            };

            \node[draw, fill=blue!20, rounded corners=5pt, anchor=south west, font=\sffamily\bfseries\small, inner sep=4pt]
            at ([xshift=1em, yshift=-1em]box.north west) {R Code: Non-linear effects};
        \end{tikzpicture}

Empirical applications in Section~\ref{subsec:application} illustrate several practical scenarios in which these types of effects are particularly useful for improving both the flexibility and the interpretability of the model.

\subsubsection{Penalised likelihood maximisation.}\label{subsubsec:penalisation}
To avoid overfitting, likelihood is usually combined with a smoothing penalty \(P^{\bm{\nu}}(\bm{\theta})\), 
\begin{equation}\label{eq:partial-penalty}
	\log{\mathcal{L}^{\bm{\nu}}(\bm{\theta})} = \log{\mathcal{L}(\bm{\theta})} - P^{\bm{\nu}}(\bm{\theta}) \quad \quad P^{\bm{\nu}}(\bm{\theta}) = \sum_{l=1}^{L} \nu_l \cdot p_l(\bm{\theta}_{\bm{j}_l}), 
\end{equation}
where the \textit{penalty term} $P^{\bm{\nu}}(\bm{\theta})$ can be expressed as a linear combination of \textit{wiggliness measures} for each smooth function according to a set of \textit{smoothing parameters}, included in $\bm{\nu}$, and controlling the \textit{fit-wiggliness trade-off} \citep{wood2017generalized}. Specifically, $L$ is the number of penalised terms, $p_l$ is the $l$th penalty function and $\nu_l$ the $l$th parameter. $\bm{j}_l \subseteq \{1,...,P\}$ is a sub-vector of the set of indices of vector $\bm{\theta}$ and \(P = |\bm{\theta}|\). Estimation of $\bm{\nu}$ is usually performed via \textit{cross-validation}, aiming at minimising the prediction error in new data. 

The estimator $\hat{\bm{\theta}}$ is obtained by maximising $\log{\mathcal{L}^{\bm{\nu}}(\bm{\theta})}$ and setting the \textit{penalised score} to $0$,
\begin{equation}\label{eq:score-penalty}
	\nabla \log{\mathcal{L}^{\bm{\nu}}(\bm{\theta})} = \frac{\partial \log{\mathcal{L}^{\bm{\nu}}(\bm{\theta})}}{\partial \bm{\theta}} = \nabla \log{\mathcal{L}(\bm{\theta})}  - \nabla P^{\bm{\nu}}(\bm{\theta}) = \bm{0} \Rightarrow \nabla P^{\bm{\nu}}(\hat{\bm{\theta}}) = \nabla \log{\mathcal{L}(\hat{\bm{\theta}})}
\end{equation}
Although there is no closed-form solution for the estimator $\hat{\bm{\theta}}$, the \textit{Newton-Raphson} algorithm \citep{hastie2009elements, dobson2018introduction} typically converges rapidly.

\subsection{Further advancements in REM inference}\label{subsec:further-inference}

There have been significant other advances in relational event modelling and inference in the past decade, such as extensions to relational hyper-events going beyond dyadic ties \citep{lerner2025relational}, the use of modern machine learning techniques \citep{filippi2024modeling,filippi2024stochastic} or latent space model approaches \citep{artico2023fast,artico2023dynamic, mulder2024latent}. In this section, we focus on two other extensions. The first focuses on an alternative inference strategy for estimating global effects, whereas the second considers the importance of accommodating heterogeneity in relational event networks by means of random effects.

\subsubsection{Efficient inference for global covariates}\label{subsec:shifting}
Partial-likelihood REM inference techniques do not allow for the estimation of the effects of global covariates that take the same value regardless of the dyad under evaluation \citep{kreiss2024testing}. Different approaches have been proposed to estimate the effects of global covariates. \citet{stadtfeld2017dynamic} rely on the full likelihood, but this approach faces possible computational limitations discussed above.
In this tutorial, we present an approach for estimating these effects that does not require additional assumptions on the data-generating process while keeping the computational cost under control. Specifically, we consider an extension proposed by \citet{lembo2025relational} of the counting process \(\bm{N}\) and the sampled counting process \(\dot{\bm{N}}\).

\begin{tcolorbox}[colback=white,           
    colframe=black,          
    coltitle=black,            
    colbacktitle=red!30,   
    fonttitle=\bfseries,
    boxrule=0.5pt,
    arc=2pt,
    left=6pt,
    right=6pt,
    top=4pt,
    bottom=4pt,
    title=What is the advantage of shifting?]
    Given a shifted event \((s, r, \tilde{t} = t + \tau_{sr}) \in \mathbb{P}^S\) corresponding to \((s, r, t) \in \mathbb{P}\), non-events \((s^*, r^*)\) in the shifted risk set \(\mathcal{R}^S_{\tilde{t}}\) have been assigned different time-shifts \(\tau_{s^* r^*}\). As a result, when reverting to the original time scale, the expression \(\tilde{t} - \tau_{s^* r^*} = t + \tau_{sr} - \tau_{s^* r^*}\) generally differs from \(t\). Therefore {\bf global covariates do not cancel in the partial likelihood}.
\end{tcolorbox} 

Their idea is to perturb the components $N_{sr}$ of the original counting process \(\bm{N}\) governed by $\mathbb{P}$ by random positive quantities: \(\mathbb{T} = \{\tau_{sr} \in \mathbb{R}^+\}_{(s,r) \in V^S \times V^R}\).
where $\mathbb{T}$ is independent of $\mathbb{P}$. The number of events generated by the \textit{shifted point process} is then counted by a \textit{shifted counting process} (\(\bm{N}^S\)), driven by the following intensity function
\begin{equation}\label{eq:model-simple-global} 
    \lambda^S_{sr}(\tilde{t}) = \mathds{1}_{\{\tilde{t} \in [\tau_{sr},\tau_{sr} + T]\}} \lambda_{sr}(\tilde{t}-\tau_{sr}), \qquad \lambda_{sr}(t; \bm{\theta}) = I_{sr}(t) \times \lambda_0 \times \exp\left\{ \bm{\theta}^\top \bm{h}_{sr}(t) \right\}.
\end{equation}
With respect to this shifted process, the \textit{shifted partial likelihood} is given as 
\begin{equation}\label{eq:shifted_partial_likelihood}
    \mathcal{L}^S(\bm{\theta}) = \prod_{i=1}^n \dfrac{\lambda^S_{s_ir_i}(\tilde{t}_i; \mathcal{H}^S_{(\tilde{t}_i-\tau_{s_ir_i})^-}, \bm{\theta})}{\sum_{(s,r) \in \mathcal{R}^S_{\tilde{t}_i}} \lambda_{sr}(\tilde{t}_i; \mathcal{H}^S_{(\tilde{t}_i-\tau_{s_ir_i})^-}, \bm{\theta})} = \prod_{i=1}^{n} \dfrac{\lambda_{s_ir_i}(t_i; \bm{\theta})}{\sum_{(s,r) \in \mathcal{R}^S_{\tilde{t}_i}}  \lambda_{sr}(t_i + \tau_{s_ir_i} - \tau_{sr}; \bm{\theta}) },
\end{equation}
By sampling \(m\) dyads from the \textit{shifted risk set}, we obtain the expressions of the \textit{shifted sampled partial likelihood} and \textit{shifted case-control partial likelihood}, which follow from \eqref{eq:shifted_partial_likelihood} in the same way as Equations~\eqref{eq:sampled_partial_likelihood} and~\eqref{partial-GAM} follow from Equation~\eqref{eq:partial_likelihood}.
\begin{figure}[t]
    \centering
    \begin{minipage}{\textwidth}
    \begin{center}
\resizebox{\linewidth}{!}{
\begin{tabular}[t]{llllll>{\raggedright\arraybackslash}p{1.4cm}lll}
\hline
time\_ev & sender\_ev & receiver\_ev & time\_non\_ev & sender\_non\_ev & receiver\_non\_ev & weekday\_ev & weekday\_non\_ev & delta\_weekday & y\\
\hline
... & ... & ... & ... & ... & ... & ... & ... & ... & ...\\
44.42 & 26 & 2 & 452.24 & 54 & 33 & 1 & 1 & 0 & 1\\
45.47 & 9 & 21 & 53.63 & 21 & 56 & 1 & 1 & 0 & 1\\
46.07 & 40 & 34 & 3.85 & 19 & 29 & 1 & 1 & 0 & 1\\
46.47 & 50 & 4 & 431.57 & 49 & 55 & 1 & 1 & 0 & 1\\
\addlinespace
46.52 & 16 & 40 & 98.45 & 41 & 8 & 1 & 0 & 1 & 1\\
... & ... & ... & ... & ... & ... & ... & ... & ... & ...\\
\hline
\end{tabular}}
    \end{center}
    \end{minipage}
    
    \vspace{0.5em}
    \begin{minipage}{\textwidth}
        \begin{center}
        \begin{tikzpicture}
            \node[draw, thick, rectangle, rounded corners=3pt, inner sep=3pt] (box) {
                \begin{minipage}{0.9\textwidth}
                    \centering
                    \begin{adjustbox}{max width=\textwidth}
                        \begin{lstlisting}[language=R]
library(mgcv)   
gam(y ~ -1 + delta_weekday, family="binomial"(link = "logit"), data = shifted_ncc_data)
                        \end{lstlisting}
                    \end{adjustbox}
                \end{minipage}
            };

            \node[draw, fill=blue!20, rounded corners=5pt, anchor=south west, font=\sffamily\bfseries\small, inner sep=4pt]
            at ([xshift=1em, yshift=-1em]box.north west) {R Code: Shifted case-control partial likelihood};
        \end{tikzpicture}
        \end{center}
    \end{minipage}
    \caption[Shifted case-control partial likelihood via GAMs.]{\textbf{Shifted case-control partial likelihood via GAMs}. \emph{Top:} Snapshot of a shifted augmented relational event dataset, formatted for fitting a logistic regression with a single covariate, \texttt{weekday}, which indicates whether the day is Monday to Friday. For each observed event, the corresponding event time (\texttt{time\_ev}) and associated sender and receiver (\texttt{sender\_ev}, \texttt{receiver\_ev}) are recorded. The same structure is used for non-events. The time of the non-event (\texttt{time\_nv}) differs from \texttt{time\_non\_ev} as it is first shifted and then transformed back to the original scale. For both events and non-events, the covariate is evaluated and stored in \texttt{weekday\_ev} and \texttt{weekday\_non\_ev}, and their difference is stored in \texttt{delta\_weekday}. Due to the shift, the difference is not always zero. \emph{Bottom:} Example code in \texttt{R} for fitting a logistic regression model using the \texttt{gam} function, based on the prepared relational event data.}
    \label{fig:logistic_regression_shift}
\end{figure}
\begin{tcolorbox}[colback=white,           
    colframe=black,          
    coltitle=white,            
    colbacktitle=black!40,   
    fonttitle=\bfseries,
    boxrule=0.5pt,
    arc=2pt,
    left=6pt,
    right=6pt,
    top=4pt,
    bottom=4pt,
    title=Shifted case-control partial likelihood via GAMs in practice]
    We perform inference on \(\bm{\theta}\) in Model \eqref{eq:model-simple-global} within a generalized additive model (for binary outcomes) setting:
    \[ Y_i|\Delta\bm{h}^S_i\stackrel{\mbox{iid}}{\sim} \text{Bernoulli}\left( \pi_i \right), \quad
    \text{logit}(\pi_i) = \bm{\theta}^\top\Delta\bm{h}^S_i, \quad \text{no intercept}, \quad y_i = 1~\forall i=1,\ldots,n \]
    Figure~\ref{fig:logistic_regression_shift} illustrates how to construct the \textit{shifted augmented dataset} and how to fit a GAM to it in \texttt{R}.    
\end{tcolorbox}
\begin{tcolorbox}[colback=white,
    colframe=black,
    coltitle=black,
    colbacktitle=yellow!30,
    fonttitle=\bfseries,
    coltext=black,
    boxrule=0.5pt,
    arc=2pt,
    left=6pt,
    right=6pt,
    top=4pt,
    bottom=4pt,
    title=Shifted case-control partial likelihood or full likelihood?]
    Also the Poisson regression implementation of the full likelihood is able to deal with global covariates, however its computational cost is high when the risk set is large. The shifted case-control partial likelihood should be considered \textbf{a computationally efficient alternative when the model includes global covariates of interest}.
\end{tcolorbox}

\subsubsection{Estimating random effects as smooth terms}\label{subsubsec:re-inference}

When dealing with random effects (Section~\ref{subsubsec:random-effects}), 
traditional \textit{variance component estimation} for mixed models relies on restricted maximum likelihood techniques. This can be computationally infeasible, even for a moderate number of random effects. 

Alternatively, it is possible to integrate out the random effects  $\bm{\gamma} = \{\bm{\gamma}_1, \ldots, \bm{\gamma}_F\}$ from the joint density, assuming a Gaussian prior for the random effects -- consistent with the model in Equation~\ref{eq:model-re}. Laplace approximations can be used to obtain the marginal likelihood \citep{breslow1993approximate}. In the case where random effects are independent, the covariance matrix simplifies to a diagonal matrix with elements $\bm{\sigma} = (\sigma_1, \dots, \sigma_F)$. Under this assumption, the Laplace approximation to the marginal likelihood becomes 
\begin{figure}[t]
    \centering
    \begin{minipage}{\textwidth}
    \begin{center}
\begin{tabular}[t]{llllll>{\raggedright\arraybackslash}p{1.4cm}l}
\hline
time & sender\_ev & receiver\_ev & sender\_non\_ev & receiver\_non\_ev & id\_ev & id\_non\_ev & y\\
\hline
... & ... & ... & ... & ... & ... & ... & ...\\
44.42 & 26 & 2 & 1 & 51 & 1 & -1 & 1\\
45.47 & 9 & 21 & 5 & 13 & 1 & -1 & 1\\
46.07 & 40 & 34 & 13 & 11 & 1 & -1 & 1\\
46.47 & 50 & 4 & 21 & 35 & 1 & -1 & 1\\
\addlinespace
46.52 & 16 & 40 & 14 & 32 & 1 & -1 & 1\\
... & ... & ... & ... & ... & ... & ... & ...\\
\hline
\end{tabular}    
    \end{center}
    \end{minipage}
    
    \vspace{0.5em}
    \begin{minipage}{\textwidth}
        \begin{center}
        \begin{tikzpicture}
            \node[draw, thick, rectangle, rounded corners=3pt, inner sep=3pt] (box) {
                \begin{minipage}{0.9\textwidth}
                    \centering
                    \begin{adjustbox}{max width=\textwidth}
                        \begin{lstlisting}[language=R]
library(mgcv)   
by_matrix = cbind(ncc_data$id_ev,ncc_data$id_non_ev)	
sender_act <- factor(c(ncc_data$sender_ev,ncc_data$sender_non_ev))
dim(sender_act) <- c(nrow(ncc_data), 2)
receiver_pop <- factor(c(ncc_data$receiver_ev,ncc_data$receiver_non_ev))
dim(receiver_pop) <- c(nrow(ncc_data), 2)
gam(y ~ s(sender_act, by=by_matrix, bs="re") + 
        s(receiver_pop, by=by_matrix, bs="re")  - 1, 
    family="binomial"(link = "logit"),
    data=ncc_data)
                        \end{lstlisting}
                    \end{adjustbox}
                \end{minipage}
            };

            \node[draw, fill=blue!20, rounded corners=5pt, anchor=south west, font=\sffamily\bfseries\small, inner sep=4pt]
            at ([xshift=1em, yshift=-1em]box.north west) {R Code: Random effects};
        \end{tikzpicture}
        \end{center}
    \end{minipage}
    \caption[Fitting random effects as smooth terms.]{\textbf{Fitting random effects as smooth terms}. \emph{Top}. Snapshot of case-control relational event data formatted for fitting a logistic regression with random effects for sender activity and receiver popularity. For each observed event, the corresponding \texttt{time} \(t_i\) as well as the associated \texttt{sender\_ev} and \texttt{receiver\_ev} are recorded. \texttt{sender\_non\_ev} and \texttt{receiver\_non\_ev} are also stored for related non-events. Two additional columns, \texttt{id\_ev} and \texttt{id\_non\_ev},  are added for weighting contributions of events and non-events. \emph{Bottom}. Code for fitting a logistic regression via \texttt{gam} in \texttt{R} including a random intercept for the sender and a random intercept for the receiver. Similarly to the code employed for the non-linear effects (see Figure~\ref{fig:distance_overview}), \texttt{id\_ev} and \texttt{id\_non\_ev} allows to incorporate the difference between the two spline functions. In this case, the spline functions are linear combinations of indicator functions, and for this reason we provide the matrix of levels of the random factor of interest.}
    \label{fig:random_effects}
\end{figure}
$
	\log \mathcal{L}(\bm{\theta}) \approx \log \mathcal{L}(\bm{\theta}, \bm{\gamma}) - \dfrac{1}{2} \sum_{l=1}^F \dfrac{1}{\sigma^2_l} \bm{\gamma}_l^\top \bm{\gamma}_l
$ \citep{schneble2021new}.
This is equivalent to a \textit{ridge penalty} on the random effects, where the penalty parameter is inversely related to the variance, $\lambda_j \propto \frac{1}{\sigma_j^2}$.
From a generalized additive model perspective, the random effects can be estimated as 0-dimensional splines: that is, we treat random effects as smooth functions whose basis functions are indicator functions.

\begin{tcolorbox}[colback=white,           
    colframe=black,          
    coltitle=white,            
    colbacktitle=black!40,   
    fonttitle=\bfseries,
    boxrule=0.5pt,
    arc=2pt,
    left=6pt,
    right=6pt,
    top=4pt,
    bottom=4pt,
    title=How to fit random effects as smooth terms in practice]
    We perform inference on \(\bm{\gamma}^{\textmd{s\_act}},\bm{\gamma}^{\textmd{r\_pop}}\) within a generalized additive model (for binary outcomes) setting:
    \begin{equation*}
        \begin{aligned}
        & Y_i|\Delta\bm{z}_i,\bm{\gamma}^{\textmd{s\_act}},\bm{\gamma}^{\textmd{r\_pop}} \stackrel{\mbox{iid}}{\sim} \text{Bernoulli}\left( \pi_i \right), \quad
    \text{logit}(\pi_i) = (\bm{\gamma}^{\textmd{s\_act}})^\top\Delta\bm{z}^{\textmd{s\_act}}_i + (\bm{\gamma}^{\textmd{r\_pop}})^\top\Delta\bm{z}^{\textmd{r\_pop}}_i, \\
    &\text{no intercept}, \quad y_i = 1~\forall i=1,\ldots,n
        \end{aligned}
    \end{equation*}
    where \(\Delta\bm{z^\square_{i}}\) includes the difference between the covariates with a random effect evaluated for the \(i\)-th event and the related sampled non-event.
    Figure~\ref{fig:random_effects} shows how to construct a case-control dataset and how to fit a REM with random effects in \texttt{R}.    
\end{tcolorbox}

\subsection{Model selection and validation}\label{subsec:selgof}

A key aspect of statistical inference -- and thus also of inference associated with relational event modelling -- is \textit{model selection and validation}, that is, the process of identifying the a suitable fitting model.

\subsubsection{Model selection in relational event models}
Several model selection approaches have been developed for REMs  \citep{quintane2014modeling}. These include both \textit{likelihood-based} and \textit{prediction-based} methods. AIC, introduced in Section~\ref{sec:overview}, is a widely used example of the former class \citep{amati2024goodness}. AIC estimates a model’s predictive risk by combining the log-likelihood, $\log \mathcal{L}(\hat{\bm{\theta}})$, with a penalty proportional to the number of parameters. Lower values of AIC indicate better model fit \citep{juozaitiene2024s}. 
The evaluation of the penalty term in AIC requires particular attention when dealing with random effects or smoothing terms. In this context, two main approaches can be distinguished \citep{wood2017generalized}: (i) \textit{marginal AIC}, which accounts only for the number of fixed effects and the number of variance or smoothing parameters; and (ii) \textit{conditional AIC}, which instead relies on an estimate of the effective number of parameters \citep{vaida2005conditional}. A corrected version of AIC has recently been proposed for use with smooth terms \citep{wood2016smoothing}. This adjustment provides a compromise between the conservative bias of marginal AIC and the anti-conservative bias of traditional conditional AIC estimators, such as that proposed in \citet{hastie1990gam}.
Model selection is particularly crucial for REMs, where the mechanisms influencing event occurrence may be numerous and complex, such as rich sets of network statistics \citep[e.g.,][]{bianchi2023ties} or random effects that capture latent characteristics of actors or event types \citep[e.g.,][]{juozaitiene2024nodal}.
\begin{figure}[tb]
        \begin{center}
        \begin{tikzpicture}
            \node[draw, thick, rectangle, rounded corners=3pt, inner sep=3pt] (box) {
                \begin{minipage}{0.9\textwidth}
                    \centering
                    \begin{adjustbox}{max width=\textwidth}
                        \begin{lstlisting}[language=R]
library(mgcv)   
rem_fit <- gam(y ~ -1 + s(x_matrix, by=by_matrix),
        family="binomial"(link = "logit"), data = ncc_data)
AIC(rem_fit)
                        \end{lstlisting}
                    \end{adjustbox}
                \end{minipage}
            };

            \node[draw, fill=blue!20, rounded corners=5pt, anchor=south west, font=\sffamily\bfseries\small, inner sep=4pt]
            at ([xshift=1em, yshift=-1em]box.north west) {R Code: Model selection};
        \end{tikzpicture}
        \end{center}
    \caption[Model selection.]{\textbf{Model selection}. Computation of corrected conditional AIC \citep{wood2016smoothing} in \texttt{mgcv} \citep{wood2011fast, wood2017generalized}.}
    \label{fig:AIC}
\end{figure}

\subsubsection{Testing for goodness-of-fit of relational event models}

Selecting the best model does not necessarily ensure that the model is adequate in itself. \textit{Goodness-of-fit} (GOF) assessment, in contrast, is not concerned with comparing models to one another but rather with evaluating how well the model assumptions align directly with the observed data \citep{amati2024goodness}. In general, a model is considered adequate if any observed lack of fit can be fully explained by the stochastic nature of the response variable. The guide by \citet{butts2017relational} and the online tutorial by \citet{statnet_relevent_workshop} provide some recommendations for assessing model adequacy in REMs.  For a comprehensive literature review of GOF in REMs, the reader is referred to \citet{bianchi2024relational}. In this tutorial, we refer to some practical approaches, highlighting their potential and weaknesses.

\textit{Simulation-based techniques} have been proposed in \citet{brandenberger2019predicting} and \citet{amati2024goodness}. 
Specifically, the former suggests simulating a segment of the event sequence using the estimated coefficients as parameters of the relational event data-generating process. The simulated events are then compared with the observed ones. This simulation relies on the piecewise constant hazard assumption, which may be inappropriate. The assumption is particularly restrictive in practical settings where global conditions evolve over calendar time or where endogenous network statistics are modelled explicitly as functions of time rather than as aggregated quantities. Furthermore, the researcher must decide which segment of the event sequence to replace, how many events to simulate, and which metrics to use for the comparison. The approach proposed by \citet{amati2024goodness} tests whether relevant omitted statistics are adequately captured by the fitted model. Like the previous method, it is simulation-based and therefore computationally intensive, as it requires computing endogenous statistics at each time point for every potential dyad at risk of interacting. The \texttt{R} package \texttt{relevent} \citep{relevent_package} allows for automatic simulation from the fitted model.

\begin{figure}[h]
    \centering
    \begin{minipage}{\textwidth}
        \begin{center}
        \begin{tikzpicture}
            \node[draw, thick, rectangle, rounded corners=3pt, inner sep=3pt] (box) {
                \begin{minipage}{0.9\textwidth}
                    \centering
                    \begin{adjustbox}{max width=\textwidth}
                        \begin{lstlisting}[language=R]
library(relevent)
# fit the relational event model
fit <- rem.dyad(edgelist=data_rem, n=p, effects="CovEvent", 
                covar=list(CovEvent=x_matrix))
# simulate according to the fitted model
sim_list <- lapply(1:n_sim_boot, function(i){ 
  simulate(fit, covar = list(CovEvent = x_matrix), verbose = TRUE, seed = i)})
# evaluate the auxiliary statistics on simulated and observed data
aux_stat <- function(dat, x_matrix) {
  mean(apply(dat, 1, function(x) x_matrix[x[2], x[3]]^2))}
sim_stats <- sapply(sim_list, function(sim) aux_stat(sim, x_matrix))
observed_stat <- aux_stat(data_rem, x_matrix)
                        \end{lstlisting}
                    \end{adjustbox}
                \end{minipage}
            };

            \node[draw, fill=blue!20, rounded corners=5pt, anchor=south west, font=\sffamily\bfseries\small, inner sep=4pt]
            at ([xshift=1em, yshift=-1em]box.north west) {R Code: Model fitting and GOF evaluation using \texttt{relevent}};
        \end{tikzpicture}
        \end{center}
    \end{minipage}
    % Row 2
    
    \vspace{1.5em}
    
    \begin{minipage}{0.45\textwidth}
        \centering
        \includegraphics[width=0.9\linewidth]{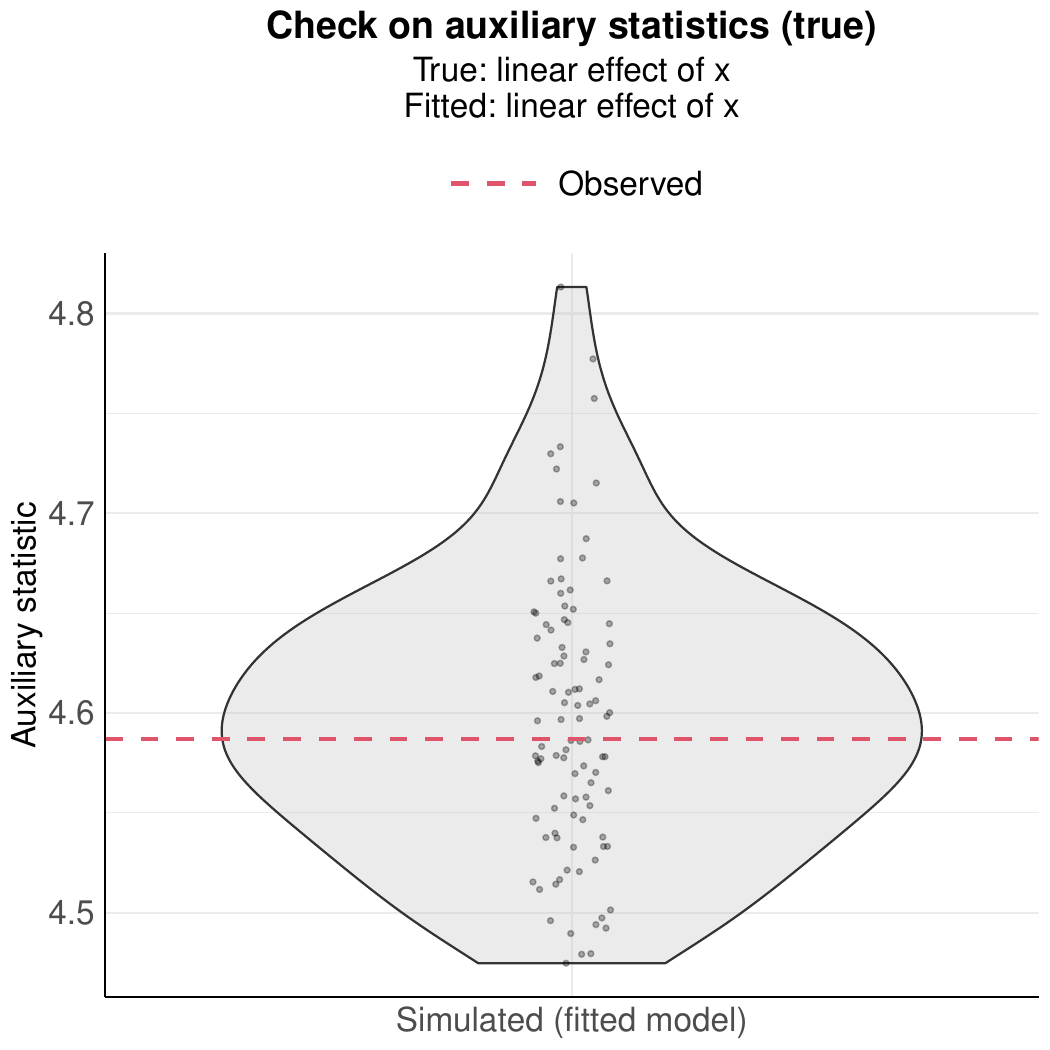}
    \end{minipage}
    \hfill
    \begin{minipage}{0.45\textwidth}
        \centering
        \includegraphics[width=0.9\linewidth]{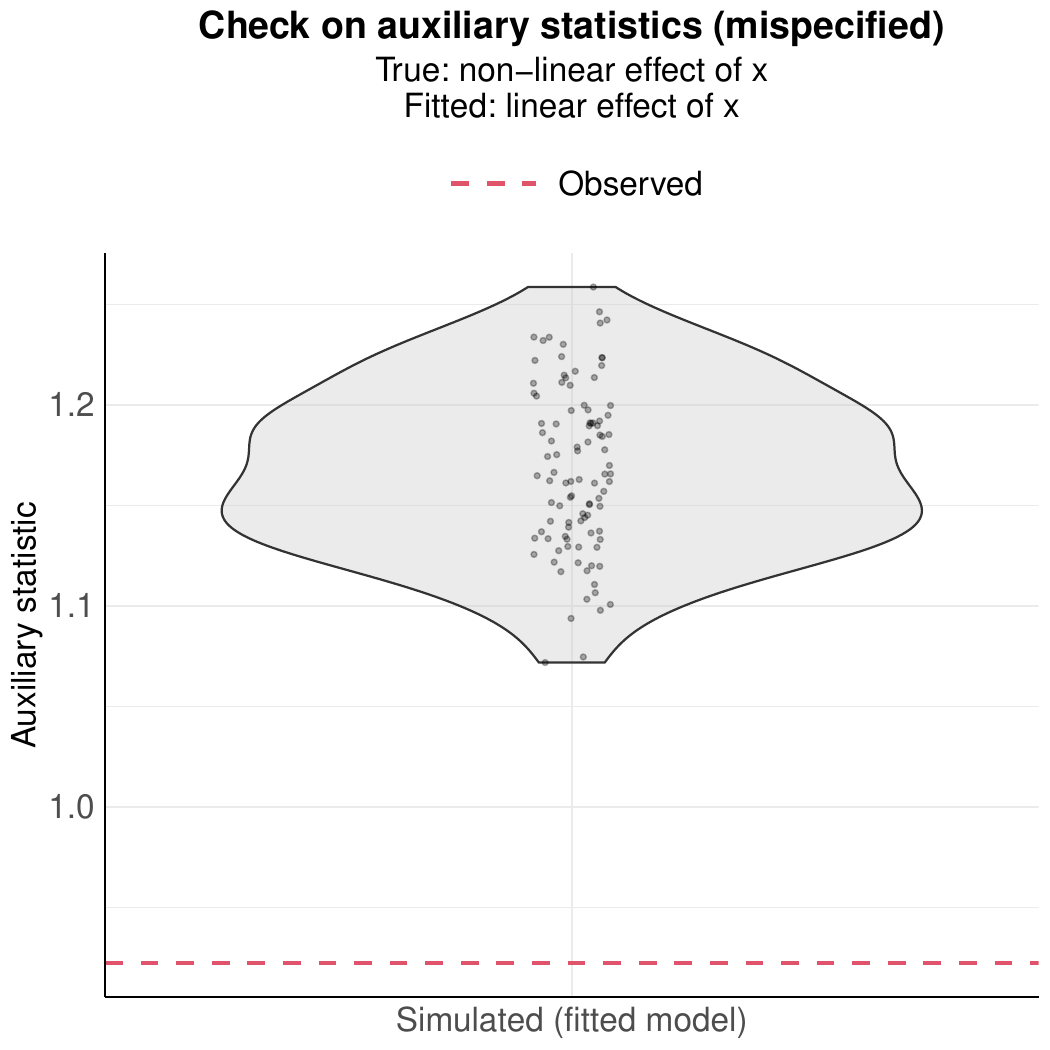}
    \end{minipage}
    \caption[Model fitting and GOF evaluation using \texttt{relevent}.]{\label{fig:relevent} \textbf{Model fitting and GOF evaluation using \texttt{relevent}.}
    \emph{Top.} \texttt{R} code used to fit a relational event model with \texttt{relevent} to the event sequence in \texttt{data\_rem} and to perform a goodness-of-fit check. \texttt{x\_matrix} contains the values of the covariate for all dyads at risk. A set of synthetic event sequences is generated under the fitted model \texttt{fit}. An auxiliary statistic is evaluated on each of the synthetic datasets and compared with the one observed on the original data. A statistic falling outside the simulated range signals a lack of fit. \emph{Bottom left.} Distribution of the auxiliary statistic across simulated datasets when the fitted model coincides with the true data-generating process.
    \emph{Bottom right.} Same comparison when the fitted model is misspecified.
    }
\end{figure}

In Figure~\ref{fig:relevent}, we fit a model using \texttt{relevent}, generate a number of synthetic datasets according to the fitted model, and evaluate an auxiliary statistic \citep{amati2024goodness} in each of the simulated datasets. We repeat this in two scenarios: first, when the model fitted to the data coincides with the true data-generating process, and second, when it does not. We see that the observed evaluation of the auxiliary statistic falls outside the range of evaluations in the misspecified case, while it does not for the correctly specified one.

The \emph{martingale residual approach} proposed by \citet{boschi2026goodness} enables GOF testing with lower computational cost than simulation-based techniques. It is also applicable to REMs that include time-varying, non-linear, and random effects. It 
relies on a cumulative sum of weighted martingale residuals, defined as a continuous process depending both on time $\tau=T_0+t(T-T_0)$ and the relative size of the order statistic $x^l_{(\lceil uN \rceil)}$,

\begin{equation}\label{eq:process-general}
\bm{G}^l(t,u;\bm{\gamma}) = \sum_{SR \in \mathcal{P}} \sum_{(s,r)\in SR} \int_{T_0}^{T_0+t(T-T_0)} \left( \bm{\phi}^l_{sr}(\tau,u) - \frac{\Phi^{l,(0)}_{SR}(\tau, u;\bm{\gamma})}{S^{(0)}_{SR}(\tau, \bm{\gamma})} \right) \mathrm{d}N_{sr,SR}(\tau), \quad t,u\in[0,1],
\end{equation}
with $S^{(0)}_{SR}(\tau,\bm{\gamma}) = \sum_{(s,r) \in SR} \exp\left( \bm{\gamma}^\top \bm{h}_{sr}(\tau) \right)$ and $\Phi^{l,(0)}_{SR}(\tau, u; \bm{\gamma}) = \sum_{(s,r) \in SR} \phi^l_{sr}(\tau,u) \cdot \exp\left( \bm{\gamma}^\top \bm{h}_{sr}(\tau) \right)$, where crucially the practitioner needs to define the statistic $\bm{\phi}^l$ for which they want to test the goodness-of-fit, 
\begin{equation*}
        \bm{\phi}^l_{sr}(\tau,u) = \mathds{1}\big(x^l_{sr}(\tau) \le x^l_{(\lceil Nu \rceil)}(\tau)\big) \times a\left(\bm{h}_{sr,\bm{j}_l}(\tau)\right)
\end{equation*}
where $N$ is the size of the pooled sample $\{x^l_{s_ir_i}(t_i)\}_{i=1}^n \cup \{x^l_{s_i^*r_i^*}(t_i)\}_{i=1}^n$, and $x^l_{(1)}\le\cdots\le x^l_{(N)}$ are its order statistics. 
Depending on the scope of the test, we define \(\bm{\phi}^l_{sr}(\tau,u)\) differently.

\paragraph{Testing for temporal misspecification.} Temporal misspecification can occur when an effect varies over time, but we fail to incorporate it in the model. We can test for temporal misspecification of the effect of covariate $x^l$ by (i) selecting the  associated columns $\bm h_{\bm{j}_l}$ of the covariate in the model matrix, (ii) setting $u\equiv 1$ and  (iii) letting  $a$ be the identity map, resulting in
\begin{equation*}
\bm\phi^l_{sr}(\tau,1) \;=\;   \bm h_{sr,\bm j_l}(\tau).
\end{equation*}
Substituting $\bm\phi^l_{sr}(\tau,1)$ into $\bm G^l(t,u;\bm\gamma)$ can be shown to be equivalent to the cumulative, time-indexed unpenalized score process of the coordinates $\bm j_l$, with convenient distributional properties exploiting the asymptotic properties of the score. Corrections are available in the case of a (smoothing) penalty in the estimation objective.

\paragraph{Testing for functional form misspecification.}
Instead, to test the misspecification of the functional form of the effect of the covariate or (unmodelled) auxiliary statistic $x^l$, we (i) fix  $t=1$ in $\bm G^l$ and (ii) set $a\equiv 1$, resulting in 
\begin{equation*}
\phi^l_{sr}(\tau ,u)  \;=\; I\big(x^l_{sr}(\tau) \le x^l_{(\lceil Nu \rceil)}(\tau)\big).
\end{equation*}
As the statistic $\bm \phi^l$ does not coincide with an element of the model matrix, we can no longer use the properties of the score. However, following \citet{borgan2015using}, it is possible to approximate the distribution of the process and simulate from it.

\begin{figure}[h]
    \centering
    \begin{minipage}{\textwidth}
        \begin{center}
        \begin{tikzpicture}
            \node[draw, thick, rectangle, rounded corners=3pt, inner sep=3pt] (box) {
                \begin{minipage}{0.9\textwidth}
                    \centering
                    \begin{adjustbox}{max width=\textwidth}
                        \begin{lstlisting}[language=R]
source("GOF.R")
library(mgcv) 
x_matrix <- cbind(ncc_data$x_ev, ncc_data$x_non_ev)
ncc_data$delta_x <- ncc_data$x_ev - ncc_data$x_non_ev
# fit the relational event models
rem_fit <- gam(y ~ -1 + s(x_matrix, by=by_matrix),
               family="binomial"(link = "logit"), data = ncc_data)
rem_fit_mis <-  gam(y ~ -1 + delta_x,
                    family="binomial"(link = "logit"), data = ncc_data)
# GOF test for correct time specification and functional form 
GOF(rem_fit = rem_fit, test = "time", time = ncc_data$tms)
GOF(rem_fit = rem_fit_mis, test = "functional_form"
    covariate_event = ncc_data$x_ev, covariate_nonevent = ncc_data$x_nv)
                        \end{lstlisting}
                    \end{adjustbox}
                \end{minipage}
            };

            \node[draw, fill=blue!20, rounded corners=5pt, anchor=south west, font=\sffamily\bfseries\small, inner sep=4pt]
            at ([xshift=1em, yshift=-1em]box.north west) {R Code: Model fitting and GOF evaluation using \texttt{martingale residual GOF test}};
        \end{tikzpicture}
        \end{center}
    \end{minipage}
    % Row 2
    
    \vspace{1.5em}
    
    \begin{minipage}{0.45\textwidth}
        \centering
        \includegraphics[width=\linewidth]{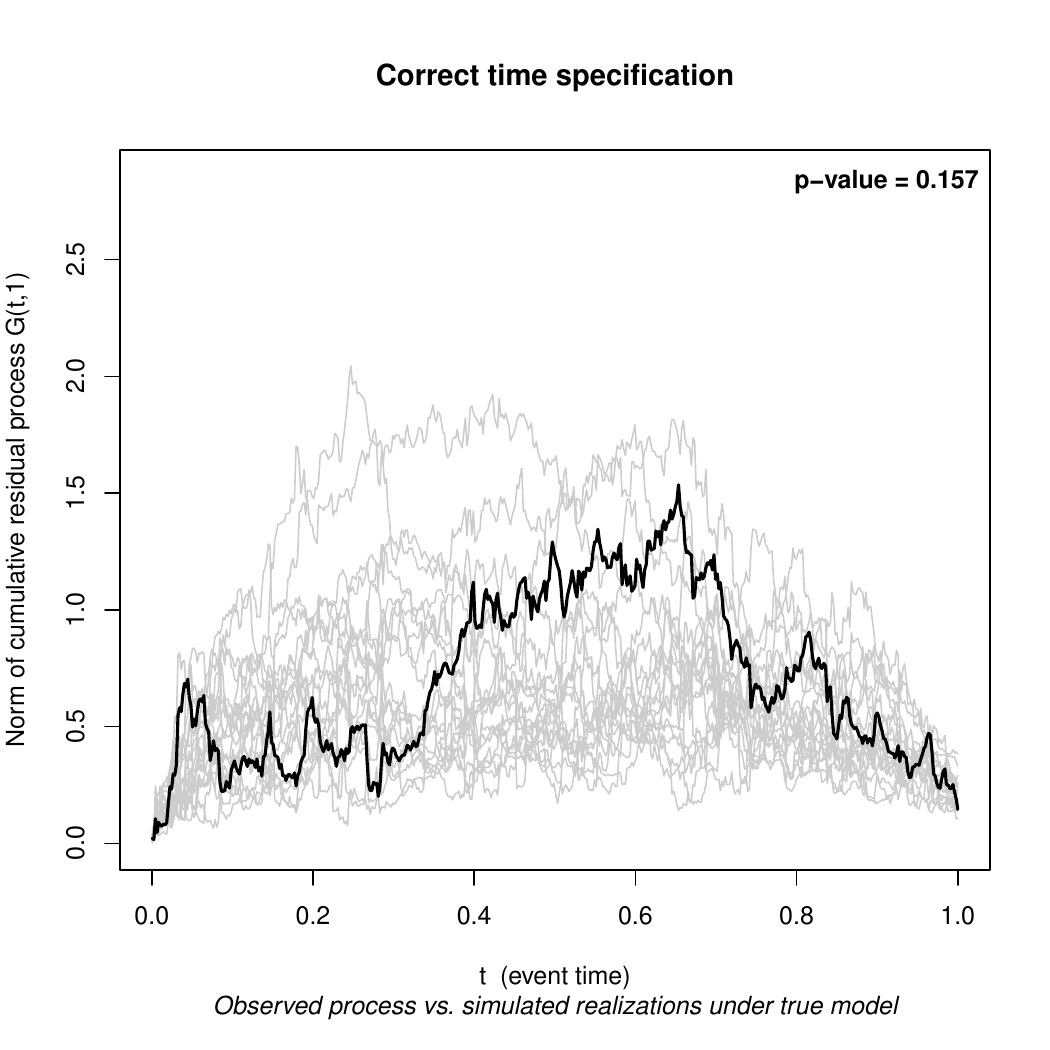}
    \end{minipage}
    \begin{minipage}{0.45\textwidth}
        \centering
        \includegraphics[width=\linewidth]{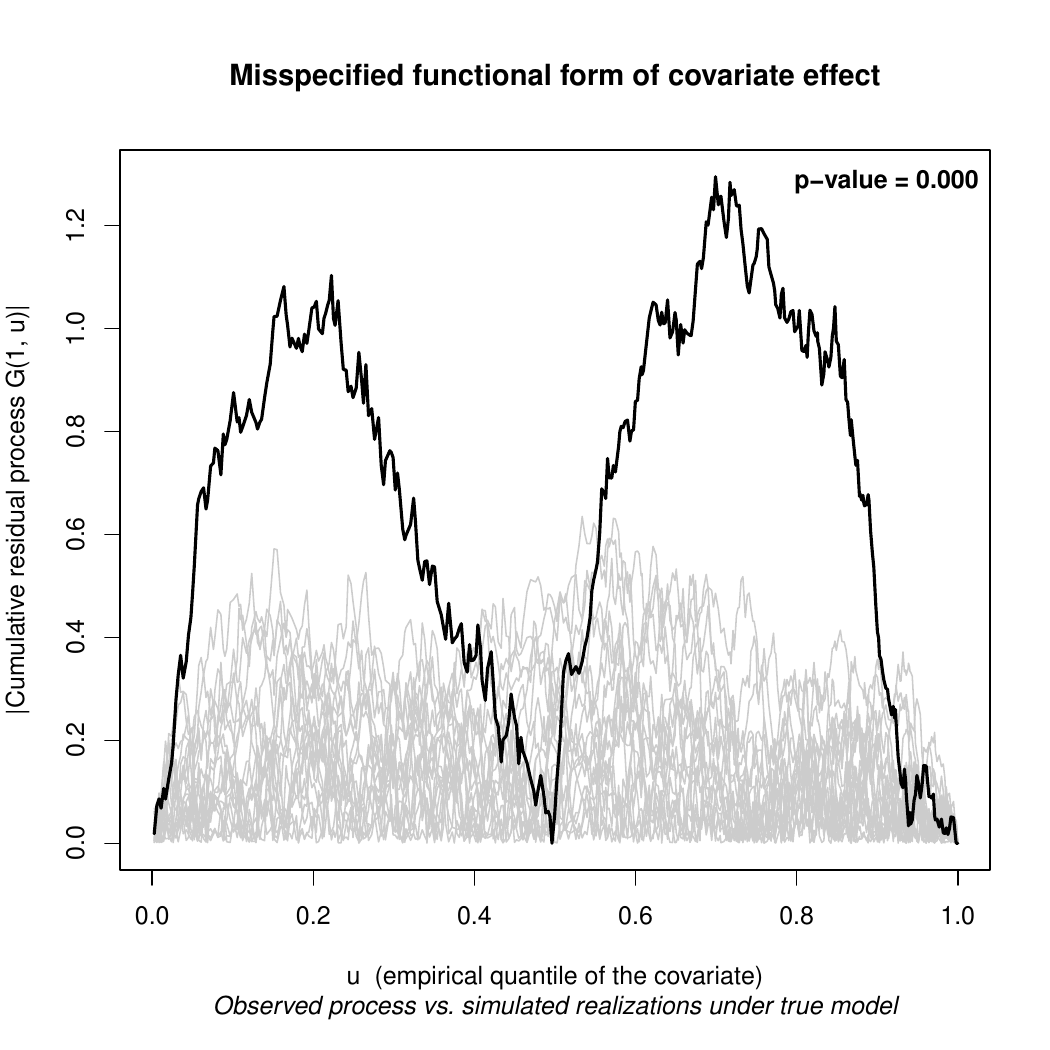}
    \end{minipage}
    \caption[Flexible model fitting and related GOF evaluation.]{\label{fig:gof} \textbf{Flexible model fitting and related GOF evaluation}. \emph{Top.} \texttt{R} code used to fit a relational event model and to perform a goodness-of-fit check. \emph{Bottom.} Observed and simulated martingale residual processes (Equation~\eqref{eq:process-general}) for a correctly specified model (\textit{Left.}) and a misspecified model (\textit{Right.}). The former is tested for correct time specification, while the latter is tested for correct functional form specification.}
\end{figure}

Figure~\ref{fig:gof} shows the application of two tests: the first tests whether time is properly modelled in a correctly specified model, while the second tests whether the functional form of the covariate is properly modelled in a misspecified model, where a linear model is fitted to data generated according to a non-linear model. The tests correctly identify the, respective, correct and incorrect model specifications via the p-values $0.157$ and $0.000$. In the Supplementary Materials, the reader can find the code necessary to implement these kinds of test. 

\clearpage

\section{Relational Event Modelling in Practice}\label{sec:applications}

This section is dedicated to the practical application of the concepts introduced earlier. First, we apply relational event modelling to synthetic data, allowing us to explore the model’s behaviour in a controlled setting. Subsequently, we turn to several empirical datasets to demonstrate the flexibility and applicability of the model across a variety of contexts.

\subsection{Generating a synthetic relational event dataset}\label{subsec:simulation}
This section serves multiple purposes. First, it explains how to generate synthetic datasets based on relational event models. In particular, we describe how the datasets used in Sections~\ref{sec:overview}, \ref{sec:modeling}, and \ref{sec:inference} were constructed. Second, it provides a framework for validating the modelling and inference procedures introduced in the latter two sections.

\subsubsection{Synthetic data generation}\label{subsubsec:data-generation}

\begin{table}[!h]
\centering
\resizebox{\linewidth}{!}{
\fontsize{9}{11}\selectfont
\begin{tabular}[t]{llllllll}
\hline
Time (Event) & Origin (Event) & Destination (Event) & Weekeday (Event) & Time (Non-Event) & Origin (Non-Event) & Destination (Non-Event) & Weekeday (Non-Event)\\
\hline
... & ... & ... & ... & ... & ... & ... & ...\\
381.29 & Iowa & Indiana & 1 & 412.13 & Guam & New Hampshire & 1\\
381.32 & Vermont & Washington & 1 & 307.6 & Maine & Hawaii & 1\\
381.85 & Colorado & Rhode Island & 1 & 212.71 & Washington & Guam & 1\\
382.91 & West Virginia & Texas & 1 & 458.89 & Washington & Delaware & 1\\
\addlinespace
383.11 & Maryland & Guam & \cellcolor[HTML]{FFE4E1}{1} & 77.74 & Maine & Louisiana & \cellcolor[HTML]{FFE4E1}{0}\\
383.29 & Arizona & Nebraska & 1 & 80.81 & Maine & New Jersey & 1\\
383.43 & New York & Hawaii & \cellcolor[HTML]{FFE4E1}{1} & 316.09 & Rhode Island & Georgia & \cellcolor[HTML]{FFE4E1}{0}\\
383.5 & Connecticut & Wisconsin & \cellcolor[HTML]{FFE4E1}{1} & 287.01 & New Jersey & West Virginia & \cellcolor[HTML]{FFE4E1}{0}\\
383.98 & New York & Texas & 1 & 47.28 & Iowa & United States Virgin Islands & 1\\
\addlinespace
384.38 & Florida & North Carolina & 1 & 362.61 & South Dakota & Montana & 1\\
384.63 & Washington & Vermont & \cellcolor[HTML]{FFE4E1}{1} & 434.26 & Connecticut & Colorado & \cellcolor[HTML]{FFE4E1}{0}\\
... & ... & ... & ... & ... & ... & ... & ...\\
\hline
\end{tabular}
}
\caption{\label{tab:synt-shifted-RCCD}\textbf{Synthetic shifted augmented dataset}. The first three columns report observed events, including the event time, origin state, and destination state. The same structure is used for non-events. The time associated with non-events differs from that of observed events due to the shifting procedure, whereby it is first shifted and then transformed back to the original scale. Due to the time shift, even though \texttt{Weekday} is a global covariate, the difference is not always zero.}
\end{table}

In many situations, it is useful to simulate data to validate the techniques a researcher intends to apply. Simulation becomes particularly valuable when models are applied to empirical contexts that may exhibit unique or atypical characteristics. In such cases, generating synthetic datasets with similar properties to the observed data can help assess whether fitting a REM remains a suitable approach. This strategy, for example, was employed in the study conducted by \citet{juozaitiene2023analysing}.

The standard approach relies on the Gillespie algorithm, which iterative samples exponential event times and multinomial events. Details, the related algorithm, and implementation code are reported in the Supplementary Materials. This procedure may become ineffective when a covariate or the effect of a covariate change continuously over time. In that case, one can refer to the \textit{$\tau$-leap method} \citep{gillespie2001approximate}. The $\tau$-leap method divides the event history into sub-intervals and recording the number of events occurring within each time window. 

Table~\ref{tab:synt-shifted-RCCD} displays sampled non-events from the shifted risk set corresponding to the events shown in Table~\ref{tab:synt-RCCD}. These non-events differ not only because of the inherent randomness of the sampling procedure, but also because of a fundamental distinction: they are drawn based on the risk set at the shifted time. This shift affects the values of global covariates. For instance, the weekday associated with some events differs from that of their corresponding non-events, reflecting the fact that the latter are evaluated at different time points.

The data reported in Table~\ref{tab:synt-shifted-RCCD} are just one instance of the 100 simulated datasets to which the model was fitted. Results are provided in the following paragraph.

\subsubsection{Validation of REM techniques}\label{subsubsec:sim-based-validation}

When the true data-generating process is known, it becomes possible to directly compare the results obtained from fitted models with the true parameter values -- either derived from statistical theory or set by the user. In such cases, and when the model is correctly specified, the resulting estimators can be shown to be unbiased. For simplicity, in Section~\ref{sec:overview}, we did not explicitly emphasise that the variable \texttt{Weekday} influences the data-generating process. However, it was indeed incorporated as a global covariate in the data simulation. We therefore revise Equation~\eqref{eq:model-overview} to reflect the rate function underlying the generation of the synthetic data, as follows:
\begin{eqnarray}\label{eq:model-simulation1}
	\lambda_{sr}(t) &=& I_{sr}(t) \times \lambda_{0}(t) \times \exp{\{ \beta^\textmd{rec} \cdot x_{\textmd{sr}}^{\textmd{rec}}(t) + \sum_{j=1}^q \theta_j \cdot b_j[x_{sr}^{\textmd{dist}}(t)]\}} \\
	\lambda_0(t) &=& \lambda_0 \times \exp\{\beta^\textmd{wd} \cdot x^{\textmd{wd}}(t)  \label{eq:model-simulation2} \}
\end{eqnarray}

\begin{figure}[h] 
	\centering
	\includegraphics[width=0.48\linewidth]{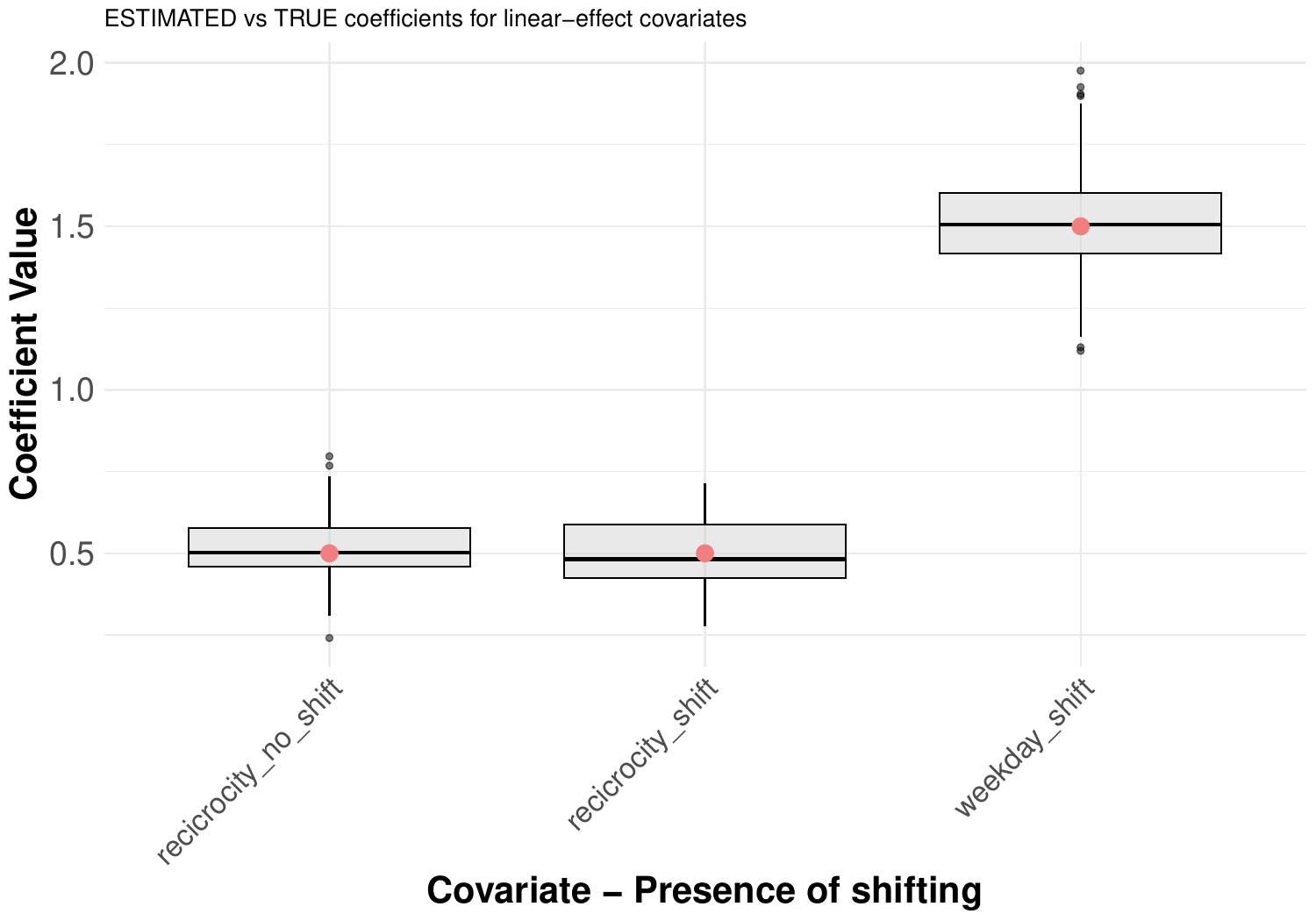}
	\includegraphics[width=0.48\linewidth]{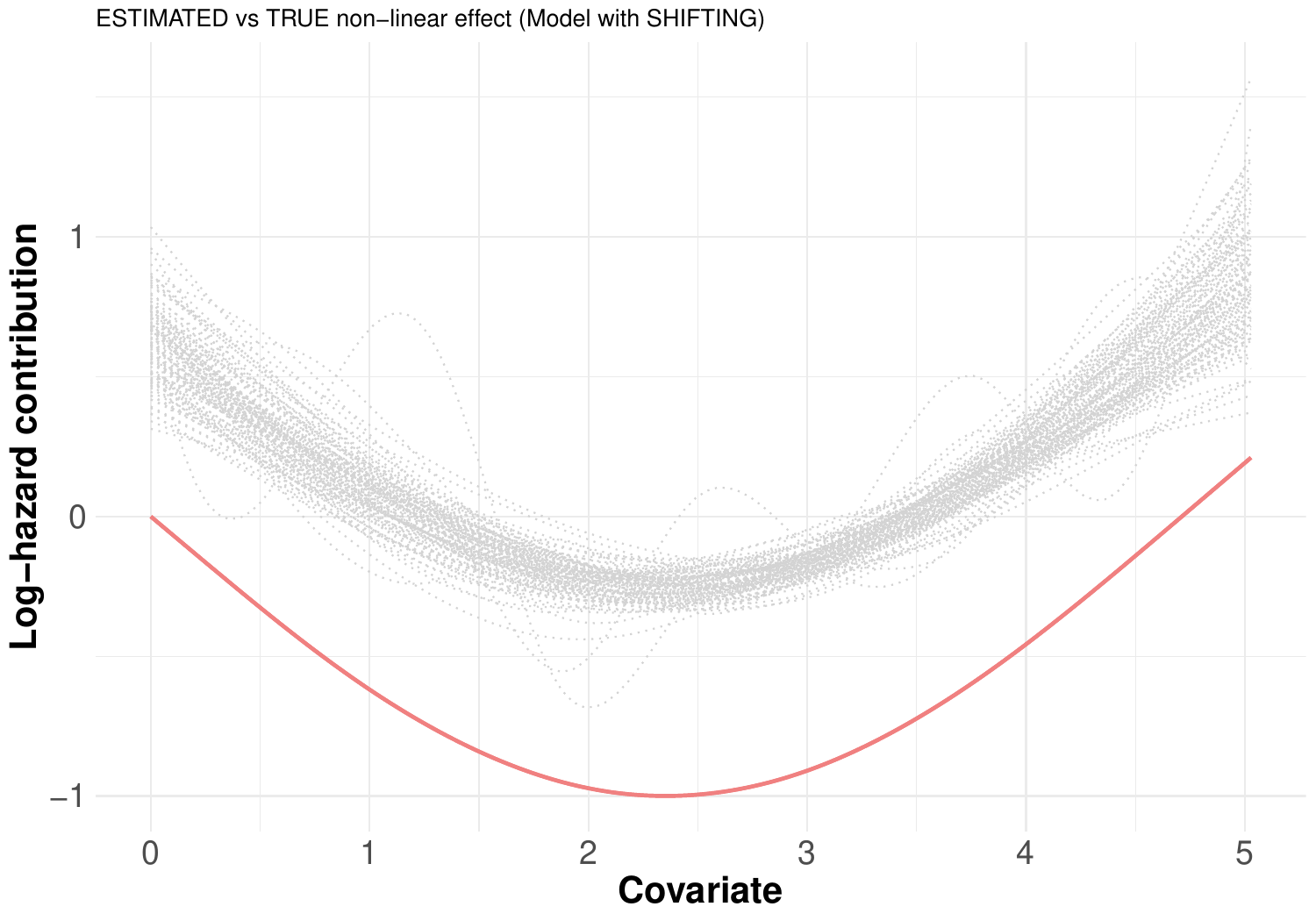}
	\caption[Simulation study: results.]{\label{fig:results_simulation} \textbf{Simulation study: results.} \textit{Left}. The estimated coefficients for fixed-effect covariates are unbiased both by fitting the model using unshifted and shifted non-events. \textit{Right}. Estimated curve for the effect of distance appropriately represents the true underlying trend of the contribution of this covariate to the log-rate. The estimated curve differs from the true one by a constant.} 
\end{figure}

Figure~\ref{fig:results_simulation} demonstrates that the estimates obtained by fitting models~\eqref{eq:model-simulation1} and~\eqref{eq:model-simulation2} are unbiased for fixed-effect covariates and successfully recover the underlying trend for covariates with non-linear effects. The estimated smooth function deviates from the true curve by a constant offset -- a difference that is expected, as discussed in Section~\ref{sec:overview}, due to the internal centring constraint applied during the inference procedure. 

\subsection{Empirical illustrations}\label{subsec:application}
One of the key strengths of REMs lies in their broad applicability across a wide range of empirical contexts. This section illustrates that versatility through several examples. We present two cases involving unimodal (or one-mode) relational graphs, where all nodes belong to the same category. We then consider an application based on a bimodal (or two-mode) relational graph, in which senders and receivers represent different types of entities.

\subsubsection{Basics: modelling radio communications in an emergency context}

This empirical example was explored in the foundational REM paper  \citep{butts2008relational}. Here, we aim to schematically recap some of the basic steps involved in fitting a REM in a setting where it remains computationally feasible to consider the complete risk set during inference.

\paragraph{Data.} The data capture sequences of radio communications among groups of emergency responders during the World Trade Center (WTC) disaster in 2001, including interactions both at the WTC site and at other related locations \citep{butts2008relational}. While exact timestamps for the communications are not available, the order of events is given, allowing the data to be represented as a temporal network. This temporal network is unimodal, as all entities involved are responders. 

The dataset we focus on comprises 481 radio communications among 37 actors, three of whom are identified as having coordinative responsibilities. Information about these role distinctions is incorporated into the analysis as an exogenous covariate.  We assume that all pairs among the 37 actors are at risk of engaging in radio communication throughout the entire observation period, with the exception of self-loops; that is, communications between an actor and itself are not considered potential events. This implies that 1,332 possible dyads are at risk at each time point, although in practice only one event is observed.

\paragraph{Model formulation.} At this stage, the aim of the analysis is not to offer a detailed sociological interpretation of the data. Rather, the focus is on demonstrating the potential, flexibility, and practical implementation of relational event modelling, along with the tools available to conduct such analyses. With this premise in mind, we use \href{https://github.com/juergenlerner/eventnet}{\texttt{eventnet}} \citep{lerner2020reliability} to compute the explanatory variables for this application.

For this application, we consider a model formulation with seven explanatory variables, five of which are endogenous and computed using the volume definition (see Table~\ref{tab:endogenous}), namely: \texttt{Sender Activity}, \texttt{Receiver Popularity}, \texttt{Reciprocity}, \texttt{Repetition}, and Transitive \texttt{Closure}. Two other explanatory variables indicate whether the sender and the receiver of the event have coordinative responsibilities (\texttt{Sender is ICR} and \texttt{Receiver is ICR}, respectively). The resulting model is:

\small
\begin{equation*}
	\lambda_{sr}(t) = \lambda_{0}(t) \times \exp{\{ 
    \beta^1x_{s}^{\textmd{deg},V}(t) + 
    \beta^2x_{r}^{\textmd{deg},V}(t) + 
    \beta^3x_{sr}^{\textmd{rep},V}(t) + 
    \beta^4x_{sr}^{\textmd{rec},V}(t) + 
    \beta^5x_{sr}^{\textmd{trs},V}(t) + 
    \beta^6x_{s}^{\textmd{ICR}}(t) + 
    \beta^7x_{r}^{\textmd{ICR}}(t)
    \}}.
\end{equation*}
\normalsize

\begin{figure}[t]
    \centering
    \includegraphics[width=0.7\linewidth]{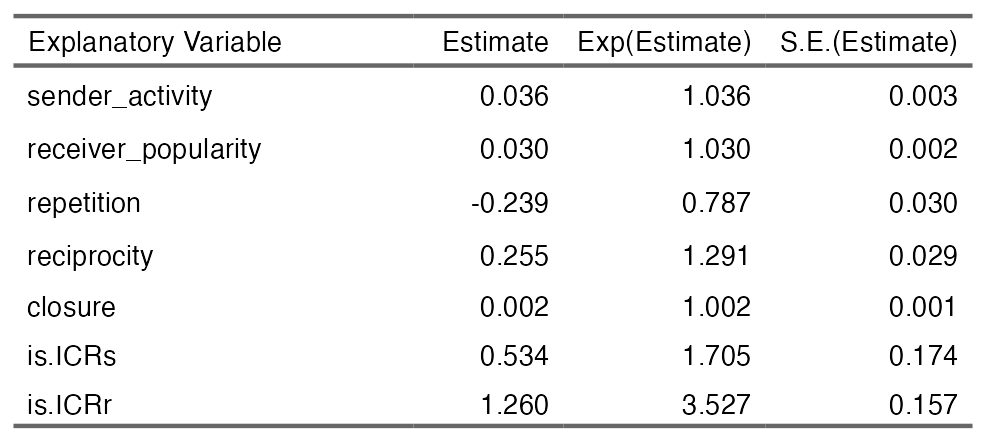}
    \caption[Summary of the relational event model fitted to the WTC radio communication data using \texttt{clogit} in \texttt{R}.]{\textbf{Summary of the relational event model fitted to the WTC radio communication data using \texttt{clogit} in \texttt{R}}. The output presents the estimated coefficients, their exponentiated values (allowing for interpretation relative to the baseline), and the corresponding standard errors for each explanatory variable. Positive coefficients indicate a higher event rate relative to the baseline risk condition, holding other factors constant.}
    \label{fig:summary_WTC}
\end{figure}

\paragraph{Inference procedure and results.}
When it comes to inference procedures, given the small size of the dataset and the absence of global covariates, we are able to compute explanatory variables for all non-events at each time point and use the partial likelihood as in Equation~\eqref{eq:partial_likelihood}. The vector of parameters \(\boldsymbol{\beta} = (\beta^1, \ldots, \beta^7)\) includes the linear contribution of the covariates described above to the log-rate, and \(|\mathcal{R}_{t_i}| = 1332~ \forall i = 1,\dots,481\). Estimates that maximize Equation~\eqref{eq:partial_likelihood} can be found using the functions \texttt{coxph} or \texttt{clogit} in the \texttt{R} package \texttt{survival}, as shown in Figures~\ref{fig:coxph_regression} and \ref{fig:clogit_regression}. The \texttt{eventnet} configuration and \texttt{R} code are available in the linked \href{https://github.com/martinaboschi/rem-tutorial/tree/main/03-Tutorial-Paper}{GitHub repository} for reproducibility. The obtained estimates and their corresponding standard errors are also reported in Figure~\ref{fig:summary_WTC}.

Regarding model interpretation, \texttt{Sender Activity} and \texttt{Receiver Popularity} have positive coefficients, indicating that higher prior activity by the sender and greater previous receptiveness of the receiver increase the instantaneous rate of an event occurring. Similarly, \texttt{Reciprocity} shows a positive effect, suggesting that communication often responds to prior communication. In contrast, \texttt{Repetition} exhibits a negative effect, meaning that calls between the same pair of actors tend not to be repeated. The positive and significant coefficient for \texttt{Closure} suggests the presence of a path-shortening dynamic in these communications. Additionally, the positive effects of the role variables for both sender and receiver highlight the important function of coordinators in this emergency context. In particular, coordinators are more than three times as likely to receive radio communications as others, holding other factors constant. \\

Whenever the risk set is not manageable, it is very useful to sample it. For sake of comparison, we consider other three estimates of \(\bm{\beta}\) using varying sample sizes. Table \ref{tab:clogit_results} reports the estimates from four conditional logistic regression models, each fitted using either the complete risk set or a subsample of 20, 5, or 1 non-events per case. All four models are based on the full specification, incorporating all computed explanatory variables. As expected, the standard errors increase as the number of sampled non-events decreases. Consistency of the estimators has been demonstrated by \citet{borgan1995methods}. However, the computational advantage achieved by sampling come at the cost of greater uncertainty in the estimates.

\begin{table}[h]
\centering
\begin{tabular}{lcccc}
  \hline
Variable & Complete Risk Set & m = 20 & m = 5 & m = 1 \\
\hline
Sender Activity & 0.036 (0.003) & 0.025 (0.003) & 0.020 (0.005) & 0.047 (0.015) \\ 
  Receiver Popularity & 0.030 (0.002) & 0.023 (0.002) & 0.024 (0.004) & 0.040 (0.013) \\ 
  Repetition & -0.239 (0.030) & -0.322 (0.083) & -0.287 (0.157) & -0.169 (0.290) \\ 
  Reciprocity & 0.255 (0.029) & 0.535 (0.084) & 1.310 (0.183) & 1.028 (0.318) \\ 
  closure & 0.002 (0.001) & 0.003 (0.001) & 0.002 (0.001) & -0.002 (0.004) \\ 
  Sender is ICR & 0.534 (0.174) & 0.355 (0.222) & 0.169 (0.335) & 0.298 (0.518) \\ 
  Receiver is ICR & 1.260 (0.157) & 1.296 (0.193) & 1.365 (0.295) & 0.901 (0.512) \\ 
   \hline
\end{tabular}
\caption[Coefficients and standard errors for decreasing \(m\).]{\textbf{Coefficients and standard errors for variable number of sampled non-events \(m\)} when modelling radio communications in an emergency context.} 
\label{tab:clogit_results}
\end{table}

\subsubsection{Beyond linearity: modelling a bimodal network of alien species invasions}

Inspired by the analyses in \citet{boschi2025mixed} and \citet{juozaitiene2023analysing}, the alien species invasions example has several goals. First, it demonstrates the potential of relational event models in fields beyond social network analysis. Second, it highlights how the logistic regression formulation increases the flexibility of these models. Third, in this application, the risk set is not a static object but varies over time.

The relational events of interest are alien species' first records, already introduced in the context of Example \ref{ex:alienspecies}. \textit{Alien species} are those introduced into a new ecosystem where they are not native, while first records refer to the first year in which a particular species is detected in a given region \citep{seebens2017no}. To define the risk set at each time point, two key aspects must be considered. First, recall that the native range is defined as the set of areas where a species is indigenous \citep{boschi2025mixed}. Second, the event of a first record is non-recurrent, since once it occurs, it cannot happen again for the same species in the same region. Therefore, to accurately define the risk set, it is necessary to have knowledge of both the native range of each species under study and alien species invasions that have already occurred by each time point. The complement of this more liberally defined native range defines the corresponding time-stamped risk set. We emphasize that the native range will also be employed in the computation of covariates, which depend on knowing which first records have already taken place.

\paragraph{Data.} The Alien Species First Record Database contains more than 47,000 first records of established alien species \citep{seebens2018global}. In this empirical application, we focus on 799 first records involving mammals, covering 186 species reaching 153 regions of the world between 1880 and 2005. Information related to the native ranges of species (including 3030 species-region pairs) comes from IUCN range maps (\url{https://www.iucnredlist.org}, accessed 08.07.2016).

\paragraph{Model formulation.} As with the other empirical applications, our goal is not to find a model that fully explains the data at hand. Rather, our objective is to demonstrate the flexibility of relational event modelling in these contexts. In this case, we consider three well-known drivers of alien species invasions -- distance, trade, and temperature -- operationalised through endogenous covariates, along with the intrinsic invasiveness of each species, modelled using a random effect. Information regarding the computation of endogenous covariates and their sources is summarized in Table~\ref{tab:explanatory_variables}. Further details can be found in \citet{juozaitiene2023analysing} and \citet{boschi2025mixed}. We propose a mixed-additive model:
\begin{equation}\label{eq:model-alien}
	\lambda_{sr}(t) = \lambda_{0}(t) \times \exp{\{ 
    \beta^{\textmd{d\_temp}} \cdot x_{sr}^{\textmd{d\_temp}}(t) + 
    \beta^{\textmd{trade}}(t) \cdot x_{sr}^{\textmd{trade}}(t) + 
    f^{\textmd{dist}}(x_{sr}^{\textmd{dist}}(t)) + 
    \gamma_{s}^{\textmd{inv}} 
    \}}, \quad \gamma_s^{\textmd{inv}} \sim \mathcal{N}(0, \phi^{\textmd{inv}}),
\end{equation}
including a linear effect for \texttt{Diff\_Temperature}, a time-varying effect for \texttt{Log\_Trade}, a non-linear effect for \texttt{Log\_Distance}, and a random intercept for species invasiveness.

\paragraph{Inference procedure and results.} Since the model does not include global covariates, inference is performed by means of non-shifted case-control partial log-likelihood \eqref{partial-GAM}. 
While fitting a linear effect on \texttt{Diff\_Temperature}, we obtain a negative estimate. This suggests that the rate of invasions tends to increase when a species has already invaded countries with similar temperature profiles. In contrast to the effect of \texttt{Diff\_Temperature}, we allow the effect of \texttt{Log\_Trade} to vary over time. Importantly, unlike non-linear effects whose sign cannot be directly interpreted, the sign of smooth time-varying effects remains interpretable. 
\begin{figure}[h]
    \centering
    \begin{minipage}{0.48\textwidth}
        \centering
        \includegraphics[width=\linewidth]{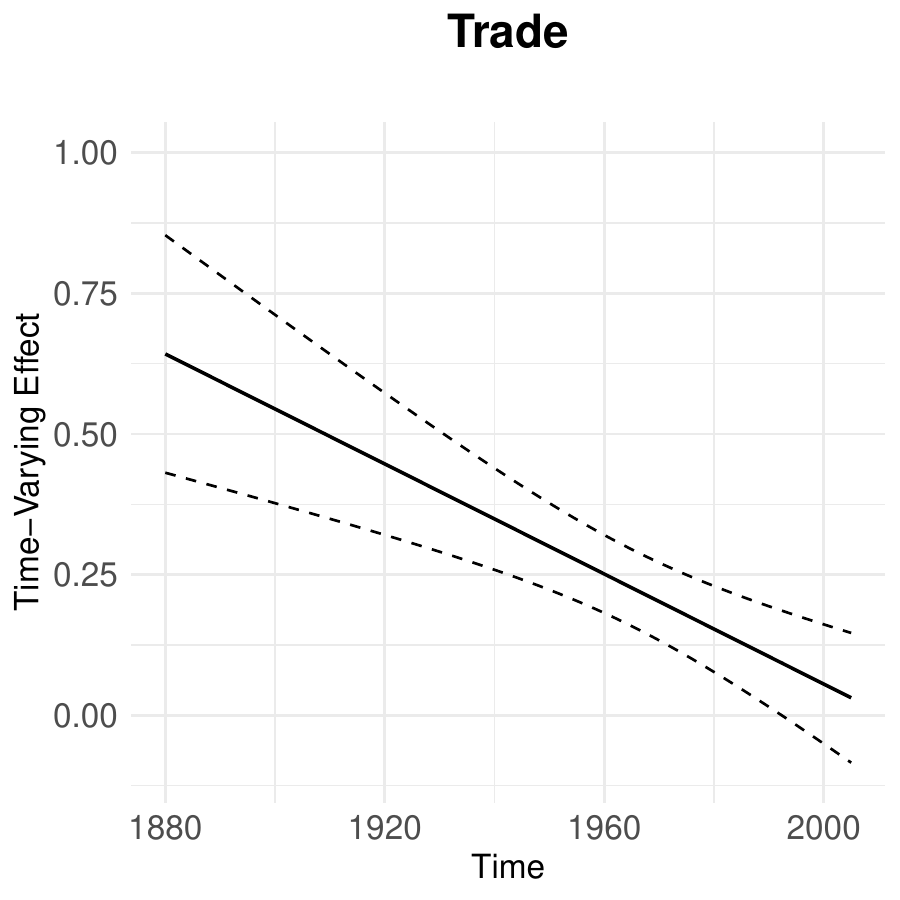}
    \end{minipage}
    \hfill
    \begin{minipage}{0.48\textwidth}
        \centering
        \includegraphics[width=\linewidth]{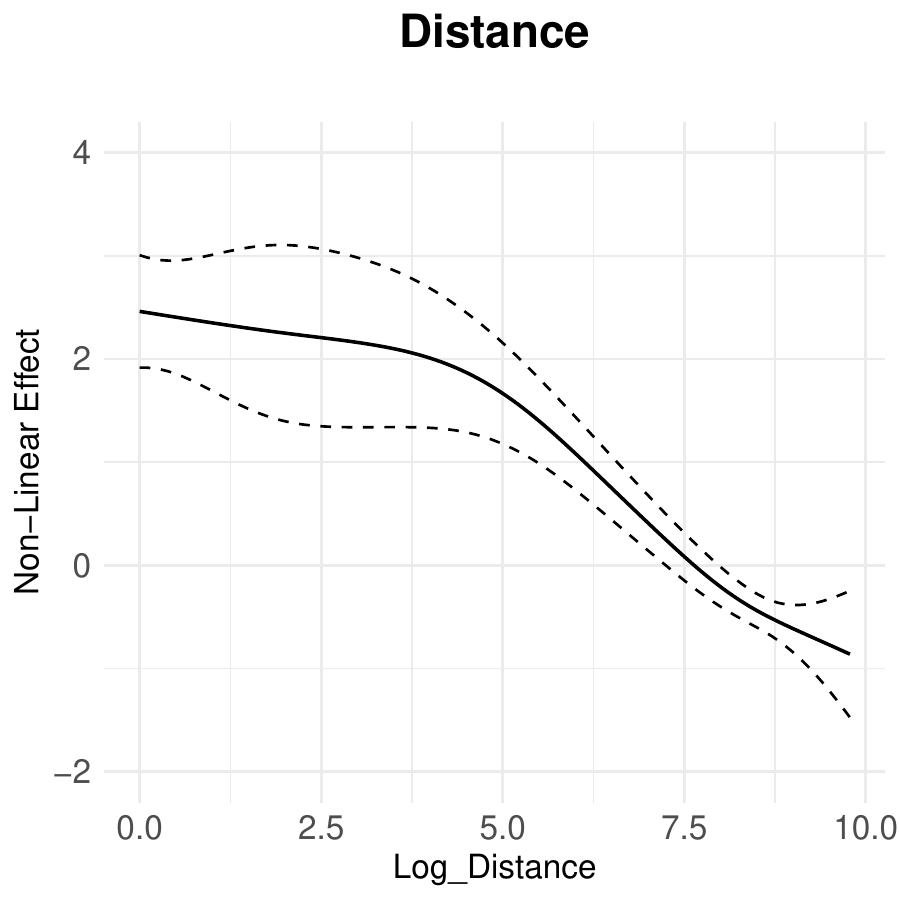}
    \end{minipage}
    \caption[Estimated time-varying and non-linear effects in alien species invasions application.]{\textbf{Estimated time-varying and non-linear effects in alien species invasions application}. The time-varying effect, estimated via a spline function of time, is interpretable in terms of its sign. In this case, a positive but decreasing effect is observed for \textit{Trade}. In contrast, non-linear effects are interpretable only in terms of their shape or trend. For \textit{Distance}, the estimated effect decreases with increasing values of distances, indicating a reduction of the rate of events for larger distances.}
    \label{fig:species-effects}
\end{figure}
Figure~\ref{fig:species-effects} \emph{Left} reports the estimated time-varying effect. The estimated positive but decreasing effect confirms that international trade has been a key driver in explaining the spread of alien species \citep{seebens2018global}. However, this influence appears to diminish over time. Practically speaking, whereas in 1880 a unit of trade between two countries made it $e^{0.63}=1.9$ times more likely for species present in one country to show up in the other country, by 1960 this effect had reduced to a mere increase of 28\% ($e^{0.25}=1.28$, to almost completely disappear by the end of the study in 2005. One possible explanation is the evolving nature of global trade, which has shifted from the exchange of high-risk raw materials to lower-risk processed goods \citep{juozaitiene2023analysing}. We note that, in the literature, the effect of distance has generally been modeled either as a fixed effect or as a time-varying linear effect, without accounting for potential non-linearities. In contrast, our analysis reveals a clear non-linear relationship between distance and the rate of invasions. As shown in Figure~\ref{fig:species-effects} \emph{Right}, the effect of \texttt{Log\_Distance} decreases as distance increases, indicating that shorter-distance invasions are more likely. For example, a region that is 150 km ($=e^5$) removed from an invaded region is $e^{1.7-0.1}=5$ times more likely to be invaded than a region that is 1,800 km ($=e^{7.5})$ removed, and $e^{1.7-(-1)}=14.9$ times more likely to be invaded than a region that is 22,000 km ($=e^{10})$ removed. Species with the largest estimated positive random effectss -- \textit{Procyon lotor}, \textit{Rattus rattus}, and \textit{Myocastor coypus} -- are widely recognized as pervasive invasive mammals \citep{salgado2018raccoon, shiels2014biology, carter2002review}.

\subsubsection{Time explains: modelling a unimodal network of emails in a company}

In this section, we show how the availability of timing information can further enhance the model. Specifically, we aim to incorporate timing information from two perspectives: first, as internal time reflecting the dynamics of relational mechanisms, and second, as calendar time capturing the temporal context of events.

\paragraph{Data.} For this analysis, we examine another common setting for relational events: email communication, already introduced in Example \ref{ex:communication}. Specifically, we analyze email exchanges among employees of a manufacturing company in Poland \citep{michalski2014seed, michalski2011matching}. We focus on a subset of the data that excludes emails sent to multiple recipients as well as self-addressed emails. This results in a dataset of 57,791 emails exchanged between January 2 and September 30, 2010. As for the first application, this forms a unimodal relational graph, where employees are the nodes and emails are time-stamped edges. All pairs among the 159 employees are assumed to be able to exchange emails throughout the entire observation period; no self-loops are permitted. Since 25,281 sender–receiver pairs are possible at each time point, it is essential to rely on case-control sampling procedures to make the analysis computationally feasible.

\paragraph{Model formulation.} The first of the two temporal dimensions under consideration is the internal temporality of relational mechanisms. This refers to the idea that mechanisms such as repetition, reciprocity, transitive closure, and cyclic closure can be captured and measured in more nuanced ways than through simple presence indicators or frequency counts. Specifically, these mechanisms can be assessed in terms of their internal timing, offering insights into the relative speed on which they influence the occurrence of events \citep{amati2024goodness}. Notably, we do not assume that these interarrival times (potentially aggregated) have a linear effect. Instead, we allow for the possibility that they contribute to the event occurrence rate non-linearly \citep{juozaitiene2024nodal}. This approach implies that we do not impose any specific decay function to account for the varying importance of events over time. Rather, we let the data determine the most appropriate functional form to describe the influence of the related interarrival times. 

Using a shifted counting process model, we further incorporate the effect of global covariates, allowing us to control for factors such as the time of day (our second temporal dimension of interest). The complete model is stated as follows:
\begin{equation}\label{eq:model-email}
	\begin{aligned}
    \lambda_{sr}(t) =& \lambda_{0} \times \exp \big\{ 
    f^{\textmd{rep}}\big(x_{sr}^{\textmd{rep},\Delta T}(t)\big) + 
    f^{\textmd{rec}}\big(x_{sr}^{\textmd{rec},\Delta T}(t)\big) + 
    f^{\textmd{trs}}\big(x_{sr}^{\textmd{trs},\Delta T}(t)\big) + 
    f^{\textmd{cyc}}\big(x_{sr}^{\textmd{cyc},\Delta T}(t)\big) \\
    &+
    f^{\textmd{dow}}\big(x_{sr}^{\textmd{dow},\Delta T}(t)\big) + 
    f^{\textmd{tod}}\big(x_{sr}^{\textmd{tod},\Delta T}(t)\big) + 
    f^{\textmd{time}}\big(x_{sr}^{\textmd{time},\Delta T}(t)\big) 
    \big\},\\
    \lambda^S_{sr}(\tilde{t}) =& \mathds{1}_{\{\tilde{t} \in [\tau_{sr},\tau_{sr} + T]\}} \lambda_{sr}(\tilde{t}-\tau_{sr}),
	\end{aligned}
\end{equation}
where \(x_{sr}^{\textmd{rep},\Delta T}(t) \), \(x_{sr}^{\textmd{rec},\Delta T}(t)\), \(x_{sr}^{\textmd{trs},\Delta T}(t)\), and \(x_{sr}^{\textmd{cyc},\Delta T}(t)\) are computed as in Table \ref{tab:endogenous}. Time of day, day of the week, and calendar time are also included as explanatory global variables. 

\begin{figure}[t]
    \centering
    \begin{minipage}{0.45\textwidth}
        \centering
        \includegraphics[width=\linewidth]{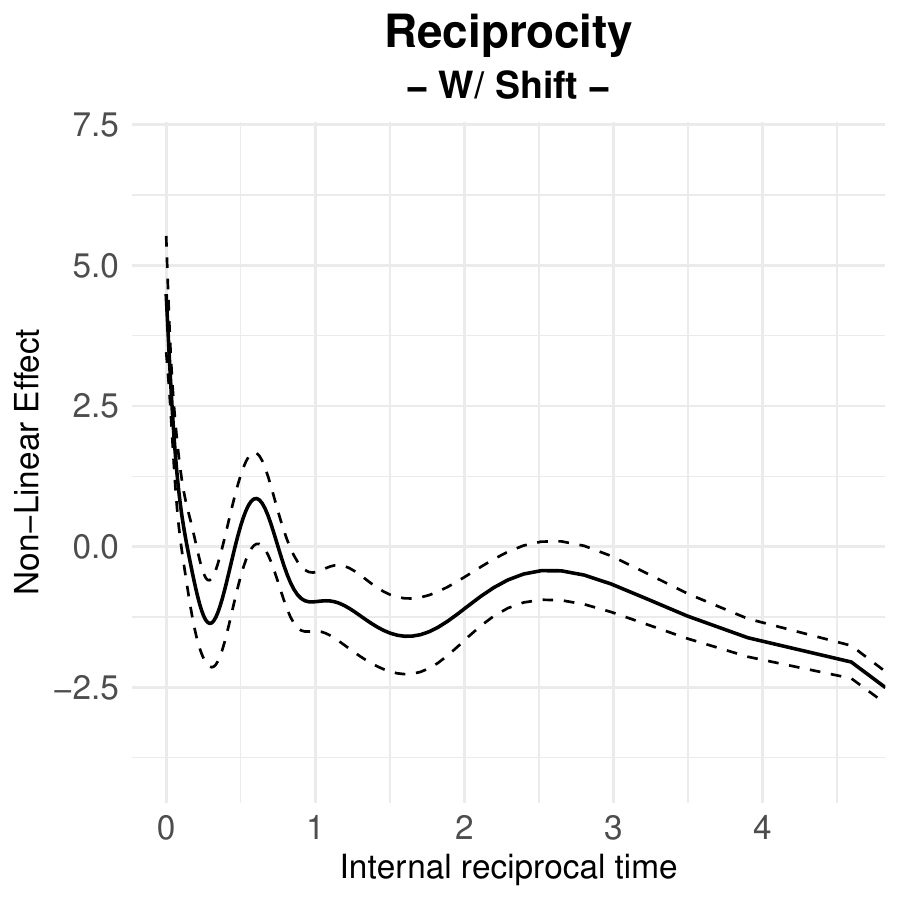}
    \end{minipage}
    \hfill
    \begin{minipage}{0.45\textwidth}
        \centering
        \includegraphics[width=\linewidth]{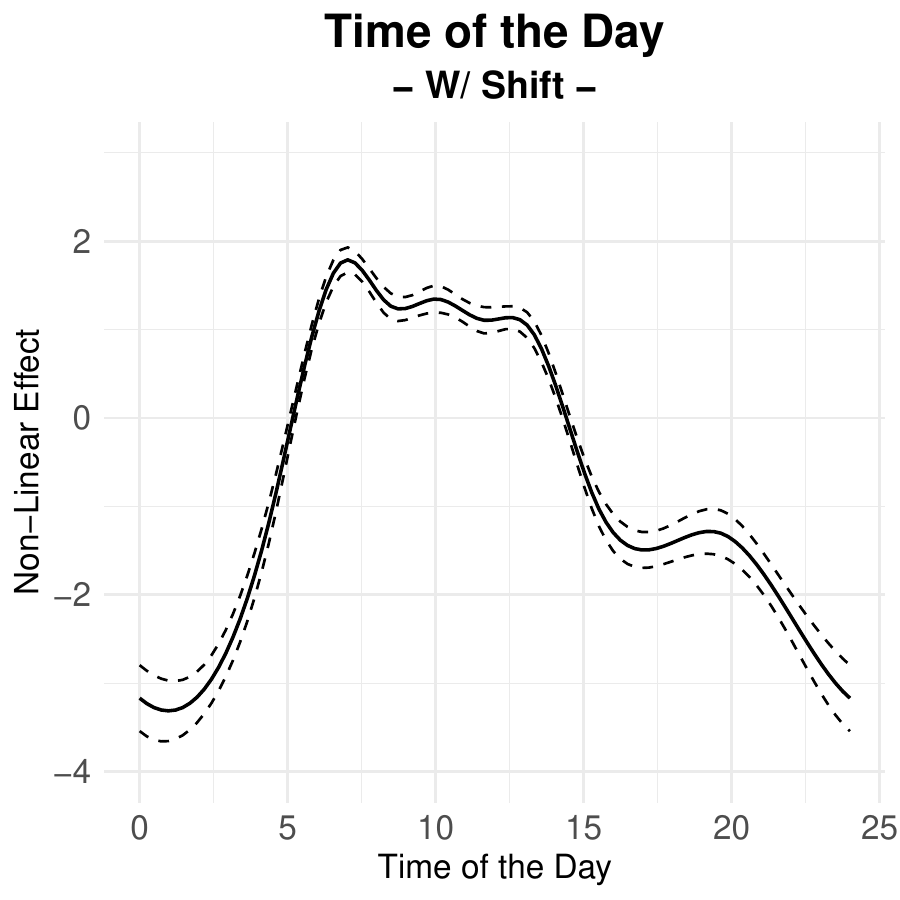}
        
    \end{minipage}
    \caption[Estimated non-linear effects of \texttt{Reciprocity} and \texttt{Time of the Day} in email communication application.]{\label{fig:reciprocity-non-linear}\textbf{Estimated non-linear effects of \texttt{Reciprocity} and \texttt{Time of the Day} in email communication application}. The reciprocity effect exhibits a general downward trend, with occasional peaks that  correspond to interactions with working hours. The time-of-day effect indicates that email activity varies over the day, with the main activity between 7am and 3pm.} 
\end{figure}

\paragraph{Inference procedure and results.} By employing the shifted case-control partial likelihood, we obtain fitted non-linear effects for all explanatory variables included in model~\eqref{eq:model-email}. All the corresponding plots are presented in the Supplementary Materials. 
We focus on the effects of \texttt{Reciprocity} and \texttt{Time of the Day}, both shown in Figure~\ref{fig:reciprocity-non-linear}. Since these are modelled as non-linear effects, we interpret only their general trends and changes over time. 

Looking at \texttt{Reciprocity}, we see a mostly decreasing pattern. In particular, if you have not yet received a response to your email within half an hour, this will reduce the instantaneous probability of a response by a dramatic factor of $e^5=150$. After that there are a series of peaks and troughs, which mainly relate to the relationship between relative reciprocity time and real time. In particular, the first trough happens at roughly 9 hours, which corresponds to the length of a working day. The peak thereafter at approximately 17 hours corresponds to people responding in the morning to emails sent at the late afternoon the day before. The plot for \texttt{Time of the Day} shows that email activity varies throughout the day. The peak email activity is around 7 and 8 in the morning, which starts to drop off after 2 in the afternoon. There seems to be a little blip after dinner time again, before dropping off for the night. When fewer or more emails are sent at certain times, this also affects the chance of sending replies. For this reason, including time helps us better understand the effect of reciprocity on its own. For completeness, the analysis without global covariates is available in the Supplementary Materials.

\section{Discussion}\label{sec:discussion}
REMs provide a flexible framework for modelling streams of relational events, which in the presence of continuous temporal monitoring systems are becoming more routinely available. A key feature of REMs is the assumption that the interactions occur in continuous underlying time. Furthermore, thanks to their recent extensions and flexible formulations, REMs enable interpretations of covariate effects that are not readily accessible through other temporal network models or deep learning approaches. Positioned at the intersection of event-history analysis and network modelling, REMs offer the added benefit of being compatible with the well-established inferential framework of generalized linear and additive models. This tutorial demonstrates how these models can be implemented with relative ease, and how they can accommodate varying levels of data granularity.

Certainly, REMs are not without challenges. Chief among them is the computational cost. Although some progress has been made in developing more efficient inference techniques, substantial work remains, particularly when incorporating random effects. Additionally, computing covariates -- often dependent on the history of past events -- remains highly computationally intensive. Moreover, as presented in this tutorial, REMs function primarily as descriptive models. They capture correlations between event occurrence rates and their drivers but do not incorporate causal inference techniques.

Relational event modelling remains an active area of research with numerous ongoing extensions from diverse perspectives. For instance, relational hyper event models constitute a significant and growing area within REM research. Increasingly, researchers are adopting relational hyper-event models as effective tools for analysing their empirical data \citep{lerner2025relational, hancean2025processed, barbagli12025co, bright2025examining}. Moreover, recent advancements aimed at increasing model flexibility at the REM level are now being incorporated into relational hyper-event models, helping to overcome their previous limitations to linear effects \citep{boschi2026beyond}. Recent work also explores the connection between relational event and relational state processes, bringing them into a single unifying framework \citep{juozaitiene2026inference}. Beyond these new methodological developments, more conceptual developments in the context of local independence \citep{thams2023local} and causality \citep{lembo2025causal} are deepening our theoretical understanding of relational event processes. 

In conclusion, the relational event model framework is a highly productive paradigm for dynamic network modelling. Across applied, methodological, and theoretical perspectives, it provides a flexible and principled way to study interaction processes as they unfold over time. Its combination of interpretability, generality, and statistical rigour makes it a valuable tool for analysing relational dynamics across a wide range of domains.

\section*{Reproducibility}
The GitHub repository (\href{https://github.com/martinaboschi/rem-tutorial/tree/main/03-Tutorial-Paper}{Link}) contains the complete \texttt{R} implementation of the methods described in Section~\ref{sec:applications}, including the source data, scripts for reproducing all figures, and code for both the synthetic and real-data analyses.

\section*{Conflict of interest}
None.

\section*{Acknowledgements}
This tutorial paper is the result of a series of practical tutorials delivered at various workshops and conferences. We are grateful to all the collaborators who contributed to the development and delivery of these tutorials: Melania Lembo, Dr. Federica Bianchi, Dr. Rūta Juozaitienė, Dr. Jürgen Lerner, Dr. Edoardo Filippi Mazzola, Dr. Francisco Javier Richter Mendoza, and Prof. Alessandro Lomi. We also sincerely thank the participants for their valuable feedback, which helped shape and improve this work.
Large language models have only been used to improve the phrasing and correct the grammar.

\FloatBarrier
\bibliography{bibliography}

@article{lerner2017third,
  title={The third man: Hierarchy formation in Wikipedia},
  author={Lerner, J{\"u}rgen and Lomi, Alessandro},
  journal={Applied network science},
  volume={2},
  number={1},
  pages={24},
  year={2017},
  publisher={Springer},
  doi={10.1007/s41109-017-0043-2}
}

@article{wood2011fast,
  title = {Fast stable restricted maximum likelihood and marginal likelihood estimation of semiparametric generalized linear models},
  journal = {Journal of the Royal Statistical Society (B)},
  volume = {73},
  number = {1},
  pages = {3-36},
  year = {2011},
  author = {S. N. Wood},
  doi = {10.1111/j.1467-9868.2010.00749.x},
}

@article{mulder2024latent,
  title={A latent variable approach for modeling relational data with multiple receivers},
  author={Mulder, Joris and Hoff, Peter D},
  journal={The Annals of Applied Statistics},
  volume={18},
  number={3},
  pages={2359--2381},
  year={2024},
  publisher={Institute of Mathematical Statistics},
  doi = {10.1214/24-AOAS1885},
}

@Manual{relevent_package,
  title = {relevent: Relational Event Models},
  author = {Carter T. Butts},
  year = {2023},
  note = {R package version 1.2-1},
  url = {https://CRAN.R-project.org/package=relevent},
}

@misc{statnet_relevent_workshop,
  title        = {Statnet Workshop: Modeling Relational Event Dynamics with Statnet},
  author = {
  Pavel N. Krivitsky and
  Mark S. Handcock and
  David R. Hunter and
  Carter T. Butts and
  Michal Bojanowski and
  Chad Klumb and
  Steven M. Goodreau and
  Martina Morris
  },
  howpublished = {\url{https://statnet.org/workshop-relevent/}},
  year         = {2025}
}

@article{borgan2015using,
  title={Using cumulative sums of martingale residuals for model checking in nested case-control studies},
  author={Borgan, {\O}rnulf and Zhang, Ying},
  journal={Biometrics},
  volume={71},
  number={3},
  pages={696--703},
  year={2015},
  publisher={Oxford University Press},
  doi={10.1111/biom.12308}
}

@misc{juozaitiene2026inference,
      title={Inference of large scale relational state processes}, 
      author={Rūta Juozaitienė and Ernst C. Wit},
      year={2026},
      eprint={2607.09363},
      archivePrefix={arXiv},
      primaryClass={stat.ME},
      doi={10.48550/arXiv.2607.09363}, 
}

@article{peduzzi1996simulation,
  title={A simulation study of the number of events per variable in logistic regression analysis},
  author={Peduzzi, Peter and Concato, John and Kemper, Elizabeth and Holford, Theodore R and Feinstein, Alvan R},
  journal={Journal of clinical epidemiology},
  volume={49},
  number={12},
  pages={1373--1379},
  year={1996},
  publisher={Elsevier},
  doi={10.1016/S0895-4356(96)00236-3}
}

@article{van2019sample,
  title={Sample size for binary logistic prediction models: beyond events per variable criteria},
  author={Van Smeden, Maarten and Moons, Karel GM and de Groot, Joris AH and Collins, Gary S and Altman, Douglas G and Eijkemans, Marinus JC and Reitsma, Johannes B},
  journal={Statistical methods in medical research},
  volume={28},
  number={8},
  pages={2455--2474},
  year={2019},
  publisher={SAGE Publications Sage UK: London, England},
  doi={10.1177/0962280218784726}
}

@article{yang2011detecting,
	author = {Yang, Tianbao and Chi, Yun and Zhu, Shenghuo and Gong, Yihong and Jin, Rong},
	doi = {10.1007/s10994-010-5214-7},
	journal = {Machine Learning},
	number = {2},
	pages = {157--189},
	title = {Detecting communities and their evolutions in dynamic social networks---a Bayesian approach},
	volume = {82},
	year = {2011}
}

@article{sarkar2005dynamic,
  title={Dynamic social network analysis using latent space models},
  author={Sarkar, Purnamrita and Moore, Andrew W},
  journal={Acm sigkdd explorations newsletter},
  volume={7},
  number={2},
  pages={31--40},
  year={2005},
  publisher={ACM New York, NY, USA}, 
  doi={10.1145/1117454.1117459}
}

@article{zhu2022quantifying,
  title={Quantifying Uncertainty for Temporal Motif Estimation in Graph Streams under Sampling},
  author={Zhu, Xiaojing and Kolaczyk, Eric D},
  journal={arXiv preprint arXiv:2202.10513},
  year={2022},
  doi={10.48550/arXiv.2202.10513}
}

@article{fox2016modeling,
  title={Modeling e-mail networks and inferring leadership using self-exciting point processes},
  author={Fox, Eric W and Short, Martin B and Schoenberg, Frederic P and Coronges, Kathryn D and Bertozzi, Andrea L},
  journal={Journal of the American Statistical Association},
  volume={111},
  number={514},
  pages={564--584},
  year={2016},
  publisher={Taylor \& Francis}, 
  doi={10.1080/01621459.2015.1135802}
}

@article{hawkes1971spectra,
  title={Spectra of some self-exciting and mutually exciting point processes},
  author={Hawkes, Alan G},
  journal={Biometrika},
  volume={58},
  number={1},
  pages={83--90},
  year={1971},
  publisher={Oxford University Press},
  doi={10.2307/2334319}
}

@article{snijders2010introduction,
  title={Introduction to stochastic actor-based models for network dynamics},
  author={Snijders, Tom AB and Van de Bunt, Gerhard G and Steglich, Christian EG},
  journal={Social networks},
  volume={32},
  number={1},
  pages={44--60},
  year={2010},
  publisher={Elsevier},
  doi={10.1016/j.socnet.2009.02.004}
}

@article{snijders2001statistical,
  title={The statistical evaluation of social network dynamics},
  author={Snijders, Tom AB},
  journal={Sociological methodology},
  volume={31},
  number={1},
  pages={361--395},
  year={2001},
  publisher={Wiley Online Library},
  doi={10.1111/0081-1750.00099}
}

@article{hanneke2010discrete,
  title={Discrete temporal models of social networks},
  author={Hanneke, Steve and Fu, Wenjie and Xing, Eric P},
  year={2010}, 
  doi={}
}

@article{faloutsos1999power, 
    author = {Faloutsos, Michalis and Faloutsos, Petros and Faloutsos, Christos}, 
    title = {On power-law relationships of the Internet topology}, 
    year = {1999}, 
    publisher = {Association for Computing Machinery},
    volume = {29}, 
    number = {4},
    doi = {10.1145/316194.316229},
    journal = {SIGCOMM Comput. Commun. Rev.},
    pages = {251–262}
}

@article{pastor2001epidemic,
  title={Epidemic spreading in scale-free networks},
  author={Pastor-Satorras, Romualdo and Vespignani, Alessandro},
  journal={Physical review letters},
  volume={86},
  number={14},
  pages={3200},
  year={2001},
  publisher={APS},
  doi={10.1103/PhysRevLett.86.3200}
}

@article{onnela2007structure,
author = {J.-P. Onnela  and J. Saramäki  and J. Hyvönen  and G. Szabó  and D. Lazer  and K. Kaski  and J. Kertész  and A.-L. Barabási },
title = {Structure and tie strengths in mobile communication networks},
journal = {Proceedings of the National Academy of Sciences},
volume = {104},
number = {18},
pages = {7332-7336},
year = {2007},
doi = {10.1073/pnas.0610245104}
}

@article{granovetter1973strength,
  title={The strength of weak ties},
  author={Granovetter, Mark S},
  journal={American journal of sociology},
  volume={78},
  number={6},
  pages={1360--1380},
  year={1973},
  publisher={University of Chicago Press},
  doi={10.1086/225469}
}

@article{sporns2006human,
    author = {Sporns, Olaf AND Tononi, Giulio AND Kötter, Rolf},
    journal = {PLOS Computational Biology},
    publisher = {Public Library of Science},
    title = {The Human Connectome: A Structural Description of the Human Brain},
    year = {2005},
    month = {09},
    volume = {1},
    doi = {10.1371/journal.pcbi.0010042},
    number = {4},
}

@article{jeong2000large,
	author = {Jeong, H. and Tombor, B. and Albert, R. and Oltvai, Z. N. and Barab{\'a}si, A. -L.},
	doi = {10.1038/35036627},
	isbn = {1476-4687},
	journal = {Nature},
	number = {6804},
	pages = {651--654},
	title = {The large-scale organization of metabolic networks},
	volume = {407},
	year = {2000},
}

@incollection{barabasi2016network-chapter1,
  author    = {P{\'o}sfai, M{\'a}rton and Barab{\'a}si, Albert-L{\'a}szl{\'o}},
  title     = {Section 1.2},
  booktitle = {Network Science},
  publisher = {Cambridge University Press},
  address   = {Cambridge, UK},
  year      = {2016},
  chapter   = {1.2},
  isbn      = {9781107076266}
}

@incollection{barabasi2016network-chapter10,
  author    = {P{\'o}sfai, M{\'a}rton and Barab{\'a}si, Albert-L{\'a}szl{\'o}},
  title     = {Section 1.2},
  booktitle = {Network Science},
  publisher = {Cambridge University Press},
  address   = {Cambridge, UK},
  year      = {2016},
  chapter   = {10.5},
  isbn      = {9781107076266}
}

@article{gillespie2001approximate,
  title={Approximate accelerated stochastic simulation of chemically reacting systems},
  author={Gillespie, Daniel T},
  journal={The Journal of chemical physics},
  volume={115},
  number={4},
  pages={1716--1733},
  year={2001},
  publisher={AIP Publishing},
  doi={10.1063/1.1378322}
}

@article{boschi2026beyond,
  title={Beyond linearity and time-homogeneity: relational hyper event models with time-varying non-linear effects},
  author={Boschi, Martina and Lerner, J{\"u}rgen and Wit, Ernst C},
  journal={Social Networks},
  volume={86},
  pages={120--137},
  year={2026},
  publisher={Elsevier}, 
  doi={10.1016/j.socnet.2026.02.001}
}

@article{harary1953notion,
  title={On the notion of balance of a signed graph.},
  author={Harary, Frank},
  journal={Michigan Mathematical Journal},
  volume={2},
  number={2},
  pages={143--146},
  year={1953},
  publisher={University of Michigan, Department of Mathematics},
  doi={10.1307/mmj/1028989917}
}

@article{thams2023local,
  title={Local independence testing for point processes},
  author={Thams, Nikolaj and Hansen, Niels Richard},
  journal={IEEE Transactions on Neural Networks and Learning Systems},
  volume={35},
  number={4},
  pages={4902--4910},
  year={2023},
  publisher={IEEE},
  doi={10.1109/TNNLS.2023.3335265}
}

@article{bauer2022smooth,
  title={A smooth dynamic network model for patent collaboration data},
  author={Bauer, Verena and Harhoff, Dietmar and Kauermann, G{\"o}ran},
  journal={AStA advances in statistical analysis},
  volume={106},
  number={1},
  pages={97--116},
  year={2022},
  publisher={Springer}, 
  doi={10.1007/s10182-021-00393-w}
}

@book{daley2003introduction,
  title={An introduction to the theory of point processes: volume I: elementary theory and methods},
  author={Daley, Daryl J and Vere-Jones, David},
  year={2003},
  publisher={Springer},
  doi={10.1007/b97277}
}

@article{vaida2005conditional,
  title={Conditional Akaike information for mixed-effects models},
  author={Vaida, Florin and Blanchard, Suzette},
  journal={Biometrika},
  volume={92},
  number={2},
  pages={351--370},
  year={2005},
  publisher={Biometrika Trust},
  doi={10.1093/biomet/92.2.351}
}

@Manual{survivalRpackage,
  title = {A Package for Survival Analysis in R},
  author = {Terry M Therneau},
  year = {2023},
  note = {R package version 3.5-3},
  url = {https://CRAN.R-project.org/package=survival},
}

@Book{therneau2000modeling,
  title = {Modeling Survival Data: Extending the {C}ox Model},
  author = {{Terry M. Therneau} and {Patricia M. Grambsch}},
  year = {2000},
  publisher = {Springer},
  address = {New York},
  isbn = {0-387-98784-3},
  doi={10.1007/978-1-4757-3294-8}
}

@Book{pebesma2023spatial,
  author = {Edzer Pebesma and Roger Bivand},
  title = {{Spatial Data Science: With applications in R}},
  year = {2023},
  publisher = {{Chapman and Hall/CRC}},
  url = {https://r-spatial.org/book/},
  doi = {10.1201/9780429459016},
}

@article{lembo2025causal,
  title={Causal drivers of dynamic networks},
  author={Lembo, Melania and Riccardi, Ester and Vinciotti, Veronica and Wit, Ernst C},
  journal={arXiv preprint arXiv:2503.03333},
  year={2025},
  doi={}
}

@article{hancean2025processed,
  title={Processed food intake assortativity in the personal networks of older adults},
  author={H{\^a}ncean, Marian-Gabriel and Lerner, J{\"u}rgen and Perc, Matja{\v{z}} and Molina, Jos{\'e} Luis and Geant{\u{a}}, Marius and Oan{\u{a}}, Iulian and Mih{\u{a}}il{\u{a}}, Bianca-Elena},
  journal={Scientific reports},
  volume={15},
  number={1},
  pages={10459},
  year={2025},
  publisher={Nature Publishing Group UK London},
  doi={10.1038/s41598-025-94969-0}
}

@article{barbagli12025co,
  title={Co-authors and Co-principal Investigators: Relational Hyperevent},
  author={Barbagli$^1$, Amin Gino Fabbrucci and Lerner, J{\"u}rgen and Boudourides, Moses},
  journal={Statistics for Innovation I: SIS 2025, Short Papers, Plenary, Specialized, and Solicited Sessions},
  pages={289},
  year={2025},
  publisher={Springer Nature},
  doi={10.1007/978-3-031-96736-8_48}
}

@article{bright2025examining,
  title={Examining co-offending and re-offending across crime categories using relational hyperevent models},
  author={Bright, David and Lerner, J{\"u}rgen and Putra Sadewo, Giovanni Radhitio},
  journal={Journal of Criminology},
  volume={58},
  number={1},
  pages={109--134},
  year={2025},
  publisher={SAGE Publications Sage UK: London, England},
  doi={10.1177/26338076241272864}
}

@article{boschi2026goodness,
  title={Goodness of fit in relational event models},
  author={Boschi, Martina and Wit, Ernst C},
  journal={Statistics and Computing},
  volume={36},
  number={1},
  pages={4},
  year={2026},
  publisher={Springer},
  doi={10.1007/s11222-025-10751-2}
}

@incollection{butts2017relational,
  title={A relational event approach to modeling behavioral dynamics},
  author={Butts, Carter T and Marcum, Christopher Steven},
  booktitle={Group processes: Data-driven computational approaches},
  pages={51--92},
  year={2017},
  publisher={Springer},
  doi={10.1007/978-3-319-48941-4_4}
}

@book{hastie1990gam,
  author    = {Trevor Hastie and Robert Tibshirani},
  title     = {Generalized Additive Models},
  year      = {1990},
  publisher = {Chapman and Hall},
}

@article{wood2016smoothing,
	author = {Simon N. Wood and Natalya Pya and Benjamin Säfken},
	title = {Smoothing Parameter and Model Selection for General Smooth Models},
	journal = {Journal of the American Statistical Association},
	volume = {111},
	number = {516},
	pages = {1548-1563},
	year  = {2016},
	publisher = {Taylor & Francis},
	doi = {10.1080/01621459.2016.1180986}
}

@article{quintane2014modeling,
  title={Modeling relational events: A case study on an open source software project},
  author={Quintane, Eric and Conaldi, Guido and Tonellato, Marco and Lomi, Alessandro},
  journal={Organizational Research Methods},
  volume={17},
  number={1},
  pages={23--50},
  year={2014},
  publisher={Sage Publications Sage CA: Los Angeles, CA},
  doi={10.1177/1094428113517007}
}

@article{breslow1993approximate,
  title={Approximate inference in generalized linear mixed models},
  author={Breslow, Norman E and Clayton, David G},
  journal={Journal of the American statistical Association},
  volume={88},
  number={421},
  pages={9--25},
  year={1993},
  publisher={Taylor \& Francis},
  doi={10.1080/01621459.1993.10594284}
}

@article{artico2023dynamic,
  title={Dynamic latent space relational event model},
  author={Artico, Igor and Wit, Ernst C},
  journal={Journal of the Royal Statistical Society Series A: Statistics in Society},
  volume={186},
  number={3},
  pages={508--529},
  year={2023},
  publisher={Oxford University Press US},
  doi={10.1093/jrsssa/qnad042}
}

@article{lerner2021dynamic,
  title={Dynamic network analysis of contact diaries},
  author={Lerner, J{\"u}rgen and Lomi, Alessandro and Mowbray, John and Rollings, Neil and Tranmer, Mark},
  journal={Social Networks},
  volume={66},
  pages={224--236},
  year={2021},
  publisher={Elsevier},
  doi={10.1016/j.socnet.2021.04.001}
}

@article{juozaitiene2024s,
  title={It’s about time: revisiting reciprocity and triadicity in relational event analysis},
  author={Juozaitien{\.e}, R{\=u}ta and Wit, Ernst C},
  journal={Journal of the Royal Statistical Society Series A: Statistics in Society},
  pages={qnae132},
  year={2024},
  publisher={Oxford University Press UK},
  doi={10.1093/jrsssa/qnae132}
}

@article{amati2019some,
  title={Some days are better than others: Examining time-specific variation in the structuring of interorganizational relations},
  author={Amati, Viviana and Lomi, Alessandro and Mascia, Daniele},
  journal={Social Networks},
  volume={57},
  pages={18--33},
  year={2019},
  publisher={Elsevier},
  doi={10.1016/j.socnet.2018.10.001}
}

@article{lerner2025relational,
  title={Relational hyperevent models for the coevolution of coauthoring and citation networks},
  author={Lerner, J{\"u}rgen and H{\^a}ncean, Marian-Gabriel and Lomi, Alessandro},
  journal={Journal of the Royal Statistical Society Series A: Statistics in Society},
  volume={188},
  number={2},
  pages={583--607},
  year={2025},
  publisher={Oxford University Press UK},
  doi={10.1093/jrsssa/qnae068}
}

@article{stadtfeld2017dynamic,
  title={Dynamic network actor models: Investigating coordination ties through time},
  author={Stadtfeld, Christoph and Hollway, James and Block, Per},
  journal={Sociological Methodology},
  volume={47},
  number={1},
  pages={1--40},
  year={2017},
  publisher={Sage Publications Sage CA: Los Angeles, CA},
  doi={10.1177/0081175017709295}
}

@article{lembo2025relational,
  title={Relational event models with global covariates: an application to bike sharing},
  author={Lembo, Melania and Juozaitien{\.e}, R{\=u}ta and Vinciotti, Veronica and Wit, Ernst C},
  journal={Journal of the Royal Statistical Society Series C: Applied Statistics},
  year={2025},
  publisher={Oxford University Press UK},
  doi={10.1093/jrsssc/qlaf058}
}

@article{cox1975partial,
  title={Partial likelihood},
  author={Cox, David R},
  journal={Biometrika},
  volume={62},
  number={2},
  pages={269--276},
  year={1975},
  publisher={Oxford University Press},
  doi={10.2307/2335362}
}

@article{arena2024bayesian,
  title={A Bayesian semi-parametric approach for modeling memory decay in dynamic social networks},
  author={Arena, Giuseppe and Mulder, Joris and Leenders, Roger Th AJ},
  journal={Sociological Methods \& Research},
  volume={53},
  number={3},
  pages={1201--1251},
  year={2024},
  publisher={Sage Publications Sage CA: Los Angeles, CA},
  doi={10.1177/00491241221113875}
}

@article{arastuie2020chip,
  title={CHIP: A Hawkes process model for continuous-time networks with scalable and consistent estimation},
  author={Arastuie, Makan and Paul, Subhadeep and Xu, Kevin},
  journal={Advances in neural information processing systems},
  volume={33},
  pages={16983--16996},
  year={2020}
}

@article{shiels2014biology,
  title={Biology and impacts of Pacific island invasive species. 11. Rattus rattus, the black rat (Rodentia: Muridae)},
  author={Shiels, Aaron B and Pitt, William C and Sugihara, Robert T and Witmer, Gary W},
  journal={Pacific Science},
  volume={68},
  number={2},
  pages={145--184},
  year={2014},
  publisher={BioOne}, 
  doi={/10.2984/68.2.1}
}

@article{carter2002review,
  title={A review of the literature on the worldwide distribution, spread of, and efforts to eradicate the coypu (Myocastor coypus)},
  author={Carter, Jacoby and Leonard, Billy P},
  journal={Wildlife Society Bulletin},
  pages={162--175},
  year={2002},
  publisher={JSTOR},
  url={https://www.jstor.org/stable/3784650}
}

@article{salgado2018raccoon,
  title={Is the raccoon (Procyon lotor) out of control in Europe?},
  author={Salgado, Iv{\'a}n},
  journal={Biodiversity and Conservation},
  volume={27},
  number={9},
  pages={2243--2256},
  year={2018},
  publisher={Springer},
  doi={10.1007/s10531-018-1535-9}
}

@Manual{geosphereRpackage,
	title = {Package geosphere: Spherical Trigonometry, , 1(7)},
	author = {Hijmans, R. J. and Karney, C. and Williams, E. and Vennes C.},
	year = {2017},
	note = {R package version 1.5-7},
	url = {https://cran.r-project.org/web/packages/geosphere/index.html}
}

@article{watanabe2011miroc,
	title={MIROC-ESM 2010: Model description and basic results of CMIP5-20c3m experiments},
	author={Watanabe, Shingo and Hajima, T and Sudo, K and Nagashima, T and Takemura, T and Okajima, H and Nozawa, Toru and Kawase, H and Abe, M and Yokohata, TJGMD and others},
	journal={Geoscientific Model Development},
	volume={4},
	number={4},
	pages={845--872},
	year={2011},
	publisher={Copernicus GmbH},
    doi={10.5194/gmd-4-845-2011}
}

@article{amati2024goodness,
  title={A goodness of fit framework for relational event models},
  author={Amati, Viviana and Lomi, Alessandro and Snijders, Tom AB},
  journal={Journal of the Royal Statistical Society Series A: Statistics in Society},
  volume={187},
  number={4},
  pages={967--988},
  year={2024},
  publisher={Oxford University Press UK},
  doi={doi.org/10.1093/jrsssa/qnae016}
}

@article{michalski2014seed,
  title={Seed selection for spread of influence in social networks: Temporal vs. static approach},
  author={Michalski, Rados{\l}aw and Kajdanowicz, Tomasz and Br{\'o}dka, Piotr and Kazienko, Przemys{\l}aw},
  journal={New Generation Computing},
  volume={32},
  pages={213--235},
  year={2014},
  publisher={Springer},
  doi={10.1007/s00354-014-0402-9}
}

@article{wit2012all,
	title={‘All models are wrong...’: an introduction to model uncertainty},
	author={Wit, Ernst and Heuvel, Edwin van den and Romeijn, Jan-Willem},
	journal={Statistica Neerlandica},
	volume={66},
	number={3},
	pages={217--236},
	year={2012},
	publisher={Wiley Online Library},
    doi={10.1111/j.1467-9574.2012.00530.x}
}

@Manual{tigris,  title = {tigris: Load Census TIGER/Line Shapefiles},  author = {Kyle Walker},  year = {2024},  note = {R package version 2.1},  url = {https://CRAN.R-project.org/package=tigris},}

@phdthesis{schneble2021new,
	author = {Marc Schneble},
	title = {New Approaches in Statistical Modeling},
	school = {Ludwig-Maximilians-Universität München},
	year = {2021},
    doi={10.5282/edoc.28868}
}

@article{vieira2024fast,
  title={Fast meta-analytic approximations for relational event models: Applications to data streams and multilevel data},
  author={Vieira, Fabio and Leenders, Roger and Mulder, Joris},
  journal={Journal of Computational Social Science},
  volume={7},
  number={2},
  pages={1823--1859},
  year={2024},
  publisher={Springer},
  doi={10.1007/s42001-024-00290-7}
}

@article{filippi2024modeling,
	title={Modeling non-linear effects with neural networks in Relational Event Models},
	author={Filippi-Mazzola, Edoardo and Wit, Ernst C},
	journal={Social Networks},
	volume={79},
	pages={25--33},
	year={2024},
	publisher={Elsevier},
    doi={10.1016/j.socnet.2024.05.004}
}

@book{hastie2009elements,
	title={The elements of statistical learning: data mining, inference, and prediction},
	author={Hastie, Trevor and Tibshirani, Robert and Friedman, Jerome H and Friedman, Jerome H},
	volume={2},
	year={2009},
	publisher={Springer}
}

@book{dobson2018introduction,
	title={An introduction to generalized linear models},
	author={Dobson, Annette J and Barnett, Adrian G},
	year={2018},
	publisher={Chapman and Hall/CRC}
}

@article{kreiss2024testing,
	title={Testing For Global Covariate Effects in Dynamic Interaction Event Networks},
	author={Kreiss, Alexander and Mammen, Enno and Polonik, Wolfgang},
	journal={Journal of Business \& Economic Statistics},
	volume={42},
	number={2},
	pages={457--468},
	year={2024},
	publisher={Taylor \& Francis},
    doi={10.1080/07350015.2023.2263537}
}

@book{wood2017generalized,
	title = {Generalized Additive Models: An Introduction with R},
	edition = {2},
	pages = {397},
	author = {Wood, Simon N},
	langid = {english},
	publisher = {Chapman and Hall/CRC.},
	year = {2017},
    doi = {10.1201/9781315370279}
}

@article{barbieri2009trading,
	title={Trading data: Evaluating our assumptions and coding rules},
	author={Barbieri, Katherine and Keshk, Omar MG and Pollins, Brian M},
	journal={Conflict Management and Peace Science},
	volume={26},
	number={5},
	pages={471--491},
	year={2009},
	publisher={SAGE Publications Sage UK: London, England},
    doi={10.1177/0738894209343887}
}

@article{pattison2002neighborhood,
	title={Neighborhood--based models for social networks},
	author={Pattison, Philippa and Robins, Garry},
	journal={Sociological methodology},
	volume={32},
	number={1},
	pages={301--337},
	year={2002},
	publisher={Wiley Online Library},
    doi={10.1111/1467-9531.00119}
}

@article{meijerink2022dynamic,
	title={Dynamic relational event modeling: Testing, exploring, and applying},
	author={Meijerink-Bosman, Marlyne and Leenders, Roger and Mulder, Joris},
	journal={PLoS One},
	volume={17},
	number={8},
	pages={e0272309},
	year={2022},
	publisher={Public Library of Science San Francisco, CA USA},
    doi={10.1371/journal.pone.0272309}
}

@article{pilny2016illustration,
	title={An illustration of the relational event model to analyze group interaction processes.},
	author={Pilny, Andrew and Schecter, Aaron and Poole, Marshall Scott and Contractor, Noshir},
	journal={Group Dynamics: Theory, Research, and Practice},
	volume={20},
	number={3},
	pages={181},
	year={2016},
	publisher={Educational Publishing Foundation},
    doi={10.1037/gdn0000042}
}

@article{hulme2021unwelcome,
	title={Unwelcome exchange: International trade as a direct and indirect driver of biological invasions worldwide},
	author={Hulme, Philip E},
	journal={One Earth},
	volume={4},
	number={5},
	pages={666--679},
	year={2021},
	publisher={Elsevier},
    doi={10.1016/j.oneear.2021.04.015}
}

@article{seebens2021around,
	title={Around the world in 500 years: Inter-regional spread of alien species over recent centuries},
	author={Seebens, Hanno and Blackburn, Tim M and Hulme, Philip E and van Kleunen, Mark and Liebhold, Andrew M and Orlova-Bienkowskaja, Marina and Py{\v{s}}ek, Petr and Schindler, Stefan and Essl, Franz},
	journal={Global Ecology and Biogeography},
	volume={30},
	number={8},
	pages={1621--1632},
	year={2021},
	publisher={Wiley Online Library},
    doi={10.1111/geb.13325}
}

@article{seebens2017no,
	title={No saturation in the accumulation of alien species worldwide},
	author={Seebens, Hanno and Blackburn, Tim M and Dyer, Ellie E and Genovesi, Piero and Hulme, Philip E and Jeschke, Jonathan M and Pagad, Shyama and Py{\v{s}}ek, Petr and Winter, Marten and Arianoutsou, Margarita and others},
	journal={Nature communications},
	volume={8},
	number={1},
	pages={1--9},
	year={2017},
	publisher={Nature Publishing Group},
    doi={10.1038/ncomms14435}
}

@article{seebens2018global,
	title={Global rise in emerging alien species results from increased accessibility of new source pools},
	author={Seebens, Hanno and Blackburn, Tim M and Dyer, Ellie E and Genovesi, Piero and Hulme, Philip E and Jeschke, Jonathan M and Pagad, Shyama and Py{\v{s}}ek, Petr and van Kleunen, Mark and Winter, Marten and others},
	journal={Proceedings of the National Academy of Sciences},
	volume={115},
	number={10},
	pages={E2264--E2273},
	year={2018},
	publisher={National Acad Sciences},
    doi={}
}

@article{snijders1996stochastic,
	title={Stochastic actor-oriented models for network change},
	author={Snijders, Tom AB},
	journal={Journal of mathematical sociology},
	volume={21},
	number={1-2},
	pages={149--172},
	year={1996},
	publisher={Taylor \& Francis},
    doi={10.1080/0022250X.1996.9990178}
}

@article{robins2001random,
	title={Random graph models for temporal processes in social networks},
	author={Robins, Garry and Pattison, Philippa},
	journal={Journal of Mathematical Sociology},
	volume={25},
	number={1},
	pages={5--41},
	year={2001},
	publisher={Taylor \& Francis},
    doi={10.1080/0022250X.2001.9990243}
}

@book{aalen2008survival,
	title={Survival and event history analysis: a process point of view},
	author={Aalen, Odd and Borgan, Ornulf and Gjessing, Hakon},
	year={2008},
	publisher={Springer Science \& Business Media},
    doi={10.1007/978-0-387-68560-1}
}

@article{stadtfeld2017interactions,
	title={Interactions, actors, and time: Dynamic network actor models for relational events},
	author={Stadtfeld, Christoph and Block, Per},
	journal={Sociological Science},
	volume={4},
	pages={318--352},
	year={2017},
	publisher={Society for Sociological Science},
    doi={10.15195/v4.a14}
}

@misc{mulder2022using,
  author       = {Goana, C. A.},
  title        = {Using Relational Event Modelling to Predict the Activity of Twitter Users},
  year         = {2022},
  month        = {June},
  day          = {30},
  note         = {Bachelor's thesis}
}

@article{artico2023fast,
	title={Fast inference of latent space dynamics in huge relational event networks},
	author={Artico, Igor and Wit, Ernst},
	journal={arXiv preprint arXiv:2303.17460},
	year={2023},
    doi={10.48550/arXiv.2303.17460}
}

@article{lerner2020reliability,
	title={Reliability of relational event model estimates under sampling: How to fit a relational event model to 360 million dyadic events},
	author={Lerner, J{\"u}rgen and Lomi, Alessandro},
	journal={Network science},
	volume={8},
	number={1},
	pages={97--135},
	year={2020},
	publisher={Cambridge University Press},
    doi={10.1017/nws.2019.57}
}

@article{lerner2023micro,
	title={Micro-level network dynamics of scientific collaboration and impact: relational hyperevent models for the analysis of coauthor networks},
	author={Lerner, J{\"u}rgen and H{\^a}ncean, Marian-Gabriel},
	journal={Network Science},
	volume={11},
	number={1},
	pages={5--35},
	year={2023},
	publisher={Cambridge University Press},
    doi={10.1017/nws.2022.29}
}

@article{bianchi2023ties,
	title={From ties to events in the analysis of interorganizational exchange relations},
	author={Bianchi, Federica and Lomi, Alessandro},
	journal={Organizational research methods},
	volume={26},
	number={3},
	pages={524--565},
	year={2023},
	publisher={SAGE Publications Sage CA: Los Angeles, CA},
    doi={10.1177/10944281211058469}
}

@article{tranmer2015using,
	title={Using the relational event model (REM) to investigate the temporal dynamics of animal social networks},
	author={Tranmer, Mark and Marcum, Christopher Steven and Morton, F Blake and Croft, Darren P and de Kort, Selvino R},
	journal={Animal behaviour},
	volume={101},
	pages={99--105},
	year={2015},
	publisher={Elsevier},
    doi={10.1016/j.anbehav.2014.12.005}
}

@article{juozaitiene2024nodal,
	title={Nodal Heterogeneity can Induce Ghost Triadic Effects in Relational Event Models},
	author={Juozaitien{\.e}, R{\=u}ta and Wit, Ernst C},
	journal={Psychometrika},
	pages={1--21},
	year={2024},
	publisher={Springer},
    doi={10.1007/s11336-024-09952-x}
}

@article{butts2023relational,
	title={Relational event models in network science},
	author={Butts, Carter T and Lomi, Alessandro and Snijders, Tom AB and Stadtfeld, Christoph},
	journal={Network Science},
	volume={11},
	number={2},
	pages={175--183},
	year={2023},
	publisher={Cambridge University Press},
    doi={10.1017/nws.2023.9}
}

@article{bianchi2024relational,
  title={Relational event modeling},
  author={Bianchi, Federica and Filippi-Mazzola, Edoardo and Lomi, Alessandro and Wit, Ernst C},
  journal={Annual Review of Statistics and Its Application},
  volume={11},
  year={2024},
  publisher={Annual Reviews},
  doi={10.1146/annurev-statistics-040722-060248}
}

@inproceedings{michalski2011matching,
	title={Matching organizational structure and social network extracted from email communication},
	author={Michalski, Rados{\l}aw and Palus, Sebastian and Kazienko, Przemys{\l}aw},
	booktitle={Business Information Systems: 14th International Conference, BIS 2011, Pozna{\'n}, Poland, June 15-17, 2011. Proceedings 14},
	pages={197--206},
	year={2011},
	organization={Springer},
    doi={10.1007/978-3-642-21863-7_17}
}

@article{juozaitiene2022non,
	title={Non-parametric estimation of reciprocity and triadic effects in relational event networks},
	author={Juozaitien{\.e}, R{\=u}ta and Wit, Ernst C},
	journal={Social Networks},
	volume={68},
	pages={296--305},
	year={2022},
	publisher={Elsevier},
    doi={10.1016/j.socnet.2021.08.004}
}

@article{uzaheta2023random,
	title={Random effects in dynamic network actor models},
	author={Uzaheta, Alvaro and Amati, Viviana and Stadtfeld, Christoph},
	journal={Network Science},
	pages={1--18},
	year={2023},
	publisher={Cambridge University Press},
    doi={10.1017/nws.2022.37}
}

@article{filippi2023drivers,
	doi = {10.1371/journal.pone.0283247},
	author = {Filippi-Mazzola, Edoardo AND Bianchi, Federica AND Wit, Ernst C.},
	journal = {PLOS ONE},
	publisher = {Public Library of Science},
	title = {Drivers of the decrease of patent similarities from 1976 to 2021},
	year = {2023},
	month = {03},
	volume = {18},
	url = {10.1371/journal.pone.0283247},
	pages = {1-13},
	number = {3},
	
}

@article{vu2017relational,
	title = {Relational event models for longitudinal network data with an application to interhospital patient transfers},
	volume = {36},
	issn = {0277-6715, 1097-0258},
	url = {https://onlinelibrary.wiley.com/doi/10.1002/sim.7247},
	doi = {10.1002/sim.7247},
	pages = {2265--2287},
	number = {14},
	journal = {Statistics in Medicine},
	author = {Vu, Duy and Lomi, Alessandro and Mascia, Daniele and Pallotti, Francesca},
	year = {2017},
}

@article{juozaitiene2023analysing,
title={Analysing ecological dynamics with relational event models: The case of biological invasions},
author={Juozaitien{\.e}, R{\=u}ta and Seebens, Hanno and Latombe, Guillaume and Essl, Franz and Wit, Ernst C},
journal={Diversity and Distributions},
volume={29},
number={10},
pages={1208--1225},
year={2023},
publisher={Wiley Online Library},
doi={10.1111/ddi.13752}
}

@article{filippi2024stochastic,
  title={A stochastic gradient relational event additive model for modelling US patent citations from 1976 to 2022},
  author={Filippi-Mazzola, Edoardo and Wit, Ernst C},
  journal={Journal of the royal statistical society series C: applied statistics},
  volume={73},
  number={4},
  pages={1008--1024},
  year={2024},
  publisher={Oxford University Press UK}, 
  doi={10.1093/jrsssc/qlae023}
}

@article{boschi2025mixed,
  title={Mixed additive modelling of global alien species co-invasions of plants and insects},
  author={Boschi, Martina and Juozaitien{\.e}, R{\=u}ta and Wit, Ernst C},
  journal={Journal of the Royal Statistical Society Series C: Applied Statistics},
  volume={75},
  number={1},
  pages={57--78},
  year={2025},
  publisher={Oxford University Press UK},
  doi={10.1093/jrsssc/qlaf034}
}

@article{butts2008relational,
	title = {A Relational Event Framework for Social Action},
	author = {Butts, Carter T.},
	volume = {38},
	issn = {0081-1750, 1467-9531},
	doi = {10.1111/j.1467-9531.2008.00203.x},
	pages = {155--200},
	number = {1},
	journal = {Sociological Methodology},
	year = {2008},
	langid = {english},
}

@article{perry2013point,
	title = {Point process modelling for directed interaction networks},
	author = {Perry, Patrick O. and Wolfe, Patrick J.},
	volume = {75},
	issn = {13697412},
	url = {https://onlinelibrary.wiley.com/doi/10.1111/rssb.12013},
	doi = {10.1111/rssb.12013},
	pages = {821--849},
	number = {5},
	journal = {Journal of the Royal Statistical Society: Series B (Statistical Methodology)},
	urldate = {2022-09-22},
	date = {2013-11},
	year={2013},
	langid = {english},
}

@article{brandenberger2019predicting,
	title = {Predicting Network Events to Assess Goodness of Fit of Relational Event Models},
	volume = {27},
	issn = {1047-1987, 1476-4989},
	url = {https://www.cambridge.org/core/product/identifier/S104719871900010X/type/journal_article},
	doi = {10.1017/pan.2019.10},
	pages = {556--571},
	number = {4},
	journal = {Political Analysis},
	author = {Brandenberger, Laurence},
	year={2019}
}

@article{borgan1995methods,
	title={Methods for the analysis of sampled cohort data in the Cox proportional hazards model},
	author={Borgan, Ornulf and Goldstein, Larry and Langholz, Bryan},
	journal={The Annals of Statistics},
	pages={1749--1778},
	year={1995},
    doi={10.1214/AOS/1176324322}
}

@article{vu2015relational,
	title = {Relational event models for social learning in MOOCs},
	journal = {Social Networks},
	volume = {43},
	pages = {121-135},
	year = {2015},
	issn = {0378-8733},
	doi = {10.1016/j.socnet.2015.05.001},
	url = {https://www.sciencedirect.com/science/article/pii/S0378873315000477},
	author = {Duy Vu and Philippa Pattison and Garry Robins}
}

\end{document}